\documentclass[twocolumn,aps,prl,showpacs,amsmath,superscriptaddress,longbibliography,notitlepage]{revtex4-1}
\usepackage{amssymb}
\usepackage{textcomp,booktabs}
\usepackage[usenames,dvipsnames]{color}
\usepackage{colortbl}
\definecolor{mygray}{gray}{.96}
\definecolor{mypink}{rgb}{.99,.91,.95}
\definecolor{mycyan}{cmyk}{.3,0,0,0}
\usepackage{multirow}
\usepackage{amsfonts}
\usepackage{amsmath}
\usepackage{amssymb}
\usepackage{amsbsy}
\usepackage{graphicx}
\usepackage{bm}
\usepackage{color}
\usepackage{mathrsfs}
\usepackage[caption=false]{subfig}
\usepackage[pdftex,colorlinks=red]{hyperref}
\usepackage{comment}
\usepackage{xcolor}
\usepackage{float}
\usepackage{epstopdf}
\usepackage[normalem]{ulem}
\usepackage{verbatim}
\usepackage{xcolor}
\usepackage{amsmath}
\usepackage[normalem]{ulem}

\usepackage{verbatim}
\usepackage{xcolor}
\usepackage{amsmath}
\usepackage{tikz}
\usetikzlibrary{positioning, shapes.geometric}
\usepackage{makecell}
\usepackage{rotating}
\usepackage{soul}

\usepackage{tabularx} % for 'tabularx' env. and 'X' col. type
\usepackage{ragged2e} % for \RaggedRight macro
\usepackage{booktabs} % for \toprule, \m

\def\red{\textcolor{red}}

\def\CH{\textcolor{magenta}}

\def\red{\textcolor[rgb]{0,0,0}}

\begin{document}

\title{Dimensional Transmutation from Non-Hermiticity}

\author{Hui Jiang}  \email{phyhuij@nus.edu.sg}
\affiliation{Department of Physics, National University of Singapore, Singapore 117551, Republic of Singapore}

\author{Ching Hua Lee}  \email{phylch@nus.edu.sg}
\affiliation{Department of Physics, National University of Singapore, Singapore 117551, Republic of Singapore}
\affiliation{Joint School of National University of Singapore and Tianjin University, International Campus of Tianjin University, Binhai New City, Fuzhou 350207, China}

\begin{abstract}
Dimensionality plays a fundamental role in the classification of novel phases and their responses. % and their possible physical properties. 
In generic lattices of 2D and beyond, however, we found that non-Hermitian couplings do not merely distort the Brillouin zone (BZ), but can in fact alter its effective dimensionality. This is due to the fundamental non-commutativity of multi-dimensional non-Hermitian pumping, which obstructs the usual formation of a generalized complex BZ. As such, basis states are forced to assume ``entangled'' profiles that are orthogonal in a lower dimensional effective BZ, completely divorced from any vestige of lattice Bloch states unlike conventional skin states. Characterizing this reduced dimensionality is an emergent winding number intimately related to the homotopy of non-contractible spectral paths. 
%We show how the effective Brillouin zone can be constructed by avoiding homotopy generators exhibiting nontrivial spectral winding, and illustrate how this 
We illustrate this dimensional transmutation through a 2D model whose topological zero modes are protected by a 1D, not 2D, topological invariant. Our findings can be readily demonstrated via the bulk properties of non-reciprocally coupled platforms such as circuit arrays, and provokes us to rethink about the fundamental role of geometric obstruction in the dimensional classification of topological states. 
\end{abstract}

\date{\today}
\maketitle
%\newpage

\noindent{\it Introduction.--} \red{Dimensionality is fundamental in determining possible physical phenomena, such as in %. At the cosmological level, our $(3+1)$-D space-time have constrained possible gravitational and string theories~\cite{bailin1987kaluza,maldacena1999large,witten1998anti,gubser1998gauge}. %%In field theory, the rich conformal symmetry in 2D has led to great advancements in percolation theory and quantum entanglement~\cite{calabrese2009entanglement,cardy2001conformal,bauer2002slekappa,vidal2003entanglement,latorre2003ground,its2005entanglement,simmons2007percolation,solodukhin2008entanglement,cardy2010unusual,saberi2015recent}. Condensed matter phenomena are also intimately tied to dimensionality, such as in % the scaling of heat capacities~\cite{marder2010condensed}, 
Anderson localization~\cite{lee1981anderson,PhysRevB.26.4735,PhysRevLett.100.036803} and critical phase transitions~\cite{aharony1976lowering,ALEKHIN200543}.} In particular, symmetry-protected topological phases can be systematically classified based on Bott periodicity in the number of dimensions, via the tenfold-way~\cite{PhysRevB.55.1142,PhysRevB.78.195125,ryu2010topological,kitaev2009periodic,stone2010symmetries,morimoto2013topological,PhysRevB.88.075142}. More recently, this classification is greatly enriched~\cite{gong2018topological,liu2019topological,li2019geometric,kawabata2019symmetry,zhou2019periodic,li2021homotopical} in non-Hermitian lattices, which %are open systems with gain/loss that 
are increasingly studied theoretically~\cite{bergholtz2021exceptional,hu2011absence,
esaki2011edge,lee2014heralded,lee2014entanglement,leykam2017edge,Hu2017EP,PhysRevB.97.045106,PhysRevA.97.052115,Hui2018nonH,PhysRevB.97.115436,shen2018topological,borgnia2020nonH,pan2020non,Yao2018nonH2D,PhysRevB.99.041406,lee2021many,shen2021non,kawabata2021nonunitary,lee2022exceptional,qin2022non} and in photonic, mechanical, electrical and cold-atom experiments~\cite{gao2015observation,zeuner2015observation,xu2016topological,xiao2017observation,feng2017non,weimann2017topologically,parto2018edge,weidemann2020topological,helbig2020generalized,hofmann2020reciprocal,xiao2020non,ghatak2020observation,zou2021observation,stegmaier2021topological,wang2021topological,gao2021non,zhang2021observation,shang2022experimental,weidemann2022topological,li2022non,zhang2022observation,rosa2022observing,zhang2023electrical}.

Usually, it is taken for granted that the dimensionality of the topological invariant~\cite{TKNN,KOHMOTO1985343,qi2008,schindler2018higher,lee2018tidal,Lee2019hybrid, Luo2019HOnonH,tuloup2020nonlinearity,jiang2022filling} coincides with that of the physical space. This is because they are defined in reciprocal (momentum) space, which should be of the same dimension as the physical lattice, at least in Euclidean space% which underlies almost all realistic systems
\footnote{This may not be the case for hyperbolic lattices~\cite{maciejko2021hyperbolic,maciejko2022automorphic,boettcher2022crystallography}
, which can be realized in circuit setups~\cite{kollar2019hyperbolic,boettcher2020quantum,lenggenhager2021electric}}. \red{Even among enigmatic non-Hermitian phenomena featured lately}%%, where the band structure and topological invariants are substantially modified by boundaries and impurities
\cite{zhang2022review,lin2023topological,yao2018edge,xiong2018does,Lee2019anatomy,song2019realspace,song2019non,longhi2019probing,zhang2019correspondence,yokomizo2019non,yang2019auxiliary,lee2020ultrafast,longhi2020non,claes2020skin,okuma2020topological,PhysRevB.104.L161106,li2021impurity, PhysRevLett.126.176601,li2021quantized,PhysRevLett.127.116801,li2022nonherm,yang2022designing,rafi2022system,xue2022non,zhang2022real,rafi2022critical,tai2022zoology,PhysRevB.104.125109,kawabata2020higher,longhi2022self,gu2022transient,PhysRevB.102.241202,PhysRevB.103.L140201,qin2023universal,lei2023mathcal,guo2023non}, the highly distorted effective Brillouin zone (BZ) is still indexed by states living in the same dimensionality.

Yet we discover, surprisingly, that in 2D and beyond, non-Hermiticity %does not merely modify the effective BZ, but 
can in fact \emph{change} the effective BZ dimensionality. This holds true for generic non-Hermitian lattices beyond the simplest monoclinic structures, whenever the lattice is bounded (as all realistic lattices should be). Hence, the effective band structure of a $D$-dim lattice may in reality live in $D'$$<$$D$ dimensions, and be classified by $D'$ instead of $D$-dim topology. 

Underlying this dimensional transmutation %i.e. breakdown of dimensional correspondence between real and momentum space lattices 
is a hitherto unnoticed geometric obstruction, specifically the non-commutativity in the equilibration of \red{states that have been directionally amplified i.e. ``pumped'' by the non-Hermitian skin effect (NHSE)} along different directions. \red{This ``equilibration process'' is the mathematical elimination of non-reciprocity upon switching to the generalized Brillouin zone, conventionally constructed one dimension at a time.} Fundamentally resulting from emergent non-locality~\footnote{which arises even if all couplings are local}\cite{yao2018edge,lee2020unraveling}, it is reminiscent of the non-commutativity of magnetic translations from the non-locality of flux threading, as epitomized by the Aharonov-Bohm effect~\cite{Berry45,bachtold1999aharonov,
gou2020tunable}. %%Below, we demonstrate the necessity of dimensional transmutation through concrete examples, and show how the effective BZ manifold can be constructed from homotopy loops in the original BZ that avoid the non-commutativity. 

\noindent{\it Non-Hermitian equilibration and its non-commutativity.--} %%We first review the equilibration of non-Hermitian pumping, and reveal how its non-commutative nature can fundamentally alter the momentum-space lattice (the BZ), including its dimensionality. 
~%\footnote{Here, non-Hermitian pumping is taken to be broader in scope than the non-Hermitian skin effect, encompassing critical~\cite{} or dimensionally transmutated cases without a well-defined skin depth.}
Consider a generic lattice Hamiltonian under open boundary conditions (OBCs)%described by the Hamiltonian 
\begin{equation}
H=\sum_{\bm x;\alpha,\beta}\sum_{\{\bf e\}}h_{\bold e}^{\alpha\beta}c^\dagger_{\bm x+\bold e,\alpha} c_{\bm x,\beta},
\end{equation}
where $\bold e$ ranges over all coupling displacements from each unit cell, and $\alpha,\beta$ are sublattice components. When the couplings have asymmetric amplitudes $|h_{\bold e}^{\alpha\beta}|\neq |h_{-\bold e}^{\beta\alpha}|$, %%which can physically arise from the combination of flux and gain or loss, 
all left/right moving states are invariably attenuated/amplified by a factor of $|h_{\bold e}^{\alpha\beta}|/|h_{-\bold e}^{\beta\alpha}|$ per unit cell shifted~\cite{HN1996prl,HN1997prb,PhysRevB.92.094204}. This leads to a dramatic density accumulation of \red{directionally NHSE amplified states} at lattice boundaries or impurities. %~\cite{Lee2019hybrid,lee2020unraveling,li2021impurity}. 
When it is just simple exponential build-up, they are NHSE eigenstates; in more esoteric critical cases, they can assume special scale-free eigenstate profiles~\cite{li2020critical,kawabata2020higher,liu2020helical,yokomizo2021scaling,li2021impurity,kawabata2022entanglement,qin2023universal}. \red{In generic higher-dimensional lattices that we focus on, such boundary accumulations have not been properly understood.} 

\begin{figure}
\includegraphics[width=.73\linewidth]{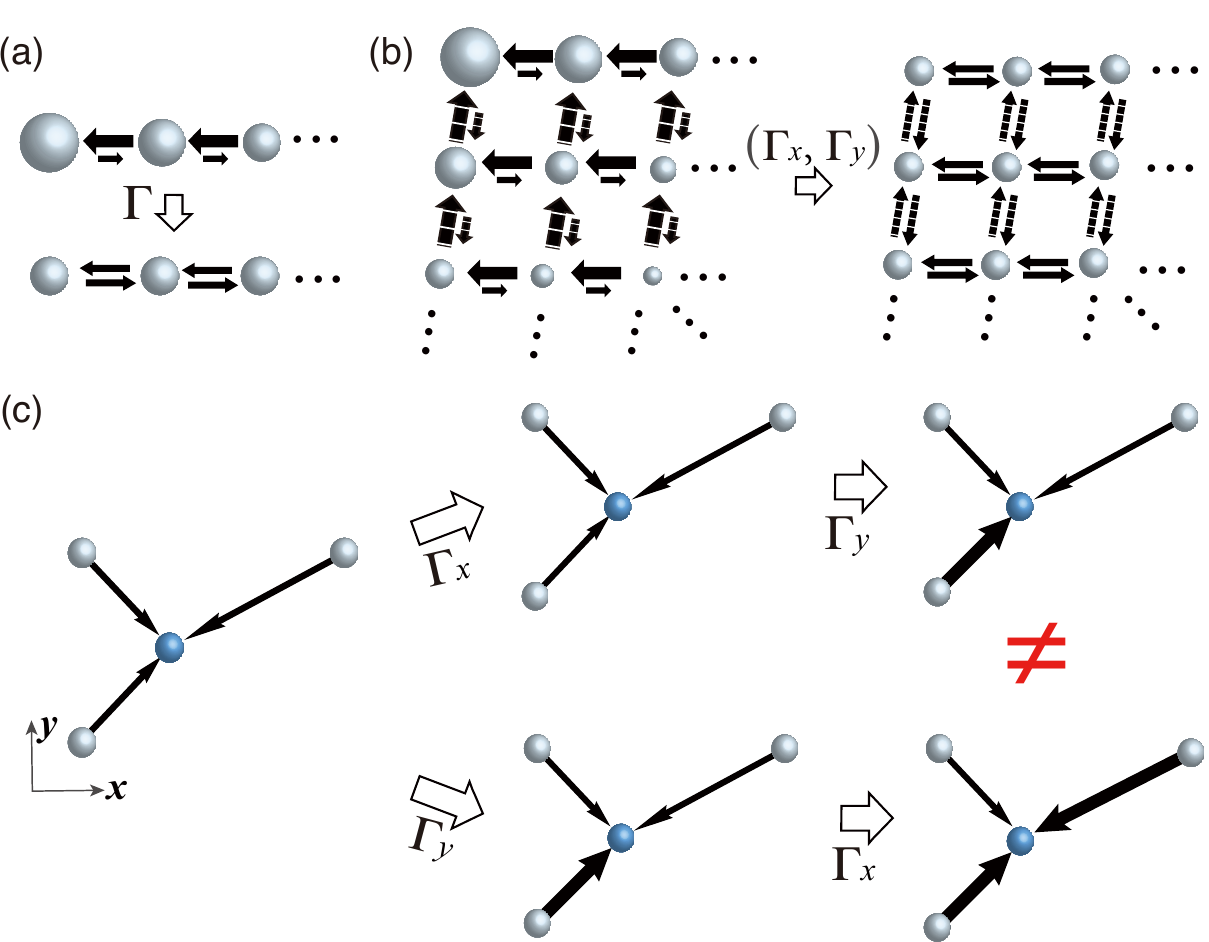}
\caption{  \red{\bf Failure of effective BZ construction in 2D through conventional basis rescaling.}\ a) To obtain the effective BZ of a \red{nearest-neighbor 1D lattice, all couplings can simply be ``symmetrized'' through a change of basis known as the equilibration operation $\Gamma$.} b) Higher-dimensional ``unentangled'' lattices can still be similarly symmetrized via independent equilibrations $\Gamma_x,\Gamma_y,...$ c) Generic ``entangled'' lattices of 2D  and beyond cannot be completely equilibrated, since equilibrations $\Gamma_j$ do not commute in general; shown is a minimal example where $\Gamma_x\Gamma_y\neq \Gamma_y\Gamma_x$\red{(non-commutativity)}, i.e. where symmetrization in one direction can un-symmetrize the coupling components in the other direction (arrow thickness depict coupling strength). Hence obtaining the effective BZ through naive \red{NHSE-inspired equilibration (change-of-basis to the generalized BZ)} is doomed to failure. \\
}\label{fig:1}
\end{figure} 

Since the Bloch eigenstates that define the original BZ are highly distorted by non-Hermitian pumping \red{(directed amplification)}, all ``bulk'' properties such as band topology, transport and geometry will be radically modified. To correctly characterize them, it is necessary to construct the effective BZ where the spatially non-uniform pumped eigenstates are \red{``equilibrated'' to approximately resemble Bloch states. This equilibration is mathematically a transformation to a basis where the NHSE is eliminated - in that basis, the couplings appear symmetrized and the NHSE no longer acts}~\cite{yang2019auxiliary,lee2020unraveling}. The simplest illustrative example, well-known in the NHSE literature, is the 1D ``Hatano-Nelson'' chain with asymmetric nearest-neighbor couplings $h_{\pm\hat x}=he^{\mp\kappa}$ (Fig.~\ref{fig:1}a)~%\cite{cite HN model}
\cite{HN1996prl,HN1997prb,PhysRevB.92.094204,claes2020skin}. Under OBCs, its eigenstates assume the boundary-localized form $\psi_\text{HN}(x)\sim e^{-\kappa x}$\red{[Fig.~\ref{fig:1}a (balls increasing in size)]}, which can be ``equilibrated'' into the bulk through a basis rescaling operator $\Gamma$: $c^\dagger_x\rightarrow e^{\kappa x} c^\dagger_x$, $c_x\rightarrow e^{-\kappa x} c_x$. \red{(We write $\Gamma_{j}$ for $\Gamma$ corresponding to a boundary in the $j$-th direction)}. At the same time, $\Gamma$ also ``balances'' the equilibrated couplings, as shown in Fig.~\ref{fig:1}a, as well as induce an effective complex deformed BZ viz. $c^\dagger_k=\sum_xe^{ikx}c^\dagger_x\rightarrow \sum_x e^{i(k-i\kappa)x}c^\dagger_x=\sum_x z(k)^xc^\dagger_x$ where $-i\log z(k)=k-i\kappa$ is the \red{complexified momentum. The assumption here is that, even though translation invariance is lost due to OBCs, the eigenmodes are still approximately labeled by appropriately discretized wavenumbers, albeit with an additional $e^{-\kappa x}$ spatial factor to account for NHSE accumulation. }

\red{In higher-dimensions $D$, only the simplest lattices i.e. monoclinic lattice for $D=2$ (Fig.~\ref{fig:1}b) can be ``unentangled'' into separate sets of 1D chains $H(k_1,k_2,...)=H^1_\text{1D}(k_1)\oplus H^2_\text{1D}(k_2)\oplus ...$ . For these, the equilibration operator $\Gamma_j$ can be analogously applied whenever OBCs are taken along the $j$-th direction. }

\red{But generically, most $D\geq 2$ lattices are ``entangled'' due to non-trivial inter-chain couplings,} and this NHSE-inspired equilibration procedure \red{(generalized BZ construction)} fails to give the correct equilibrated lattice couplings and hence effective BZ. Consider the minimal model with three non-orthogonal asymmetric hoppings from each site non-trivially ``entangling'' the two lattice directions (Fig.~\ref{fig:1}c). \red{Let's derive the boundary-accumulated eigenstates when its lattice (not explicitly shown) is under OBCs in both $x$ and $y$ directions.} At each equilibration step $\Gamma_j$, the combined coupling strength component in the $j$-th direction are to be ``balanced'': in Fig.~\ref{fig:1}c, the $\Gamma_x$ operation modifies the original couplings negligibly because the x-components are already approximately equal, but not so for $\Gamma_y$. \red{But therein lies the paradox: exchanging the order of performing the equilibrations $\Gamma_x,\Gamma_y$ yield different equilibrated couplings, even though the effective lattice should of course \emph{not} depend on the order in which the $x$,$y$-OBCs are taken.} %performing the equilibrations $\Gamma_x$ followed by $\Gamma_y$ yields different results from having $\Gamma_y$ followed by $\Gamma_x$. 
This non-commutativity of $\Gamma_x$ and $\Gamma_y$, even for such a minimal example, suggests that physical states are pumped in a peculiar non-local manner, and an entirely new approach is needed for correctly characterizing the effective BZ whenever a multi-dimensional lattice cannot be trivially decoupled into 1D chains, as further explained in the Appendix.
\begin{figure}
\includegraphics[width=.76\linewidth]{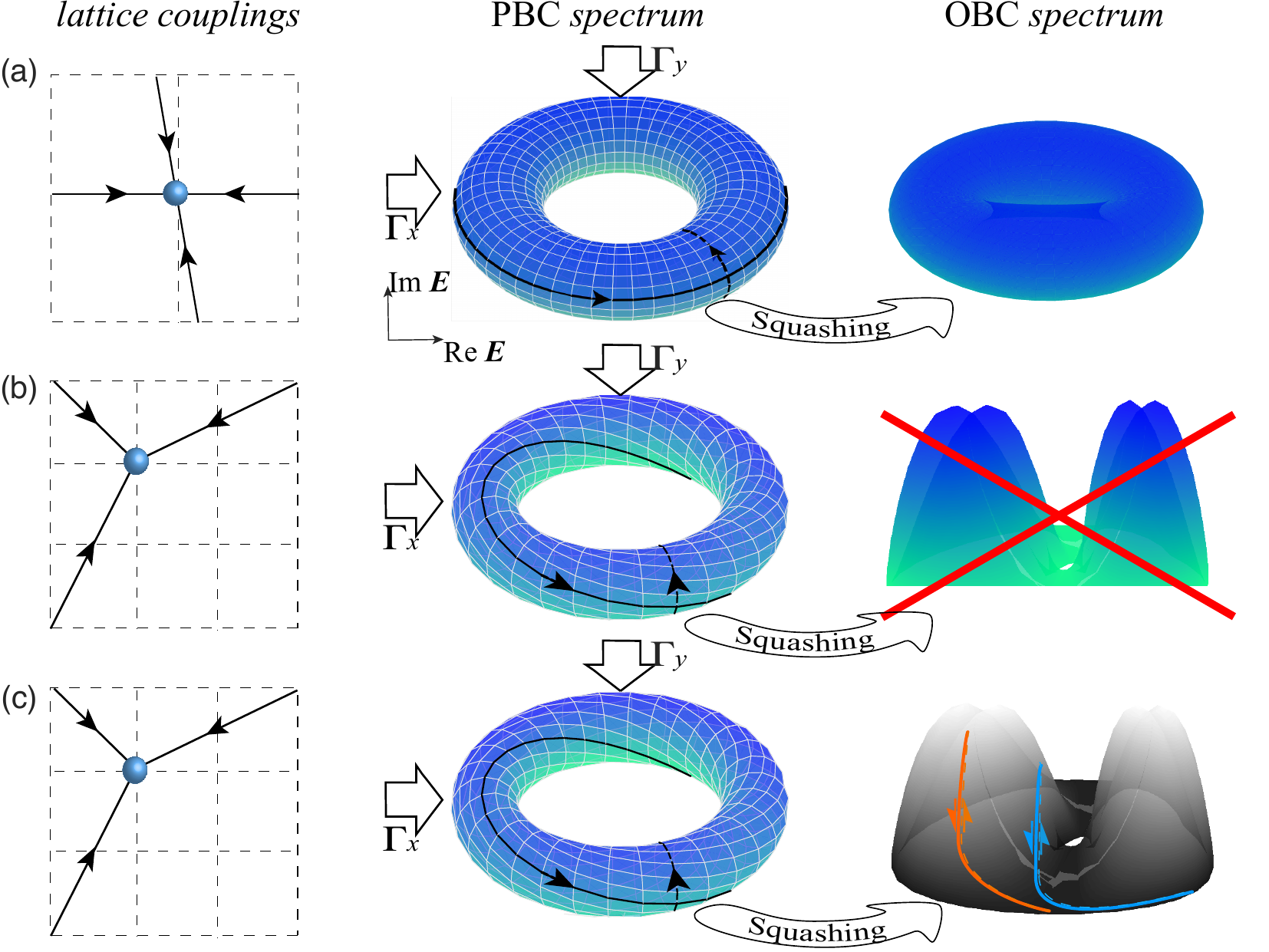}
\caption{\red{\bf Non-commutativity of NHSE equilibration violates the requirement of vanishing OBC spectral winding.} a) An ``unentangled'' lattice admits fully commuting equilibration operators $\Gamma_x,\Gamma_y$ that completely ``squashes'' \red{(flattens)} its PBC spectral torus $E_\text{2D}$ into a ``flattened'' OBC torus $\bar E_\text{2D}$, \red{reminiscent of 1D cases, where the OBC spectrum consists of PBC spectral loops ``squashed'' into interior curves~\cite{yang2019auxiliary}.} b) An ``entangled'' lattice is subject to non-commuting equilibrations $\Gamma_x\Gamma_y\Gamma_x^{-1}\Gamma_y^{-1}\neq \mathbb{I}$, such that its PBC spectrum can no longer be completely ``squashed'' into a valid OBC spectrum with no spectral winding, akin to a filled balloon. c) \red{The correct OBC spectrum of the ``entangled'' 2D lattice is traced out by up to two 1D homotopy paths (blue, orange)} on the incompletely squashed spectral torus that avoid any spectral winding. \red{The tori so illustrated do not live in 3D, but are projections on the 2D energy plane, being composed of collections of 1D spectral loops.}
}
\label{fig:2}
\end{figure} 

\noindent{\it Dimensional transmutation from non-commutative equilibration.--} We next show how multi-dimensional non-Hermitian \red{directed NHSE amplification i.e. pumping} on the energy spectrum advocates an effective BZ of a different, lower dimension. Consider a 2D model $H_\text{2D}(k_x,k_y)$ in momentum space. Under periodic boundary conditions (PBCs), its spectrum $E_\text{2D}(k_x,k_y)$ generically resembles a deformed torus projected onto a 2D plane (Fig.~\ref{fig:2}), since it takes complex values and is parametrized by two periodic momenta. Going from PBCs to OBCs, this spectrum $E_\text{2D}\rightsquigarrow \bar E_\text{2D}$ must necessarily be ``squashed'' \red{i.e. flattened into lines or curves} in the complex plane by non-Hermitian pumping, since %%it is known that 
under OBCs, any 1D subsystem i.e. any 1D loop traced by $E_\text{2D}(k_x,k_y)$ with fixed $k_x$ or $k_y$ must \red{enclose zero area (be degenerate)} in the complex energy plane: $\oint \partial_{k_{j}}\log (\bar E_\text{2D}(\bm k)-E_0)\,\text{d}k_j=0$ for all $E_0\in \mathbb{C}$, $j=x,y$ \red{\cite{zhang2019correspondence}}. \red{Intuitively, this is because nontrivial spectral winding requires non-reciprocity, but OBC eigenstates are fully ``equilibrated'' at the boundaries and are no longer pumped non-reciprocally}~\footnote{This can also be seen via a complex flux threading Gedankan experiment~\cite{xiong2018does,Lee2019anatomy,li2021quantized}. Gauge transforming all flux $\phi$ onto the boundary coupling, the latter shrinks when we ramp up $\text{Im}(\phi)$; correspondingly, the effect of cycling $\text{Re}(\phi)$ across $[0,2\pi]$ diminishes. In the OBC limit, further spectral flow is not possible, and only degenerate spectral loops can exist.}.

\begin{figure}
\includegraphics[width=0.83\linewidth]{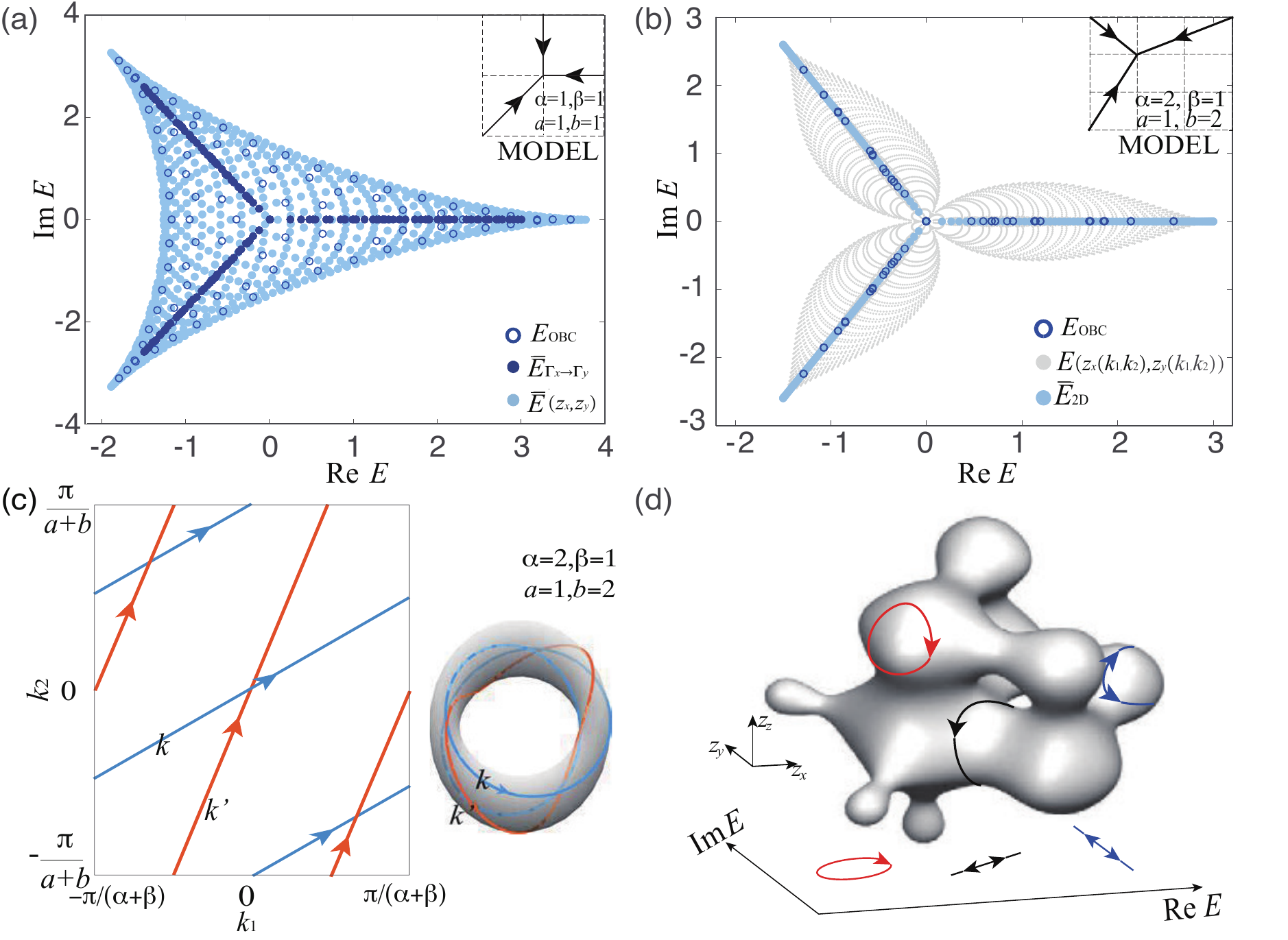}
\caption{\red{\bf Dimensionally transmutated effective BZ gives the correct OBC spectrum.}\ a) Sequentially applying $\Gamma_x$ and then $\Gamma_y$ ($x$-OBCs and then $y$-OBCs) yields an incorrect OBC spectrum $\bar E_{\Gamma_x\rightarrow \Gamma_y}$ (Dark blue) for the illustrative ``entangled'' 2D lattice $H=\sum_{\bm x}\red{2}c^\dagger_{\bm x + \hat x}c_{\bm x}+c^\dagger_{\bm x + \hat y}c_{\bm x}+c^\dagger_{\bm x - \hat x-\hat y}c_{\bm x}$\red{ (no dimensional transmutation )}, at odds with the symmetrically obtained $\bar E(z_x,z_y)$ (light blue), which reproduces the exact numerical $E_\text{OBC}$ (blue circles). b) Necessity of dimensional transmutation of the BZ: For our model $H_\text{2D}$ (Eq.~\ref{H2D}), the effectively 1D $\bar E_\text{2D}$ %%of Eq.~\ref{barE} 
(light blue) agrees with the numerical $E_\text{OBC}$ (blue circles), while the unconstrained $E_\text{2D}$ from Eqs.~\ref{E2D} and~\ref{zxzy1},\ref{zxzy2} gives extraneous eigenenergies (gray). \red{The systems of 3a and 3b belong to scenarios depicted in Figs. 2a and 2b,c respectively.} c) The effective 1D BZ is given by the union of 1D winding paths (blue, red for $k,k'$ respectively) on the $k_1$-$k_2$ 2-torus. 
d) $\mathscr{M}$ (gray blob) of an illustrative 3D model, with effective BZ given by its blue and black loops that correspond to degenerate spectral loops in the complex $E$ plane below. 
}
\label{fig:3}
\end{figure}

However, the spectral squashing in 2D is often not straightforward like in 1D, where equilibration always amounts to a complex BZ deformation $e^{ik}\rightarrow z(k)\rightarrow e^{i(k-i\kappa(k))}$ that completely squashes $E_\text{1D}(k)\rightarrow E_\text{1D}(k)(k-i\kappa(k))=\bar E_\text{1D}(k)$ into a degenerate spectral loop \red{with no spectral winding i.e. $\bar E_\text{1D}(k)=\bar E_\text{1D}(k')$ for some $k\neq k'$.} %: $\oint \partial_k \log(\bar E_\text{1D}(k)-E_0)dk=0$ for all $E_0\in \mathbb{C}$. 
As sketched in Fig.~\ref{fig:2}a for an ``unentangled'' 2D lattice, \red{the Hamiltonian can be written by $H({\bm k})=\sum_n A_n(k_x)\exp(ink_y)$ with  the solution $z_y$ of $\sum_n A_n(k_x)z_y^n \sin (q_y)=0$ independent of $k_x $,} $\Gamma_x$ and $\Gamma_y$ is allowed to successively ``squash'' the spectral torus until it contains no non-degenerate loops \red{enclosing nonzero area}, since the lattice trivially decouples into two non-parallel 1D chains. However, for an ``entangled'' 2D lattice (Fig.~\ref{fig:2}b,c),\red{ $|A_n(k_x)/A_{-n}(k_x)|$ dependent of $k_x$,} $\Gamma_x\Gamma_y\Gamma_x^{-1}\Gamma_y^{-1}\neq \mathbb{I}$ and the ``squashing'' cannot be complete - picture a filled balloon which can be compressed in one direction, but not squashed in all directions simultaneously. \red{As the incompletely ``squashed'' spectral torus still contains non-degenerate loops, the only solution is to exclude them from the effective BZ itself.} In this case, the effective BZ can only be spanned by the homotopy generator independent from any non-degenerate spectral loop, and can only be of 1D despite the physical lattice being of 2D. \red{Fig.~\ref{fig:2}c shows two possible loops (blue, orange) that enclose zero area on the complex $E$ plane, and either (or both) of them would rightly span the effective BZ.} Fig.~\ref{fig:3}a shows an example where successive application of $\Gamma_x$ followed by $\Gamma_y$ gives the incorrect spectrum (dark blue), different from the numerically obtained spectrum (blue). As such, even though effective 1D BZs possess well-defined complex momenta viz. $z(k)=e^{i(k-i\kappa(k))}$ in 2D or higher, \red{in general $z_j(\bm k)\neq e^{i(k_j-i\kappa_j(\bm k))}$, $j=x,y,...$, defying the well-established NHSE framework.}

\noindent{\it Construction of dimensionally transmutated effective BZ.--}  %%Having explained why the effective BZ may have lower dimensionality than the physical lattice, 
\red{We now construct the effective BZ of a 1-component example} of the type in Fig.~\ref{fig:2}b,c:\footnote{We refer to it as the type-III  model in Sect. III of our Supplement, where we also explained why it undergoes nontrivial dimensional transmutation, but not related models}:
\begin{small}
\begin{equation}
H_\text{2D}=\sum_{\bm x}t_1c^\dagger_{\bm x}c_{\bm x+\alpha\hat x +a\hat y}+t_2c^\dagger_{\bm x}c_{\bm x-\beta\hat x +a\hat y}+t_3c^\dagger_{\bm x}c_{\bm x-\beta\hat x -b\hat y}.
\label{H2D}
\end{equation}
\end{small}
%\begin{equation}
%H_\text{2D}=\sum_{x,y}t_1c^\dagger_{x,y}c_{x+\alpha,y+a}+t_2c^\dagger_{x,y}c_{x-\beta,y+a}+t_3c^\dagger_{x,y}c_{x-\beta,y-b}
%\end{equation}
Applying the ansatz $\psi_{2D}(x,y)\propto z_x^x z_y^y$ for an eigenstate, we obtain the energy relation
\begin{equation}
E_\text{2D}(z_x,z_y)=t_1z^\alpha_xz^a_y+t_2z^{-\beta}_xz^a_y+t_3z^{-\beta}_xz_y^{-b}.
\label{E2D}
\end{equation}
Here, no assumption is made about the boundary conditions, and the assertion is that $E_\text{2D}(z_x,z_y)$ yields the correct eigenenergies given appropriate forms of $z_x,z_y$. 

To correctly obtain the effective BZ from $E_\text{2D}(z_x,z_y)$, we would need to treat the effects of both $x$ and $y$-OBCs \red{on equal footing, such the order of opening up OBCs in different directions do not matter, as physically expected}. This can be achieved by \red{alternately} implementing the two OBCs one at a time, by considering the other momentum as a parameter. Given a quasi-1D energy function $E_\text{1D}(z)$, we determine the effective BZ by finding a complex effective momentum function $-i\log z(k)$, $k\in [0,2\pi)$, such that every energy eigenvalue $E=E_\text{1D}(z(k))$ corresponds to at least two different $k$ solutions with identical $|z(k)|$~\cite{Lee2019anatomy,yokomizo2019non}. In a trivial case without non-Hermitian pumping, we simply have $z(k)=e^{ik}$, such that the effective and original BZs coincide. 
%%In this work, we shall not delve into the intricacies of finding the effective BZ of various 1D models; rather, we just quote the new result relevant to our 2D example, as derived in our Supplement: 
For $E_\text{1D}(z)=Az^{p}+Bz^{-q}$ corresponding to left(right) hoppings over $p$ ($q$) sites, we have \red{from Sect. I of}~\cite{suppmat}
\begin{equation}
z(k)=\left(\frac{B\sin qk}{A\sin pk}\right)^{\frac1{p+q}}e^{e^{i\frac{2\pi \nu}{p+q}}}e^{ik}=e^{-\kappa_\text{1D}(k)}e^{ik}\ ,
\label{zk}
\end{equation}
for $k\in(-\pi/(p+q),\pi/(p+q)]$, $\nu=1,2,...,p+q$ labeling the solution branch. The decay function $e^{-\kappa_\text{1D}(k)}$ encodes how non-Hermitian {directed amplification} distorts the Bloch phase factor $e^{ik}$. %%Notice that the mapping $e^{ik}\rightarrow z(k)$ is only piecewise-smooth -- this non-analyticity stems from the non-locality of non-Hermitian pumping, even though the hopping processes are clearly local.

By applying Eq.~\ref{zk} on $z_x,z_y$ of Eq.~\ref{E2D} separately, we obtain 
$z_x^{\alpha+\beta}=(t_2+t_3z_y^{-(a+b)}) /(t_1\sin \alpha k_1)\sin \beta k_1\,\text{e}^{i(\alpha+\beta)k_1}$ and $z_y^{-(a+b)}=(t_2+t_1z_x^{\alpha+\beta})/(t_3\sin bk_2)\sin ak_2 \,\text{e}^{-i(a+b)k_2}$, where we have used $k_1,k_2$ instead of $k_x,k_y$ to emphasize that they may not be conjugate momenta to the $x,y$ coordinates. %  with $k_1\in (-\pi/(\alpha+\beta),\pi/(\alpha+\beta)]$ and $k_2\in(-\pi/(a+b),\pi/(a+b)]$.
We can simultaneously solve these to obtain 
\begin{subequations}
\begin{equation}\label{zxzy1}
z_x^{\alpha+\beta}=\frac{t_2}{t_1}\frac{(\sin ak_2 +e^{i(a+b)k_2}\sin bk_2)e^{i(\alpha+\beta)k_1}\sin \beta k_1}{e^{i(a+b)k_2}\sin \alpha k_1 \sin bk_2-e^{i(\alpha+\beta)k_1}\sin\beta k_1 \sin a k_2},
\end{equation}
\begin{equation}\label{zxzy2}
z_y^{a+b}=\frac{t_3}{t_2}\frac{e^{i(a+b)k_2}\sin \alpha k_1 \sin bk_2-e^{i(\alpha+\beta)k_1}\sin\beta k_1 \sin a k_2}{(\sin \alpha k_1 +e^{i(\alpha+\beta)k_1}\sin \beta k_1)\sin a k_2}.
\end{equation}
\end{subequations}

We reiterate a major distinction between the $z_x,z_y$ above and the effective ``generalized'' BZ of NHSE systems: In the latter, the BZ is ``generalized'' in the sense that $z_j$, $j=x,y$ encapsulates complex momentum via $-i\log z_j=  k_j-i\kappa_j(\bm k)$, with $\kappa_j(\bm k)$ representing the complex deformation. But in Eqs.~\ref{zxzy1}, \ref{zxzy2}, $-i\log z_j$ manifestly do not correspond to any single well-defined complex momentum (recall that $\psi_{2D}(x,y)\propto z_x^x z_y^y$). Even though $k_1,k_2$ are the \red{individual ``momenta'' associated with quasi-1D chains within $H_\text{2D}$, they are now ``entangled''}, as evident in the highly nonlinear functional form of Eqs.~\ref{zxzy1},\ref{zxzy2}.

Importantly, the $z_x,z_y$ from Eqs.~\ref{zxzy1} and \ref{zxzy2} still do \emph{not} describe the correct effective BZ unless $k_1,k_2$ are further constrained, since we have not eliminated the possibility of $E(z_x,z_y)$ exhibiting nontrivial windings as one of $k_1$ or $k_2$ is varied over a period (Sect. II and III of~\cite{suppmat}). Indeed, from Fig.~\ref{fig:3}b, naive substitution of the unconstrained $z_x,z_y$ into Eq.~\ref{E2D} gives extraneous eigenenergies across the complex plane (gray), different from the numerical OBC spectrum (blue circles) which exhibits no spectral winding.

For our model%%Eq.~\ref{H2D}
, all spectral windings vanish along the two 1D parametrization paths $(k_1,k_2)=(bk,\beta k)$ and $(k_1,k_2)=(ak',\alpha k')$, \red{as rigorously shown in Sect. III of}~\cite{suppmat}. Indeed, \red{in Fig.~\ref{fig:3}b, the union of these energies $\bar E_\text{2D}(k)=E_\text{2D}(z_x(k),z_y(k))$ and $\bar E'_\text{2D}(k')=E_\text{2D}(z_x(k'),z_y(k'))$ also agree with the numerical OBC spectrum. }
\begin{comment}%%
Their analytic expressions are given by 
\begin{equation}\begin{split}
&\left[\bar E'_\text{2D}(k')\right]^{\alpha+\beta}=t_1^{\beta}t_2^{\frac{\alpha b-\beta a}{a+b}}t_3^{\frac{a(\alpha+\beta)}{a+b}}\text{e}^{2\pi i \left(a\nu\frac{\alpha+\beta}{a+b}+\alpha\nu'\right)}\\
&\quad\ \times\left(\frac{\sin(a\alpha+\alpha b)k'}{\sin a\alpha k'}\right)^\alpha \left(\frac{\sin(\alpha a+\beta a) k'}{\sin \alpha ak'}\right)^{\frac{b(\alpha+\beta)}{a+b}}\\
&\quad\ \times\left(\frac{\sin \alpha ak'}{\sin\beta ak'}\right)^\beta\left(\frac{\sin(\alpha b-a\beta ) k'}{\sin a\alpha k'}\right)^{\frac{\beta a-\alpha b}{a+b}} \ ,
\end{split}\label{barE}\end{equation}
and analogously for $E_\text{2D}(k)$. 
\end{comment}
The union of the 1D loops traced by $k$ and $k'$  forms the dimensionally transmutated effective BZ, as illustrated in Fig.~\ref{fig:3}c and the Appendix.

\red{Interestingly, this effectively 1D BZ reveals a new avenue of topological winding}, with winding numbers $\text{GCD}(a,\alpha)$ and $\text{GCD}(b,\beta)$ describing how the sectors $k'$ and $k$ loop around the $k_1$-$k_2$ torus.  (both windings $=2$ in Fig.~\ref{fig:3}c). Physically, $k_1,k_2$ represent the non-Bloch wavenumbers from separately taking OBCs in each direction; yet, when both OBCs are simultaneously applied, the effective BZ collapses into closed 1D paths that mixes $k_1$ and $k_2$. \red{As such, these winding numbers capture the amount of ``entanglement'' caused by 2D non-Hermitian pumping.}

\noindent{\it Generalizations.--} %Having discussed the effective 1D BZ construction of our particular $H_\text{2D}$ lattice, 
\red{The construction of the dimensionally-transmutated effective BZ from our particular $H_\text{2D}$ lattice can be generalized to a generic model $H$ in $D$ dimensions.} First, acting on the ansatz eigenstate $\psi_{D}(\bm x)\propto \prod_j^D z_j^{x_j}$, we express the model as a multivariate polynomial $E(\bm z)=\sum_\mu t_\mu\prod_j z_j^{l_{\mu j}}$, where $l_{\mu j}$ is the range of the $\mu$-th hopping $t_\mu$ in the $j$-th direction. Next, we apply the $D$ equilibrations $\Gamma_j$, $j=1,...,D$ separately on $E(\bm z)$, such that each becomes a quasi-1D problem in $z_j$, with all the components of $\tilde z = (z_1,...,z_{j-1},z_{j+1},...,z_D)$ %$z_{i\neq j}$'s 
as spectator parameters. Solving for the effective 1D BZs for each of them~\cite{suppmat,yokomizo2019non,lee2020unraveling,tai2022zoology} i.e. replacing each $z_j$ by appropriate $e^{-\kappa_j(\tilde z)}e^{ik_j}$ (of which Eq.~\ref{zk} is a special case), we obtain $D$ relations \red{(Sect. III of~\cite{suppmat})} $\mathscr{F}_j(\tilde z;k_j)=0$. Inverting these relations, we will in principle obtain $D$ expressions $z_j=\mathcal{F}_j(\bm k)$ where $\bm k\in \mathbb{T}^D$, which generalize Eqs.~\ref{zxzy1},~\ref{zxzy2}. In general, this nonlinear inversion may have to be performed numerically, and yields a highly complicated $D$-dimensional base manifold $\mathscr M$ in $\bm z$-space, possibly with cusps and singularities that give rise to higher dimensional esoteric gapped transitions~\cite{li2020critical}.

The effective dimensional-transmutated BZ depends crucially on the topology of $\mathscr{M}$. Specifically, it is $\mathscr{M}/\{\mathcal{L}\}$, where $\{\mathcal{L}\}$ is the span of homotopy loops $l$ on $\mathscr{M}$ in which $E(\bm z(l))$ exhibits nontrivial spectral winding i.e. the effective BZ is union of submanifolds of $\mathbb{T}^D$ parametrized by $(k'_1,...,k'_d)$, $d<D$, such that the recovered OBC spectrum $\bar E(\bm k')=E(\bm z(\bm k'))$ exhibits trivial spectral winding in all directions, \red{as detailed in Sect. III of}~\cite{suppmat} %%(Eq.~\ref{barE} gives a special case). 
As schematically sketched in Fig.~\ref{fig:3}d for a 3D model, the effective BZ consists of the blue and black loops which wind around $\mathscr{M}$ (gray), not the red loop which encloses nonzero spectral area.

\noindent{\it Dimensional transmutated topology.--} The fundamental dimensional modification of the effective BZ by \red{non-Hermitian pumping (directed amplification)} is not just a mathematical subtlety, but a very physical phenomenon with experimentally observable consequences. In the following, we illustrate a 2D lattice whose topological zero modes are protected by a 1D, not 2D, topological invariant due to dimensional transmutation of its BZ. %Building upon our previous $H_\text{2D}$ example (Eq.~\ref{E2D}), 
We consider the 2-component 2D model 
\begin{comment}
\begin{equation}
 H_\text{topo}(\bm z)=\left(
      \begin{array}{cc}
        0 &H_{12}\\
        H_{21}& 0 \\
      \end{array}
    \right)=\left(
      \begin{array}{cc}
        0 &\left(E_\text{2D}(z_x,z_y)+c\right)/z_y^a\\
        z_y^a& 0 \\
      \end{array}
    \right)
\end{equation}
\end{comment}
\begin{equation}
 H_\text{topo}(\bm z)=%\left(
      %\begin{array}{cc}
      %  0 &H_{12}\\
      %  H_{21}& 0 \\
     % \end{array}
   % \right)=
	\left(\begin{array}{cc}
        %0 &\left(z_x^{\alpha}z_y^{a}+z_x^{-\beta}z_y^{a}+z_x^{-\beta}z_y^{-b}+c\right)/z_y^a\\
				 0 &z_x^{\alpha}+z_x^{-\beta}+z_x^{-\beta}z_y^{-a-b}+cz^{-a}_y\\
        z_y^a& 0 \\
      \end{array}
    \right)\ ,
\end{equation}
with constant $c$ introduced such that the PBC spectrum $E_\text{topo}(e^{ik_x},e^{ik_y})=\pm\sqrt{E_\text{2D}(e^{ik_x},e^{ik_y})+c}%=t_1z_x^{\alpha}z_y^{a}+t_2z_x^{-\beta}z_y^{a}+t_3z_x^{-\beta}z_y^{-b}+c$.
$ is gapped. 

\begin{figure}
\includegraphics[width=.8\linewidth]{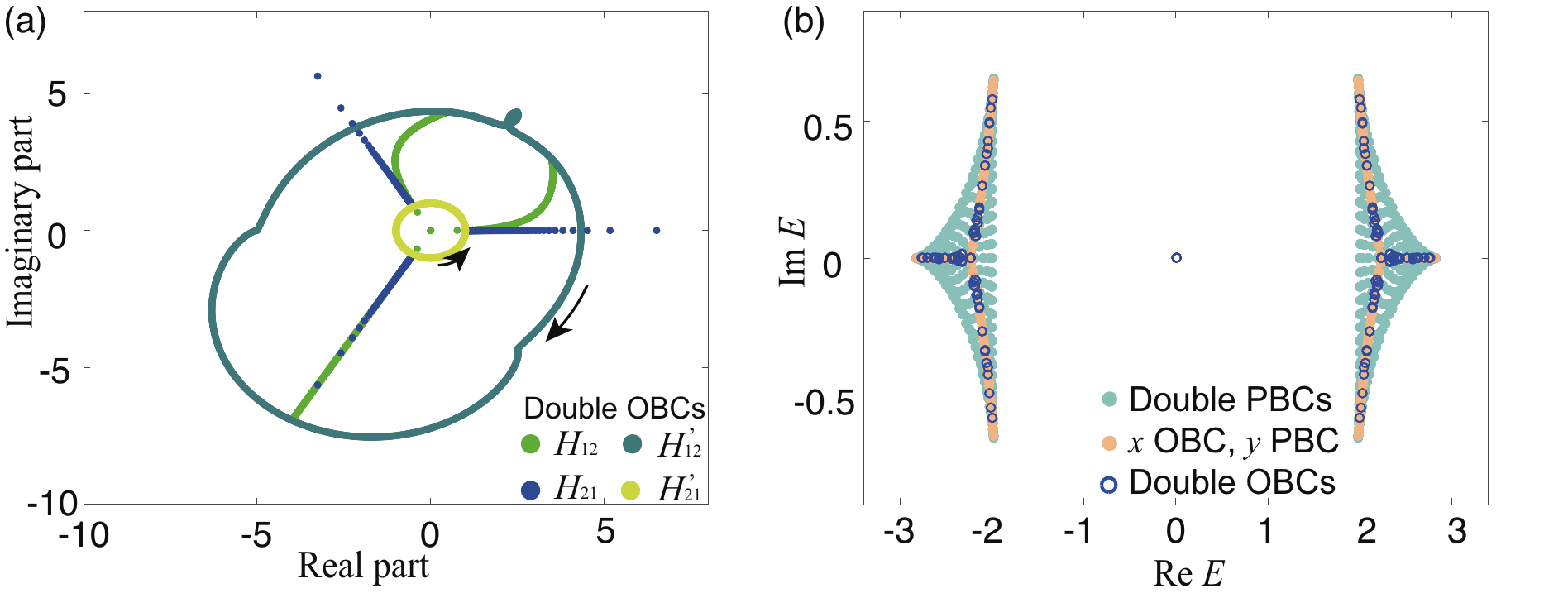}
\caption{\red{\bf Dimensional transmutated topology in 2-band model}\  a) Despite being a 2D model, $H_\text{topo}$ exhibits nontrivial topological winding in its effectively 1D BZ, as seen from the zero windings of $H_{12}(k)$ and $H'_{12}(k')$ summing to $-1$, and that of $H_{21}(k)$ and $H'_{21}(k')$ summing to $1$. 
b) Although protected by 1D topological winding, in-gap zero modes for $H_\text{topo}$ appear under double OBCs (black), and not quasi-1D single OBC (light blue). Parameters are $t_1=t_2=t_3=1$ and $c=5$. 
}
\label{fig:4}
\end{figure}

When regarded as a 2D model, $H_\text{topo}$ is topologically trivial by construction, as can be seen from its Pauli decomposition $H_\text{topo}=\left[(H_{12}+i H_{21})\sigma_x + (H_{12}-i H_{21})\sigma_y\right]/2$, which contains only two Pauli matrices and is thus of trivial 2nd homotopy. However, the effective bulk description of $H_\text{topo}$ is actually 1D, not 2D, since $E_\text{topo}(z_x,z_y)$ and $E_\text{2D}(z_x,z_y)$ are conformally related and must therefore possess identical effective 1D BZs~\cite{lee2020unraveling,tai2022zoology}. Under OBCs, an effectively 1D Hamiltonian %of the form $H_{12}(z)\sigma_++H_{21}(z)\sigma_-$, $\sigma_\pm=(\sigma_x\pm i\sigma_y)/2$ 
possesses topological zero modes if the phase windings of $H_{12}(z)$ and $H_{21}(z)$ around $z=0$ are both nonzero and of opposite signs~\cite{yao2018edge,Lee2019anatomy}; if there is more than one BZ sector, the windings should be added, \red{as performed in Sect IV of}~\cite{suppmat}. This is indeed the case in Fig.~\ref{fig:4}a, with the windings of $H_{12}$ and $H'_{12}$ summing to $-1$, and that of $H_{21}$ and $H'_{21}$ summing to $1$. Correspondingly, these windings protect the isolated zero modes in the double OBCs spectrum (black diamond in Fig.~\ref{fig:4}b); these modes are topological since they appear in the double PBCs bandgap. \red{Despite being protected by 1D topological winding, they do not appear in the quasi-1D scenario with only $x$-OBCs (light blue). }
  
\noindent{\it Discussion.--} \red{Existing higher-dimensional non-Hermitian studies i.e. Chern or higher-order skin-topological characterizations~\cite{Ezawa2018HOnonH,lee2014lattice,Yao2018nonH2D,Luo2019HOnonH,Lee2019hybrid,kawabata2020higher,groenendijk2021universal} have mostly been based on simple hyperlattices. Beyond that, in generic lattices with ``entangled'' couplings, we discover that non-Hermitian pumping does not commute, transmuting the momentum-space lattice (BZ) to an effectively lower dimension. %%This geometrical obstruction is rooted in a nonlinear inversion problem that is visualized as an incompletely ``squashed'' base BZ manifold. The classification of how the lower-dimensional effective BZ winds around it gives rise to a novel instance of topological winding. 
As a fundamentally dynamical phenomenon, this dimensional transmutation contrasts with the dimensional reduction in topological classification~\cite{qi2008,ryu2010topological,PhysRevB.96.195101}, as well as the emergence of an extra scaling dimension in lattice-based holography approaches~\cite{qi2013exact,gu2016holographic,Lee2017generalized,hu2020machine}.}

Physically, the dimensional transmutation can be manifested through bulk response and topological properties. Topological states protected by lower-dimensional invariants can be constructed and observed in open non-reciprocal arrays with sufficiently versatile engineered couplings, such as lossy photonic resonator arrays~\cite{botten1997periodic,yariv2002critical,feng2017non,bergholtz2021exceptional}, electrical circuits~\cite{suppmat,lee2018topolectrical,imhof2018topolectrical,helbig2019band,hofmann2019chiral,kotwal2019active,circuitPhysRevB.100.184202,ni2020robust,helbig2020generalized,lee2019imaging,li2019emergence,circuitPhysRevLett.124.046401,hofmann2020reciprocal,lenggenhager2021electric,bergholtz2021exceptional,circuitliu2021non,zou2021observation,stegmaier2021topological,circuitliu202001,hohmann2022observation,zhang2022observation,wu2022non,zhang2023electrical,zhu2023higher} or even quantum computers~\cite{choo2018measurement,smith2019simulating,behera2019designing,azses2020identification,koh2022stabilizing,koh2022simulation,smith2022crossing,koh2023observation,shi2022observing}. %%Additionally, quantum setups with well-defined state occupancy offer the tantalizing prospect for observing transport properties with altered dimensionality.

\noindent{\it Acknowledgements.--} This work is supported by the Ministry of Education, Singapore (MOE) Tier-I grant iRIMS no. A-8000022-00-00 and the MOE Tier-II grant (Award No. MOE-T2EP50222-0003).\\

\newpage
\onecolumngrid
\begin{center}\label{appendix}
\textbf{\large Appendix: Details on the dimensional transmutation approach }\end{center}
\twocolumngrid
\setcounter{equation}{0}
\setcounter{figure}{0}
\renewcommand{\theequation}{A\arabic{equation}}
\renewcommand{\thefigure}{A\arabic{figure}}
\renewcommand{\cite}[1]{\citep{#1}}

%\CH{CH: Please put this entire appendix section as two new pages in the main text, after the acknowledgements but before the references.}\\

Here we present a pedagogical summary of our new dimensional transmutation approach and clarify the differences between our approach and the conventional generalized Brillouin zone (GBZ) approach ~
\cite{yao2018edge,xiong2018does,Lee2019anatomy,song2019realspace,song2019non,longhi2019probing,zhang2019correspondence,yokomizo2019non,yang2019auxiliary,lee2020ultrafast,longhi2020non,lee2020unraveling}. For ease of notation, we shall specialize to two dimensions (2D), and readers may refer to Sect V of~\cite{suppmat}  for the generalization of our approach to arbitrarily high dimensions.

\begin{figure}[H]
\centering
\includegraphics[width=0.8\linewidth]{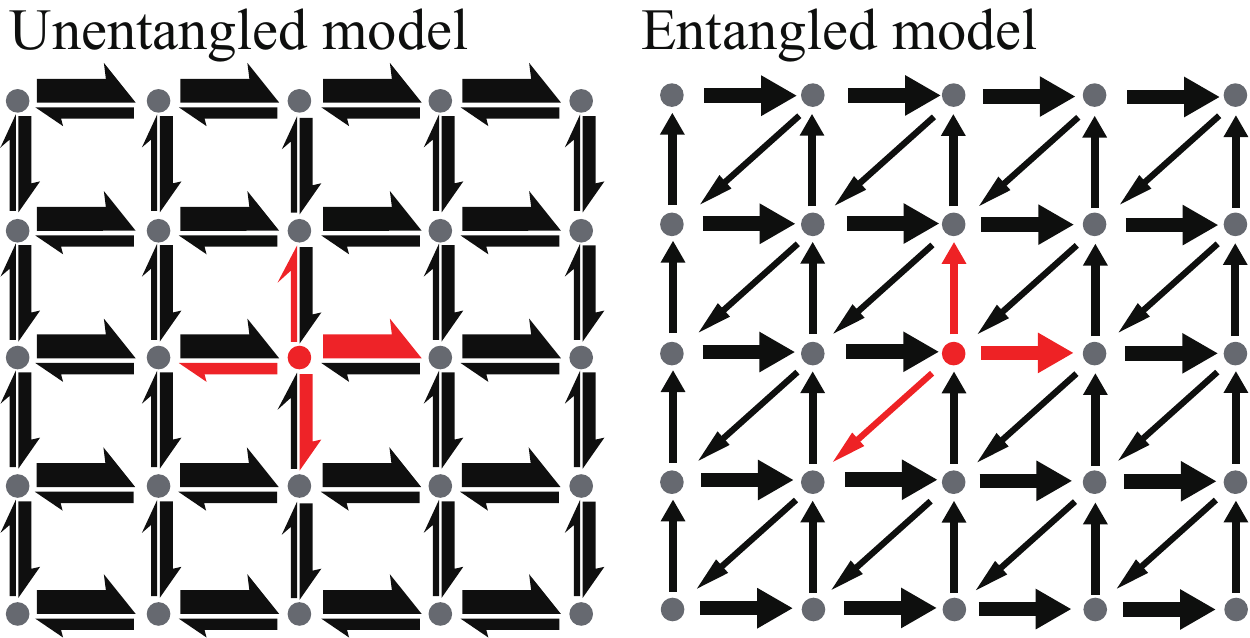}  
\caption{
(Left) An ``unentangled'' lattice model $H$ can be decomposed into arrays of 1D chains %i.e. $H=H^1_{\text{1D}}(k)\oplus H^2_{\text{1D}}(k)\oplus\cdots $
, in this case into a vertical and a horizontal array of Hatano-Nelson models. As such, its full OBC properties can be correctly predicted with conventional GBZ theory, well-established for effectively 1D models. 
(Right) With additional couplings between different arrays of 1D chains, the lattice becomes ``entangled'' -- the scenario for most realistic systems with longer-ranged effective couplings (shown here is the simplest possible case). Our dimensional transmutation approach is required to correctly characterize the full OBC system, as explained below and summarized in Fig.~\ref{tableapp}.
}
\label{fig:ent}		
\end{figure}

Our approach is motivated by the fact that the conventional GBZ approach cannot predict the correct $\bar E$ under full open boundary conditions (OBCs) whenever the lattice is ``entangled'' in 2D or higher [Fig.~\ref{fig:ent}]. %for instance in Fig.~3(b) above. 
This is because (i) %the complex momentum deformations in obtaining the GBZ for each direction can mutually interfere, 
sequentially obtaining GBZs for each OBC direction can lead to inconsistent results, %as summarized in the left column of TABLE I below
; (ii) it may not be possible [Fig.~\ref{fig:2}] to ensure zero spectral winding in all momentum directions (a necessary condition for all OBC spectra\cite{zhang2019correspondence,PhysRevB.104.125109,kawabata2020higher}), unless the effective BZ itself is of a lower dimensionality than the physical system. 

Our approach first treats all OBC directions on equal footing, obtaining a simultaneously-solved provisional effective BZ $(z_x,z_y)$, and then dimensionally transmutes (reduce) it such that zero spectral winding is respected. This yields an effective 1D GBZ in which $\bar E$ agrees with the numerically obtained full OBC spectrum.

\subsection{Detailed walkthrough}
\begin{figure}[H]
  \centering
\caption{  \label{tabtab} Summary of the key differences between our dimensional reduction approach and the conventional GBZ approach, accompanied by an illustrative example (Here we specialized to 2D, see Sect. V of ~\cite{suppmat} for higher-dimensional generalizations).}\label{tableapp}
%\begin{tabular}{c}
  \centering \includegraphics[width=0.99\linewidth]{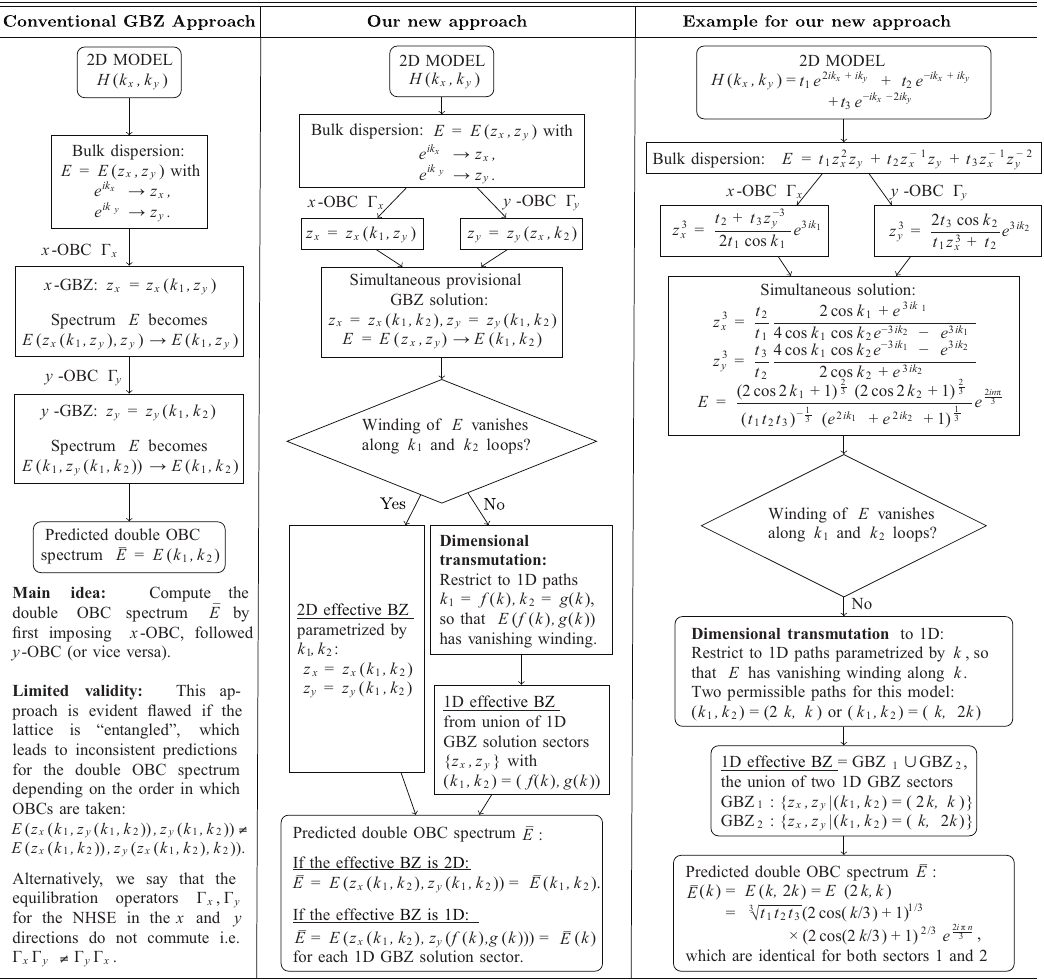}
%   \end{tabular}
\end{figure}

We now walk through our general approach in detail, illustrating it with the model of Eq.~\ref{E2D} with $\alpha=b=2,\beta=a=1$, and summarized with flowcharts in Fig.~\ref{tableapp}. The starting point for a generic 2D model is its energy dispersion $E(z_x,z_y)$, where $z_{x}=\exp(ik_{x})$, $z_{y}=\exp(ik_{y})$ under periodic boundary conditions (PBCs), but would be complex deformed under OBCs. 

Under $x$-OBC, we treat $E(z_x,z_y)$ as a 1D model with parameter $z_y$, and obtain the $x$-GBZ $z_x(k_1,z_y)$ via the condition\cite{yao2018edge,zhang2019correspondence,yokomizo2019non}. that every OBC energy $E(z_x(k_1,z_y),z_y)$ corresponds to at least two different $k_1$ solutions with identical inverse localization length $-\log |z_x(k_1,z_y)|$. To obtain the full OBC spectrum, the conventional approach would be to next implement $y$-OBCs, yielding $E(z_x(k_1,z_y(k_1,k_2)),z_y(k_1,k_2))$ (left column of Fig.~\ref{tableapp}). However, this may not correctly predict the full OBC spectrum in generic ``entangled'' lattices [Fig.\ref{fig:2}]. 

Instead, in our approach (middle \& right columns of Fig.~\ref{tableapp}), we simultaneously obtain the $y$-GBZ $z_y(z_x,k_2)$ by treating $z_x$ as a parameter, and then obtain the provisional GBZ by simultaneously solving for $z_x,z_y$ in terms of $k_1,k_2$. Explicitly for our example described by
 \begin{equation}\label{de1}
E(z_x,z_y)= t_1z_x^{2}z_y+t_2z_x^{-1}z_y+t_3z_x^{-1}z_y^{-2},
\end{equation}
the provisional GBZ is given by
\begin{small}
 \begin{equation}
 \begin{split}
 \left\{
 \begin{array}{ll}
z_x^3=\displaystyle\frac{t_2}{t_1}\frac{2\cos k_1 +\text{e}^{3ik_1}}{4\cos k_1\cos k_2\text{e}^{-3ik_2}-\text{e}^{3ik_1}},\\
z_y^3=\displaystyle\frac{t_3}{t_2} \frac{4\cos k_1\cos k_2\text{e}^{-3ik_1}-\text{e}^{3ik_2}}{2\cos k_2 +\text{e}^{3ik_2}},
\end{array}\right.\ 
 \end{split}
\end{equation}
\end{small}
such that the spectrum is deformed as $E(z_x,z_y)\rightarrow$
\begin{small}
\begin{equation}\label{Eqdede1}
 E(k_1,k_2)=\sqrt[3]{t_1t_2t_3}\frac{(2\cos 2k_1+1)^{\frac{2}{3}}(2\cos 2k_2+1)^{\frac{2}{3}}}{\left(\text{e}^{2ik_1}+\text{e}^{2ik_2}+1\right)^{\frac{1}{3}}}\text{e}^{{2i n\pi}/{3}}
\end{equation}
\end{small}
with real $k_1,k_2$ and solution branches $n=1,2,3$. Importantly, $E(k_1,k_2)$ should never possess nonzero spectral winding~\cite{okuma2020topological,zhang2019correspondence}, being an OBC spectrum. For many cases such as Eq.\ref{Eqdede1}, it is however complex with nontrivial winding. Yet, $E(k_1,k_2)$ can be rigorously verified to satisfy all the model hopping constraints, and thus cannot be incorrect. Hence we conclude that \ul{the correct effective BZ consists of 1D subspaces of the provisional 2D GBZ}. For generic $E(k_1,k_2)$ with nontrivial spectral winding, we stipulate that the 1D effective GBZ consists of paths parametrized by $k_1=f(k),k_2=g(k)$, such that $\bar E=E(f(k),g(k))$ has vanishing $k$-winding. Numerically, it indeed predicts the correct full OBC spectrum (bottom right of Fig.~\ref{tableapp}).
%The dimensionally-transmuted $\bar E$ obtained satisfy the Hamiltonian eigenequations and has vanishing spectral winding, and correctly predicts the full OBC spectrum.% with the range of $k_1,k_2$ prescribed above.\\

For our example, 1D paths given by $(k_1,k_2)=(2k,k)$ or $(k_1,k_2)=(k,2k)$, $k\in [-\pi,\pi)$ yield zero spectral winding, leading to two effective 1D GBZ sectors 
\begin{small}
\begin{equation}\label{eqde2}
\begin{split}
&\text{GBZ}_1=\left\{z_{x,1}^3=\frac{t_2}{t_1}\text{e}^{ik}, z_{y,1}^3=\frac{t_3}{t_2}\frac{1}{2\cos \left(k/3\right)-1}\right\},\\
&\text{GBZ}_2=\left\{z_{x,2}^3=\frac{t_2}{t_1}\left(2\cos \left(k'/3\right)-1\right),z_{y,2}^3=\frac{t_3}{t_2}\text{e}^{-ik'}\right\} 
%&\qquad \qquad \qquad \text{effective\ BZ}=\text{GBZ}_1\cup \text{GBZ}_2\ .
\end{split}
\end{equation}
\end{small}
whose union $\text{GBZ}_1\cup \text{GBZ}_2$ forms the full effective BZ.
		
%In other words, the set of effective BZ eigenstates consists of eigenstates from both GBZ sectors.  The effective BZ Hamiltonian is hence given by
%\begin{equation}\label{eqS320}
%\begin{split}
%H&=\bar H={H}_1\oplus {H}_2\ , \\
%\bar{E}_1(k)=\bar{E}_2(k)&= t_1^{1/3}t_2^{1/3}t_3^{1/3} (2\cos (k/3)+1)^{1/3}(2\cos (2k/3)+1)^{2/3}\text{e}^{2i\pi v/3}\ ,
%\end{split}
%\end{equation}
%with $k,k'\in(-\pi,\pi]$, $v=1,2,3$. $ \bar E_1$ and $\bar E_2$ are the corresponding eigenenergies of $ H_{1,2}$ from GBZ$_{1,2}$.  That is, the effective BZ is 1D, which is different from its physical dimensionality, which we called ''dimensional transmutation''.\\

\onecolumngrid
\newpage

\twocolumngrid
Instead of sequentially eliminating boundary conditions in the different directions, as in the conventional GBZ method,  our new approach computes the double OBC spectrum $\bar E$ by first simultaneously imposing $x$- and $y$-OBCs, and obtaining their simultaneous solution. Then we check if the spectral winding vanishes: If yes, we are done; if not, perform the additional step of dimensional transmutation, reducing the 2D effective BZ to the union of 1D GBZ sectors consistent with vanishing spectral winding. As shown in [Fig.\ref{fig:3}(b)], the 1D-transmuted $\bar E(k)$ (light blue) agrees with the numerically obtained 2D OBC spectrum $E_\text{OBC}$ (blue circles), while the unconstrained $E(z_x(k_1,k_2), z_y(k_1,k_2))$ in the 2D GBZ gives the incorrect spectrum with extraneous eigenenergies (gray).

Our new approach is valid for all 2D lattices, whether entangled or unentangled. For its extension to higher dimensions, please refer to Sect. V of~\cite{suppmat}.

%Our illustrative example (column 3 of Fig~.\ref{tableapp}),  

\bibliography{references}

\bigskip

\clearpage

\onecolumngrid
\begin{center}
\textbf{\large Supplementary Materials}\end{center}
\setcounter{equation}{0}
\setcounter{figure}{0}
\setcounter{section}{0}
\renewcommand{\theequation}{S\arabic{equation}}
\renewcommand{\thefigure}{S\arabic{figure}}
\renewcommand{\thesection}{S\arabic{section}}
\renewcommand{\cite}[1]{\citep{#1}}

\section{I. Analytic GBZ results for generic 1D systems with two hopping terms}\label{sec1}
As a foundation for the analysis of the effective BZ of higher-dimensional lattices, we develop in this section a general analytic derivation for the generalized Brillouin zone (GBZ) results for 1D systems with 2 hopping terms, and compare them with numerics. In this supplement, we shall refer to ``GBZ'' and ``effective BZ'' interchangeably. Note that these results may no longer hold when there are more than 2 hopping terms, which is the more interesting scenario which this work is concerned with. \\

\subsection{Preliminaries}

Consider the following 1D single-band model with dissimilar left/right hoppings:
\begin{equation}\label{eqS1}
\begin{split}
H_\text{1D}=\sum_{n} A|n\rangle\langle n+\alpha|+ B|n\rangle\langle n-\beta|
\ ,
\end{split}
\end{equation}
whose eigenvalues are $E(z)=Az^{\alpha}+Bz^{-\beta}$, where $z = \text{e}^{ik}$ describes the momentum component normal to the boundary of interest. This effective 1D single-band system has only 2 hoppings,
one is to move $\alpha$ sites to the left by  amplitude $A$, the other with amplitude $B$,  $\beta$ sites to the right. In higher-dimensional contexts, $A$ and $B$ can depend on the momenta from the other directions.

Corresponding to this eigenvalue $E(z)=Az^{\alpha}+Bz^{-\beta}$ is the eigenstate $|\psi(z)\rangle$. In deference to Bloch's theorem, we assume that it takes the form of a \emph{generalized} Bloch state given by the ansatz $|\psi(z)\rangle=(...,\psi_n(z),...)^{\text{T}}$, $\psi_n(z)=z^{n}$, position index $n$, which satisfies the bulk equation
\begin{equation}\label{eqS2}
A\psi_{n+\alpha}(z)+B\psi_{n-\beta}(z)=E(z)\psi_n(z)\ .
\end{equation}
At a particular energy $E(z)$,  there exists other wavefunction solutions $|\psi(z')\rangle$ which satisfies the same eigenvalue $E(z)$, i.e. 
\begin{equation}\label{z'}
\begin{split}
&Az'^{\alpha}+Bz'^{-\beta}-E(z)=0\ .
\end{split}
\end{equation}
This equation above is a polynomial relation of order $\alpha+\beta$ in $z$, and it's easy to get its solutions $z_1,z_2,...,z_{\alpha+\beta}$, which can all be expressed in terms of $z$. These solutions shall provide information about how $z$ is controlled by hopping amplitudes $A$ and $B$.

Hereupon, an eigensolution $ |\Psi \rangle $ with eigenenergy $E(z)$ can be written as
 \begin{equation}
|\Psi \rangle=\sum^{\alpha+\beta}_{i=1}c_i|\psi(z_i)\rangle\  ,
\end{equation}
 with coefficients $c_i$ $(i=1,2,...,\alpha+\beta)$. To determine how they are constrained, we apply open boundary conditions onto $|\Psi \rangle$, arriving at
   \begin{equation}
\Psi_{n'}=\sum^{\alpha+\beta}_{i=1}c_i z_i^{n'}=0\ ,\quad\quad n'=0,-1,-2,...1-\beta\quad \text{or}\quad L+1,L+2,...L+\alpha\ ,
\end{equation}
that is,  there have $\alpha+\beta$ constraints from the open boundary conditions (OBCs), which together combine to form the GBZ characteristic equation
  \begin{equation}\label{eqS2}
\text{det}\ M=\sum^{\alpha+\beta}_{n_1,n_2,...n_{\alpha+\beta}=1}\varepsilon_{n_1,n_2,...,n_{\alpha+\beta}} z_{n_1}^{1-\beta}\times z_{n_2}^{2-\beta}\times...\times z_{n_{\beta}}^{0}\times z_{n_{\beta+1}}^{L+1} \times z_{n_{\beta+2}}^{L+2}\times...\times z_{n_{\alpha+\beta}}^{L+\alpha}=0\ ,
\end{equation}
where the matrix $M$ is $(\alpha+\beta)\times(\alpha+\beta)$ square array , $\varepsilon_{n_1,n_2,...,n_{\alpha+\beta}}$ is $\alpha+\beta$-order arrangement, and $\varepsilon_{n_1,n_2,...,n_{\alpha+\beta}}=-1$, when  $n_1,n_2,...,n_{\alpha+\beta}$ is odd arrangement,   $\varepsilon_{n_1,n_2,...,n_{\alpha+\beta}}=1$ if  $n_1,n_2,...,n_{\alpha+\beta}$ is even arrangement. Since $z_1,z_2,...,z_{\alpha+\beta}$ are all functions of $z$,  the GBZ characteristic equation Eq.\eqref{eqS2} is in actuality only a function of $z$. 

 \begin{figure}[t]
    \begin{centering}
    \includegraphics[width=.65\linewidth]{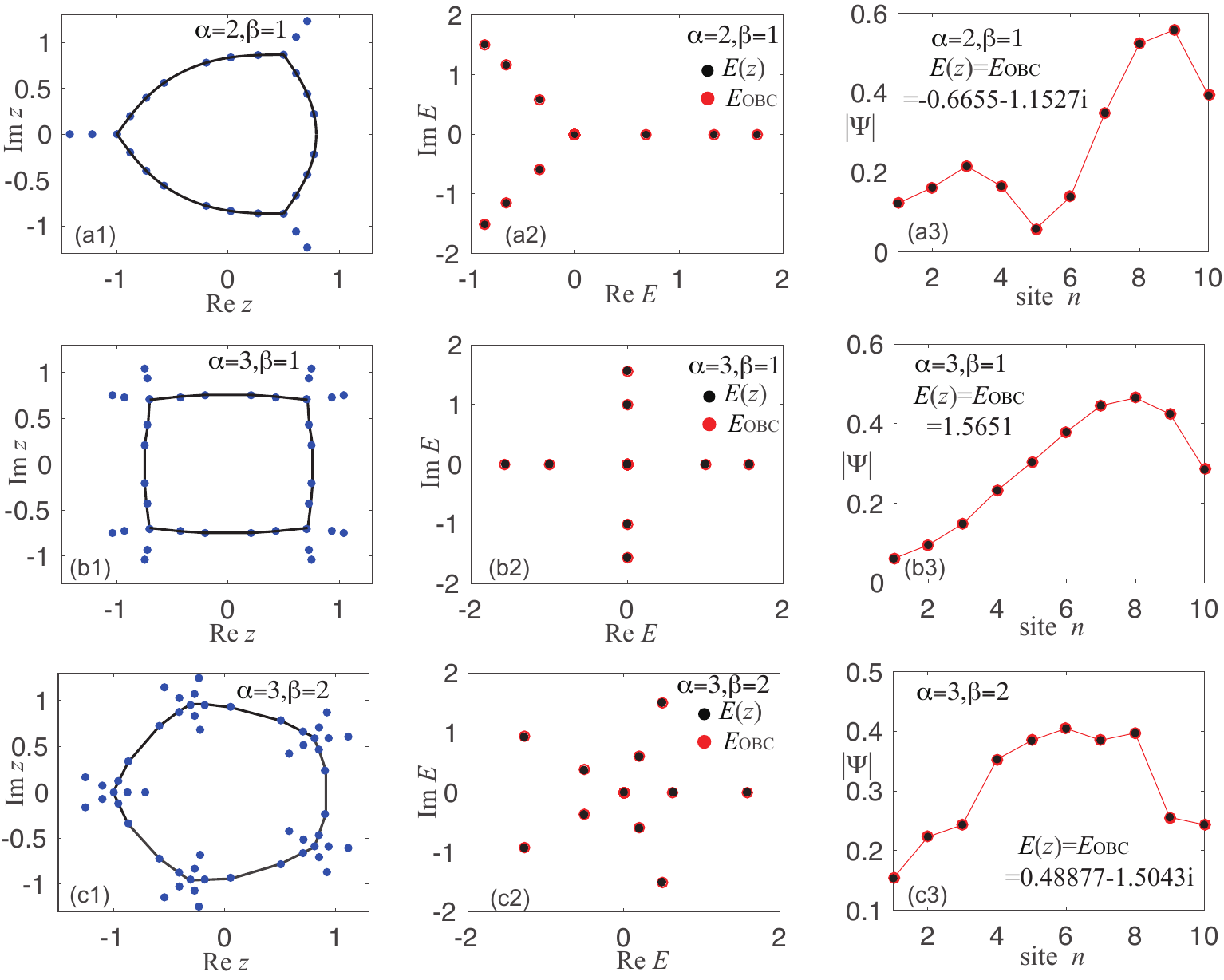}
    \par\end{centering}
    \protect\caption{ 
	(a1--c1) Solutions $z$ to the characteristic equation (Eq.\eqref{eqS2}) for a few illustrative values of $\alpha,\beta$, with $A=B=1$ and system size of $L=10$ sites. Blue dots represent the full set of solutions, but only those lying on the black constraint curves constitute the actual GBZ solutions for OBCs. (a2--c2)  Their corresponding eigenenergies $E(z)=Az^{\alpha}+bz^{-\beta}$ (black) are the set of $z$ points belonging to the GBZ, which agree excellently with $E_{\text{OBC}}$ (red) of Eq.\eqref{eqS1}. (a3--c3) shows the excellent agreement between numerical eigenstate solutions (red) and their corresponding GBZ solutions (black).}
	\label{fig:S1} 
\end{figure}

Taking the model Eq.\eqref{eqS1} with $\alpha=2,\beta=1$ as a sample,  the eigenvalues $E(z)=Az^{2}+B/z$ can be obtained by bulk equation Eq.\eqref{eqS2}. Considering the same eigenenergy $E(z)$,  the eigensolution $ |\Psi \rangle $ can be written as $ |\Psi \rangle =\sum_{i=1}^{3}c_i|\psi(z_i)\rangle $, $\psi_n(z_i)=z_i^n$, and  $z_i$  which satisfy $Az_i^2+B/z_i-E(z)=0$, are constrained by $z$ and hopping amplitude $A,B$
 \begin{equation}\label{eqS7}
z_1=z\ ,\  z_2=-\frac{z}{2}+\sqrt{\frac{B}{Az}+\frac{z^2}{4}}\ ,\ z_3=-\frac{z}{2}-\sqrt{\frac{B}{Az}+\frac{z^2}{4}}\ .
\end{equation}
 Applying open boundary conditions into $|\Psi \rangle$, we can get GBZ characteristic equation
 \begin{equation}\label{EqS8}
 \begin{split}
\text{det}\ M= \text{det}\begin{pmatrix} 1&1&1 \\ z_1^{L+1}&z_2^{L+1}&z_3^{L+1}\\ z_1^{L+2}&z_2^{L+2}&z_3^{L+2}\end{pmatrix}=
(z_1z_2)^{L+1}(z_2-z_1)+(z_1z_3)^{L+1}(z_1-z_3)+(z_2z_3)^{L+1}(z_3-z_2)=0\ ,
\end{split}
\end{equation} which is only a function of $z$. The determinant must vanish so that we have a nontrivial eigenstate solution $(c_1,c_2,c_3)^{\text{T}}$ of the matrix $M$ with eigenvalue $0$. 
It is worth noting that not all the  solution $z$ of  GBZ characteristic equation Eq.\eqref{EqS8}  is  GBZ results of the system, as show in FIG.~\ref{fig:S1}(a1---a3).
To put it simply, for fixed $z$, $z_1, z_2, z_3$ which are from Eq.\eqref{eqS7} 
can be rearranged by $|z_1|\leq|z_2|\leq|z_3|$. If the absolute value $|z_3|$ is not equal to $|z_2|$, we have $\lim\limits_{L\gg1} c_3=0,  (c_1=-c_2)$, that is, the solution $z_3$ does not belong to GBZ. And in other cases, we  can also select GBZ solutions from the set of solution $z$ by the values of the coefficients  $c_1,c_2,c_3$.

\subsection{Determination of the GBZ for OBCs}

However, not all solutions $z$ of the characteristic equation Eq.\eqref{eqS2} contribute to the actual OBC solutions. Those that do define the GBZ. In Figs.~\ref{fig:S1}(a1---c1), we see that of all solutions $z$ (blue dots), only those that lie on the black curves correspond to values of $E(z)$ that coincide with actual numerically obtained eigenvalues (Figs.~\ref{fig:S1}(a2---c2)). \\

%(FIG.~\ref{fig:S1}(a1---c1)),  we must combine the distribution of the wave function (FIG.~\ref{fig:S1}(a3---c3)) together to make sure $z$ corresponding to the system (The blue points $z$ on the black line in FIG.~\ref{fig:S1}(a1---c1)). This is the way to get the exact $z$ in limit size system, and the $z$ could describe the detailed informations of eigenstates, rather than  only the trend of distribution, as shown in FIG.~\ref{fig:S1}.Taking the  $z$ corresponding to the system into energy equation $E(z)=Az^{\alpha}+bz^{-\beta}$, the accuracy of GBZ (=$\{z\}$)is verified by the comparison between $E(z)$ and the energy $E_{\text{OBC}}$ of the numerical diagonalized open boundary system  (Eq.\eqref{eqS1}), in FIG.~\ref{fig:S1}(a2---c2).\\

From known results on GBZ construction~\cite{Lee2019anatomy,yang2019auxiliary,lee2020unraveling}, solutions $z$ that belong to the GBZ (i.e. black curves) are those $z$ such that there exists another $z'\neq z$ such that $|z|=|z'|$ and $E(z)=E(z')$. If more than one such pair of solutions exist, the GBZ is defined by the pair with $|z|=|z'|$ closest to unity. This is because $|z|$ determines the spatial decay rate of the eigensolution, and the GBZ pair is given by the pair of solutions that are mostly slowly decaying, and yet able to satisfy OBCs at both boundaries (which can be arbitrarily far) simultaneously. In other words, the GBZ is obtained through the pair among
%Next, we show how to determine the generalized Brillouin zone with``momentum" $k$, which determines continuum bands ($L\rightarrow\infty$). Let $z_1,z_2,...,z_{\alpha+\beta}$ be the solutions of the eigenvalue equation Eq.\eqref{eqS2}. We find that the condition to get continuum bands is  two of 
$z_1,z_2,...,z_{\alpha+\beta}$ with the same absolute value.  %Putting these  constraints into  expression of solutions $z_1,z_2,...,z_{\alpha+\beta}$, we also can know the form of $z$ . 
\\

Let's now solve for the GBZ of our 1D lattice $H_\text{1D}$, with $z$ parametrized by a wavenumber $k$. First, we write the two degenerate  {(equal energy)} solutions of $z$ as $z_{\pm}=z_0\text{e}^{ik'\pm k }$, with real $k', k$ and possessing the same energy $E(z_\pm)$. Here $z_0$ is a parameter that is unspecified for now. Taking $z_{\pm}$ into the energy equation,
\begin{equation}
\begin{split}
E=E(z_{+})=E(z_{-})=Az_0^{\alpha}\text{e}^{i\alpha (k'+k)}+Bz_0^{-\beta}\text{e}^{i\beta (k'+k)}=Az_0^{\alpha}\text{e}^{i\alpha (k'-k)}+Bz_0^{-\beta}\text{e}^{-i\beta (k'-k)}\ .
\label{Epm}
\end{split}
\end{equation}
As such, we can solve for $z_{\pm}$:
\begin{equation}\label{aaa}
\begin{split}
z_{\pm}&=\left(\frac{B\sin  \beta k}{A\sin \alpha k}\right)^{\frac{1}{\alpha+\beta}}\text{e}^{\pm i k+i\frac{2\pi v}{\alpha+\beta}} \ ,\\
\end{split}
\end{equation}
with $v=1,2,...,\alpha+\beta$. Note that $|z_\pm|\neq 1$ in general, and only when $k=n\pi/(\alpha+\beta)$ do we have $|\sin  \beta k|/|\sin  \alpha k|=1$. % And the distributions of the wave function ask the GBZ of the model Eq.\eqref{eqS1} with  $|A|=|B|$ always have  $|z|>1,<1$ which depends on the value of $\alpha,\beta$, that is, $ |\sin  \beta k|/|\sin  \alpha k|$ must take the value $>1$, or $<1$ in $k$ value range, 
Due to the periodicity of $2\pi/(\alpha+\beta)$, without loss of generality, we shall from now on set the range of parameter $k$ to be $[n\pi-\pi/(\alpha+\beta), n\pi+\pi/(\alpha+\beta))$, with integer $n$. For definiteness, we take $n=0$.\\

In all, the GBZ of our model $H_\text{1D}$ (Eq.\eqref{eqS1}) is given by
\begin{equation}\label{eqS3}
\text{GBZ}_\text{1D}=\left \{z=\left(\left.\frac{B\sin  \beta k}{A\sin \alpha k}\right)^{\frac{1}{\alpha+\beta}}\text{e}^{ i k+i\frac{2\pi v}{\alpha+\beta}}\,\right| k=\left(-\frac{\pi}{\alpha+\beta},\frac{\pi}{\alpha+\beta}\right],v=1,2,..\alpha+\beta \right\}\ ,
\end{equation}
giving energies
\begin{equation}\label{eqS4}
\begin{split}
\bar E(k) =E(z)&=Az^{\alpha}+Bz^{-\beta} =\left(\frac{A^{\beta}B^{\alpha}}{(\sin (\alpha k))^{\alpha}(\sin (\beta k))^{\beta}}\right)^{\frac{1}{\alpha+\beta}} \sin \left((\alpha+\beta)k\right) \text{e}^{\frac{2i\pi \alpha v}{\alpha+\beta}}\ ,
\end{split}
\end{equation}
which agrees closely with numerical OBC eigenenergies in Figs.~\ref{fig:S1}(a2---c2). Here, the bar above $\bar E(k)$ indicates that the energy function is evaluated on the GBZ, i.e. it depends on $k$ not on the ordinary BZ, but through the GBZ. Previously, it was already conceived that complex momentum can be used to represent state decay~\cite{kohn1959analytic,he2001exponential,lee2015free}, but in this GBZ formalism, the imaginary part of the momentum is specifically solved to give the profile of OBC eigenstates.
%\indent Thus, as for the model  $E=Az^{\alpha}+Bz^{-\beta}$, we get  $z=\left({B\sin  \beta k}/({A\sin \alpha k})\right)^{\frac{1}{\alpha+\beta}}\text{e}^{i k+i\frac{2\pi v}{\alpha+\beta}}$ with $k\in \left (-\frac{\pi}{\alpha+\beta},\frac{\pi}{\alpha+\beta} \right]$.
Eq.~\ref{eqS3} is a main result that will be used in the GBZ computation of higher-dimensional cases later on. 
As consistent with the fact that OBC spectra cannot undergo any further NHSE (in the same direction), the spectral winding of $E(z)$ above is always zero, that is, the distribution of $E(z)$ in the complex plane cannot form closed curves.\\
 
In generic 1D lattices with multiple hopping terms, Eq.~\ref{Epm} will have to be replaced by a simultaneously polynomial relations which has to be solved numerically. The resultant GBZ is still well-defined, although it is likely obtainable only numerically.

 %The method introduced in this section are not just for the model with 2 terms, which is still applicable to for the multi-hopping model.\\

\section{II. 2D non-Hermitian lattices with 2D GBZs}\label{sec2}
This section introduces how the GBZ can be obtained for 2D lattices through a few concrete examples with different hopping terms.\\

 {The  Schr\"odinger equation in periodic 2D  potential $U(\bm r+\bm r'_{nm})=U(\bm r)$ with $ \bm r'_{nm}=(nd,md')$ and lattice period $d$,$d'$  reads
\begin{equation}
H \psi(\bm r) =( -\partial^2 +U(\bm r)) \psi(\bm r)=E \psi(\bm r)\ , 
\end{equation}
where $\psi(\bm r)=\sum_{nm} c_{nm}\phi(\bm r-\bm r'_{nm})$, $\bm r=(x,y)$. Assume  $\phi(\bm r)$ is the eigen-wave-function of Hamiltonian $H_0$ with single atom potential $V(\bm r)$(consider only one state)
\begin{equation}
H_0\phi(\bm r)= (-\partial^2 +V(\bm r))\phi(\bm r)=E_0\phi(\bm r)\ ,
\end{equation}
Define $\delta U(\bm r)=U(\bm r)-V(\bm r)$, and substitute following equation
\begin{equation}
H \psi(\bm r) =(H_0+\delta U(\bm r)) \psi(\bm r)=E \psi(\bm r)\ , 
\end{equation}
and 
\begin{equation}
\sum _{n,m}\langle  \phi_{n'm'} |\delta U(\bm r)| \phi_{nm}\rangle c_{nm}=(E-E_0)c_{n'm'}\ ,
\end{equation}
with $\phi_{nm}=\phi(\bm r-\bm r'_{nm}) $. Define: $\langle  \phi_{n'm'} |\delta U(\bm r)| \phi_{nm}\rangle=-J(\bm r'_{nm}-\bm r'_{n'm'})$ and 
\begin{equation}
-\sum_{n,m}J(\bm r'_{nm}-\bm r'_{n'm'}) c_{nm}=(E-E_0)c_{n'm'}.
\end{equation}
Because of the tranformation symmetry of the Hamiltonian, the resulting wavefunction $\psi(\bm r)$  should take Bloch form, which means we should have the solution $c_{nm}=\exp(i\bm k \bm r'_{nm})$
\begin{equation}
E-E_0=-\sum_{n,m}J(\bm r'_{nm})\exp(-i\bm k \bm r'_{nm})\ .
\end{equation}
if consider the specific hoppings,  define $J_0=J(\bm 0), J_{nm}= J(\pm \bm r'_{nm})$, then we have 
\begin{equation}
E-E_0+J_0=\sum_{nm} J_{nm}\exp(-i\bm k \bm r'_{nm})
\end{equation}
In the second quantization language, the expectation value of energy becomes a operator, set $H= -\partial^2 +U(\bm r)$, we have
\begin{equation}
\hat{H}=\sum_{nm,n'm'} \hat{c}^\dagger_{nm}  H_{nm,n'm'}\hat{c}_{n'm'}
\end{equation}
with $\psi\rightarrow \hat{\psi}=\sum \hat{c}_{nm}\psi_{nm}$, $H_{nm,n'm'}=\langle  \phi_{n'm'} |H| \phi_{nm}\rangle$, $ \phi_{nm}$ is a orthonormal and complete basis in Hilbert space, like plane-waves $\exp(i \bm k \bm r)$ or energy eigen-states of  $H_0$, $H$  is the energy operator in single particle first quantization picture, which can only act on Hilbert space, while the second quantization energy operator $\hat H$ acts on Fock space. Here, in tight-binding method, $\phi_{nm}$ is the wave-function of site $\bm r'_{nm}$ of the energy eigen-state $H_0$. Consider the specific hoppings, we have 
\begin{equation}
\hat{H}=\sum_{nm} J_{\alpha,\beta} \hat{c}^{\dagger}_{n+\alpha,m+\beta} \hat{c}_{n,m}\ ,
\end{equation}
results in a simple Hamiltonian  and allows for quick computations. And in our manuscript, all the hamiltonian second quantization formulation.}

\subsection{2D lattice model with 2 hopping terms}
To connect with our previous treatment of 1D lattices with 2 hopping terms, we first consider the simplest case of 2D lattices with 2 hopping terms: 
\begin{equation}\label{eqS21}
H_\text{2D,2}= \sum_{m,n}t|m,n\rangle\langle m+\alpha,n+a|+t'|m,n\rangle\langle m-\beta,n-b|\ ,
\end{equation}
with hoppings of amplitudes $t$ (or $t'$) corresponding to transitions of $(-\alpha,-a)$ sites ( or $(\beta,b)$ sites) on the lattice.\\

We consider the ansatz $|\psi\rangle\propto\sum_{m,n}\psi_{mn}|m,n\rangle$ with $\psi_{mn}=z_x^mz_y^n$. % into the bulk and boundary conditions, 
In the bulk, it gives eigenenergies $E(z_x,z_y)$
  \begin{equation}
E(z_x,z_y)=t z_x^{\alpha}z_y^a+t' z_x^{-\beta}z_y^{-b}.
\label{Ezxzy}
\end{equation}
For a fixed $z_x$, there are $a+b$ solutions $z_{y,i}$, ($i=1,2,...,a+b$)  corresponding to the same energy $E(z_x,z_y)$. Similarly, we can also find $\alpha+\beta$ solutions $z_{x,j}$, ($j=1,2,...,\alpha+\beta$) corresponding to fixed $E(z_x,z_y)$ and $z_y$. Both sets of solutions $z_{y,i}$, ($i=1,2,...,a+b$) and $z_{x,j}$, ($j=1,2,...,\alpha+\beta$) can be written entirely in terms of $z_x$ and $z_y$.\\

We next show a detailed for derivation of the GBZ of $H_\text{2D,2}$. For a quick heuristic approach, the reader may directly skip to Eqs.~\ref{eqS22} and~\ref{eqS23}. 

Implementing open boundary conditions on the wavefunction $|\Psi\rangle$ for a fixed eigenenergy $E(z_x,z_y)$ gives
    \begin{equation}\label{eqS21.5}
    \begin{split}
|\Psi\rangle=|\Psi^1\rangle=|\Psi^2\rangle\ \propto \sum_{m,n}\Psi_{mn}|m,n\rangle=\sum_{m,n}\Psi^1_{mn}|m,n\rangle=\sum_{m,n}\Psi^2_{mn}|m,n\rangle\ ,\\
\quad \Psi^1_{mn}=\sum_{z_y} \left( \left(\sum_{j}^{\alpha+\beta}f_{z_{x,j}}z_{x,j}^m\right)g_{z_y} z_y^n\right)\ , \quad \Psi^2_{mn}=\sum_{z_x}\left( f_{z_x} z_x^m \left(\sum_{i}^{a+b}{g}_{z_{y,i}}z_{y,i}^n\right)\right)\ ,\\
\end{split}
\end{equation}
Here, we have written $|\Psi\rangle$ either as $|\Psi^1\rangle$ or $|\Psi^2\rangle$, depending on the order of expansion in terms of the $x$ and $y$ coefficients. % are the different manifestations of the wave function $|\Psi\rangle$, 
These two ways to expand are of course equivalent, and note the relation $f_{z_x}g_{z_{y,i}}=g_{z_{y,i}}f_{z_x}$. The $\sum_i,\sum_j$ summations in $\Psi^{1}_{mn},\Psi^{2}_{mn}$ refers to sums over all sets $z_x,z_y$ which have same energy $E$. To treat boundary conditions along different direction (i.e.,${\bm x, \bm y}$ directions), we can choose the appropriate expansion and proceed like in the 1D case (see Sec.~\ref{sec1}), paying careful attention to the indices i.e. for OBCs along the ${\bm x}$ direction, we have  $\Psi_{m'{n}}=\Psi^1_{m'{n}}=0 $ for any $n$ and $m'=-\beta+1,-\beta+2,...0,L+1,L+2,..L+\alpha $ (assuming $\alpha,\beta>0$). %and any values $n$, like `1D' system with paramters $z_y$ in Sec.~\ref{sec1}.
Thus we obtain%,  for the same energy, we have
\begin{equation}\label{eqS22}
\begin{split}
z_x^{\alpha+\beta}=\frac{t'}{t }\frac{z_1^{\alpha+\beta}}{z_y^{a+b}}, \quad z_1^{\alpha+\beta}=\frac{\sin \beta k_1}{\sin \alpha k_1}\text{e}^{i(\alpha+\beta)k_1}\ ,
\end{split}
\end{equation}
with momentum parameter $k_1\in  \left(-\frac{\pi}{\alpha+\beta},\ \frac{\pi}{\alpha+\beta}\right]$ as in a quasi-1D case (see Sec.~\ref{sec1}. for details). Likewise, considering the expansion via $|\Psi\rangle=|\Psi^2\rangle$  with boundary conditions along the ${\bm y}$ direction, we obtain the alternative expressions 
\begin{equation}\label{eqS23}
\begin{split}
z_y^{a+b}=\frac{t'}{t }\frac{z_2^{a+b}}{z_x^{\alpha+\beta}}, \quad z_2^{a+b}=\frac{\sin b k_2}{\sin a k_2} \text{e}^{i(a+b)k_2}\ ,
\end{split}
\end{equation}
with $k_2\in  \left(-\frac{\pi}{a+b},\ \frac{\pi}{a+b}\right]$. To solve for the GBZs of simultaneous x-OBCs and y-OBCs in 2D, we simultaneously solve Eqs.~\ref{eqS22} and~\ref{eqS23}. Note that heuristically, Eqs.~\ref{eqS22} and~\ref{eqS23} can be simply written down by treating $H_\text{2D,2}$ as a 1D chain with hoppings dependent on the transverse momentum. For instance, we can write Eq.~\ref{Ezxzy} either as $E(z_x)=(tz_y^a)z_x^\alpha+(tz_y^{-b})z_x^{-\beta}$ or $E(z_y)=(tz_x^\alpha)z_y^a+(tz_x^{-\beta})z_y^{-b}$, and apply Eq.~\ref{eqS3} to obtain Eqs.~\ref{eqS22} and~\ref{eqS23} respectively. Importantly, we shall see that this approach fails in general 2D lattices, even though it is valid in this case where there are only two hopping terms from each site.
%The  different manifestations $|\Psi^{1,2}\rangle$  are just to make it easier to explain the results of boundary conditions in different directions, since the two wave functions $|\Psi^{1,2}\rangle$ are equivalent, the results $z_x,z_y$, that satisfy all boundary conditions at the same time, can be obtained by connecting equation Eq.(\ref{eqS22},\ref{eqS23}). \\
\begin{figure}[h]
\centering
\includegraphics[width=.75 \linewidth]{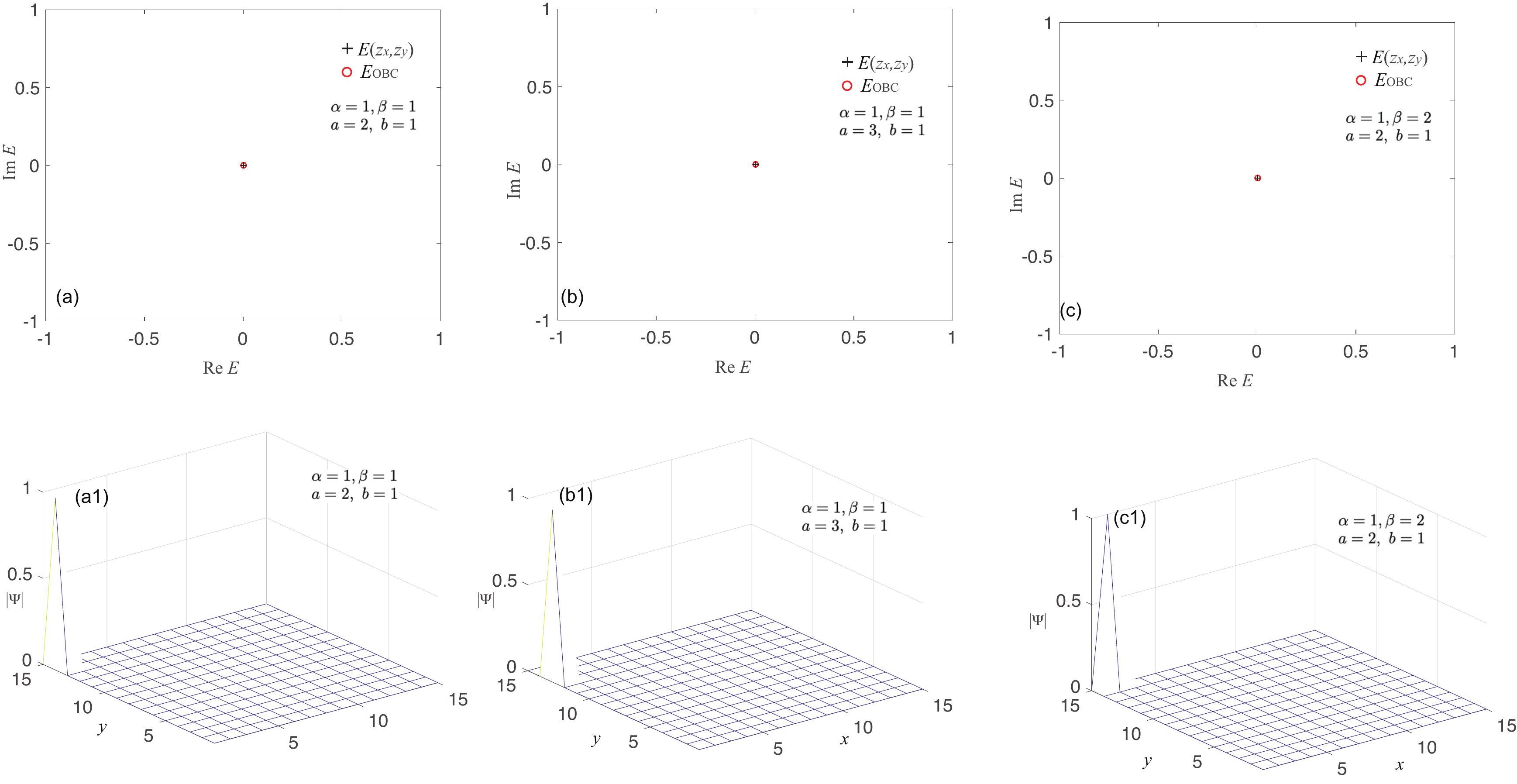}
\caption{Spectra and corner-localized eigenstates for $H_\text{2D,2}$ in the $\alpha/\beta\neq a/b$ case. (a--c) Exactly zero (flatband) OBC eigenenergies $E_{\text{OBC}}$ due to non-Bloch collapse, which coincide with $E(z_x,z_y)$ from Eq.\eqref{eqS25}. (a1---c1) Corresponding illustrative eigenstates are perfectly localized at the boundary dictated by the direction of localization. The model parameters are $L_x=L_y=15$, $t=2,\quad t'=1$  and the specific values of $\alpha,\beta,a,b$ are indicated in the figure.}\label{fig:S22}
\end{figure}  

\begin{figure}[h]
\centering
\includegraphics[width=.75 \linewidth]{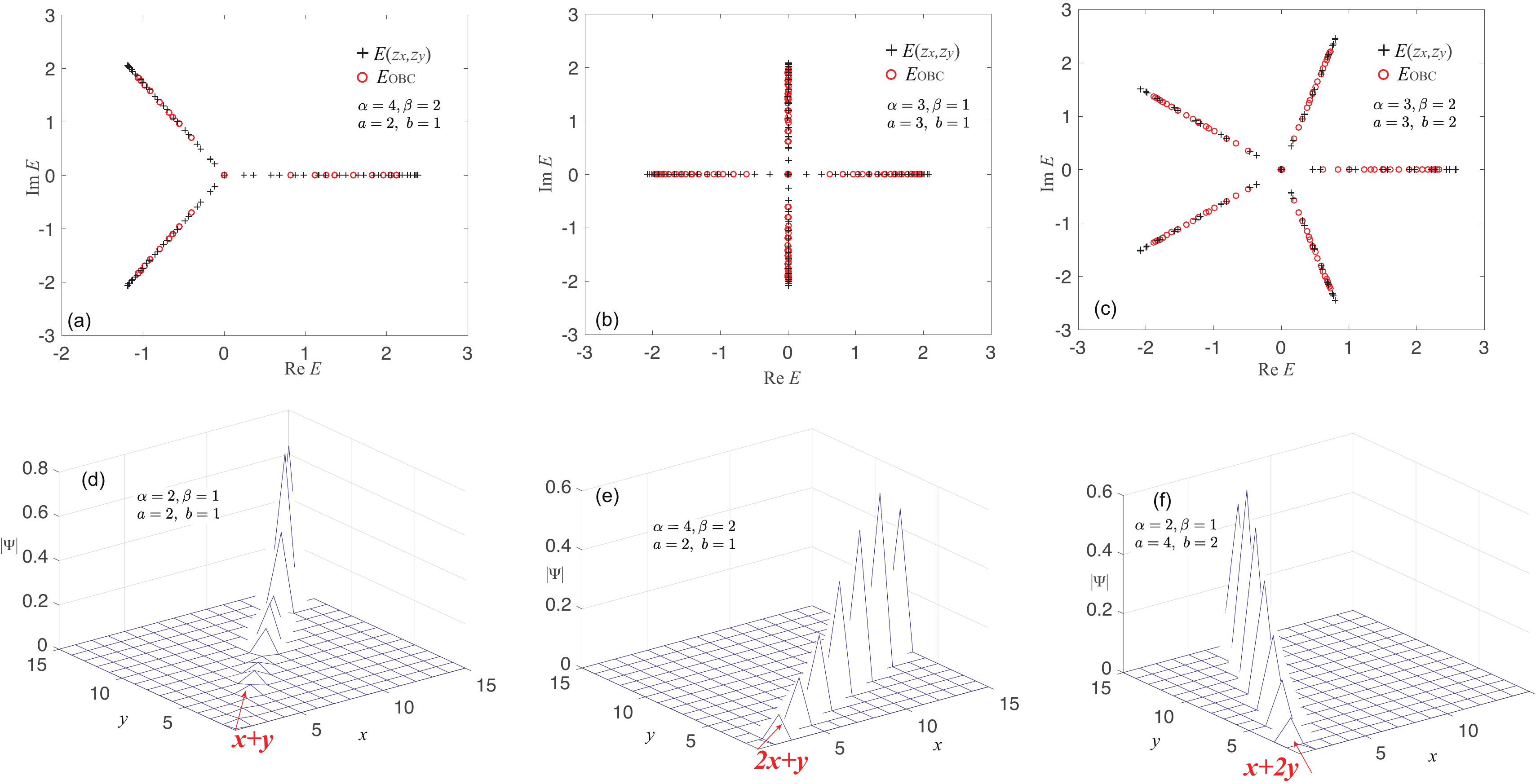}
\caption{Spectra and illustrative eigenstates for $H_\text{2D,2}$ in the $\alpha/\beta=a/b$ case. (a--c) Excellent agreement between the GBZ eigenenergies $E(z_x,z_y)$ from Eq.\eqref{eqS24} and OBC eigenenergies $E_{\text{OBC}}$ of Eq.\eqref{eqS21}. (d--f) Spatial profiles for the $E_{\text{OBC}}=0$ eigenstate of a few other illustrative cases, clearly showing that the skin states are aligned along the $\alpha \hat x +a \hat y$ (or $\beta \hat x +b \hat y$) direction. The model parameters are $L_x=L_y=15$, $t=1,\  t'=2$  and the specific values of $\alpha,\beta,a,b$ are indicated in the figure.}\label{fig:S21}
\end{figure}

In the following, we shall specialize to the case where the two hoppings are pointing in the same direction i.e. $\alpha /\beta=a/b$, for reasons that will be become evident soon. In this case, the GBZ and energy satisfy 
\begin{equation}\label{eqS24}
\begin{split}
&\text{GBZ}_\text{2D,2}=\left\{\left.z_x^{\alpha+\beta}z_y^{a+b}=\frac{t'\sin b k}{t\sin a k} \text{e}^{i(a+b)k}\right|k=\left(-\frac{\pi}{a+b},\frac{\pi}{a+b}\right]\right\}\ ,\\
&\bar E(k)=E(z_x,z_y)=\left(\frac{t^{b}t'^{a}}{(\sin (a k))^{a}(\sin (b k))^{b}}\right)^{\frac{1}{a+b}} \sin \left((a+b)k\right) \text{e}^{\frac{2i\pi  av}{a+b}}\ ,
\end{split}
\end{equation}
with $k=\left(-\frac{\pi}{a+b},\frac{\pi}{a+b}\right]$ and $v=1,2,...,a+b$, as plotted in FIG.~\ref{fig:S21} (a---c). As this case only contains the combination $z_x^{\alpha+\beta}z_y^{a+b}$,  rather than $z_x^{\alpha+\beta}$ and $z_y^{a+b}$, it is essentially a 1D model along the $\alpha\bm x+a\bm y$ direction, which is consistent with the results of numerical diagonalization as shown in FIG.~\ref{fig:S21}(a1---a3). \\

We next discuss the other case with $\alpha /\beta\neq a/b$, where $|z_1|$ and $|z_2|$ can only coincide at $z_1^{\alpha+\beta}=z_2^{a+b}=-1$. Hereby, its GBZ and energy are simply given by
\begin{equation}\label{eqS25}
\begin{split}
&\text{GBZ}_\text{2D,2}=\left\{z_x^{\alpha+\beta}z_y^{a+b}=-\frac{t'}{t}\right\}\ ,\\
&\qquad\qquad \quad  E(z_x,z_y)=0\ ,
\end{split}
\end{equation}
As confirmed in FIG.~\ref{fig:S22}(a--c), the eigenenergies are indeed zero, and the eigenstates (FIG.~\ref{fig:S22}(a1--c1)) are perfectly localized at a corner determined by $\alpha/\beta$, $a/b$ and $t/t'$. This is because of uncompensated unbalanced hoppings along at least one direction, which leads to non-Bloch collapse~\cite{longhi2020non,crespi2013dynamic}. Since there is no nontrivial dynamics to speak of in this case, we shall not discuss it further.

\begin{figure}[h]
\centering
\includegraphics[width=.65 \linewidth]{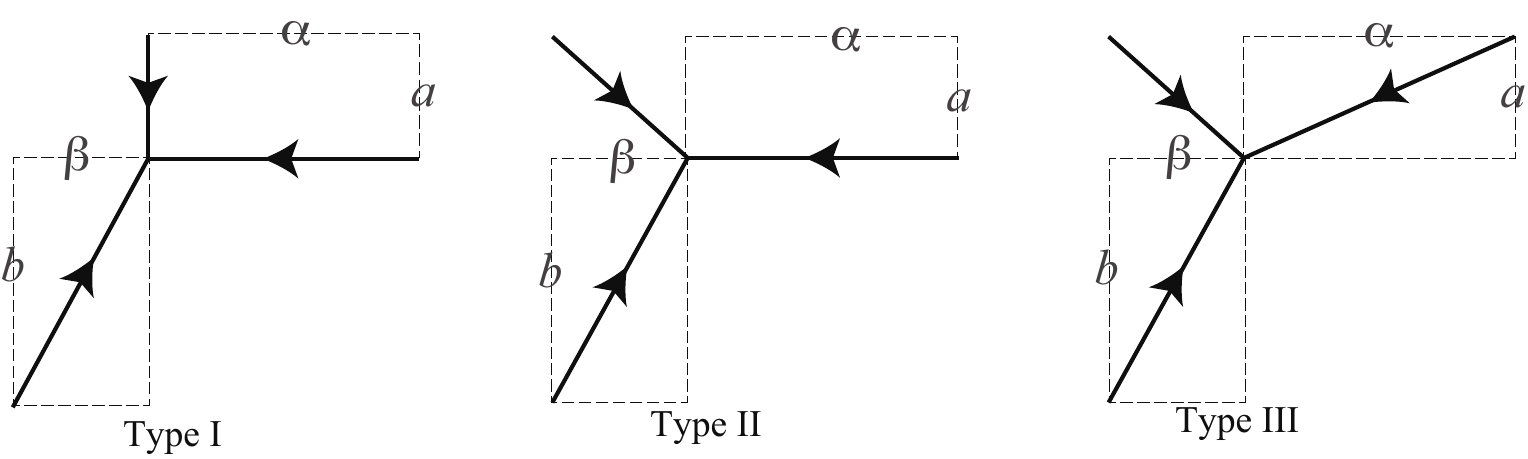}
\caption{The three directed hopping configurations in each of our type I,II and III models. For type I, two hoppings are orthogonal; for type II, two hoppings have common $\beta$ displacements in the direction antiparallel to the third hopping; for type III, there exists two common $\beta$ component displacements orthogonal to the two common $a$ displacements. }\label{fig:S221}
\end{figure}  

%For these type of model Eq.\eqref{eqS21}, the bulk, boundary conditions only give us the distribution of $z_x^{\alpha+\beta}z_y^{a+b}$ rather than $z_x$ and $z_y$ separately. Both eigenvalues and eigenstates can prove the correctness of the results  $z_x^{\alpha+\beta}z_y^{a+b}$.\\

% And this type of model is also a good example that we can not get 2D GBZ results from the single-OBC method.  The single-OBC method refers to we can get the `1D' GBZ from the single-OBC model, then opening boundary in all directions to get the other GBZ and the distribution of energies. It is worth noting that when the other GBZ is obtained, its boundary conditions cannot be guaranteed to correspond to the real system, especially for the model in this section and the results of energies are not the eigenvalues of the model Eq.\eqref{eqS21}.\\
   
%   The method introduced in this section is suitable for all the models , not only for models with only 2 hopping terms.

\subsection{2D lattice model with 3 hopping terms}

We next discuss 2D lattices with 3 unbalanced hopping terms, such that their combined effect is no longer either trivial or just that along a 1D subspace. We shall study two types of hopping configurations here, and reserve the third type, which turns out to interestingly requires a GBZ of lower dimensionality, to the next section on its own.

%This subsection introduces 2 types of  2D model with 3 terms to get their GBZ results. 
  \subsubsection{Type I: 2D lattice model with 3 hopping terms, 2 perpendicular to each other} 
We first consider type I models, the simplest of ``entangled'' 2D lattice models. They contain 2 perpendicular terms $t_1,t_2$ in directions $(\alpha,0)$ and $(0,a)$, and a third one in an oblique direction $-(\beta,b)$. We have 
  \begin{equation}\label{eqS210}
H_\text{2D,I}= \sum_{m,n}t_1|m,n\rangle\langle m+\alpha,n|+ t_2|m,n\rangle\langle m,n+a|+t_3|m,n\rangle\langle m-\beta,n-b|\ .
\end{equation}
%that 2 of hopping terms are along the $\bm x, \bm y$ directions, the others is not. 
Similarly as before, by substituting the ansatz $|\psi\rangle\propto\sum_{m,n}\psi_{mn}|m,n\rangle$ with $\psi_{m,n}=z_x^mz_y^n$ into the bulk equations,  we obtain the energy $E(z_x,z_y)$
  \begin{equation}\label{EqS211}
E(z_x,z_y)=t_1 z_x^{\alpha}+t_2z_y^a+t_3z_x^{-\beta}z_y^{-b}\ .
\end{equation}\\
By heuristically regarding $E(z_x,z_y)$ as 1D models $E(z_x)$ or $E(z_y)$ with nonconstant hoppings and using Eqs.~\ref{eqS3} and~\ref{eqS4}, or by considering boundary conditions on the wave function $|\Psi\rangle$ (Eq.\eqref{eqS21.5}), we obtain the following relationships between $z_x$ and $z_y$:
 \begin{equation}\label{EqS212}
 \begin{split}
z_x^{\alpha+\beta}z_y^b&=\frac{t_3}{t_1} z_1^{\alpha+\beta},\quad z_1^{\alpha+\beta}=\frac{\sin \beta k_1}{\sin \alpha k_1} \text{e}^{i(\alpha+\beta)k_1},\  k_1\in  \left(-\frac{\pi}{\alpha+\beta},\ \frac{\pi}{\alpha+\beta}\right]\ ,\\
z_x^{\beta}z_y^{a+b}&=\frac{t_3}{t_2} z_2^{a+b},\quad\  z_2^{a+b}=\frac{\sin b k_2}{\sin a k_2} \text{e}^{i(a+b)k_2},\quad k_2\in  \left(-\frac{\pi}{a+b},\ \frac{\pi}{a+b}\right]\ .
 \end{split}
\end{equation}
Substituting these $z_x$ and $z_y$ into the energy expression Eq.\eqref{EqS211}, we obtain the GBZ energies (Note the slight abuse of notation - below, we write $E(z_x,z_y)=E(z_1,z_2)$ to convey that the set of $E$ depends on $z_1,z_2$ through $z_x,z_y$, even though $E$ of course takes different functional dependencies in either case. )
 \begin{equation}\label{eqS213}
 \begin{split}
\bar E(k_1,k_2)=E(z_x,z_y)&=\left( \frac{t_1^{\beta a}t_2^{\alpha b}t_3^{\alpha a}}{z_1^{(\alpha+\beta)a\beta}z_2^{(a+b)b\alpha}} \right)^{\frac{1}{(\alpha+\beta)a+\alpha b}}\times\left(1+z_1^{\alpha+\beta}+z_2^{a+b}\right)\ , \\
 \end{split}
\end{equation}
and
 \begin{equation}\label{eqS214}
 \begin{split}
 \text{GBZ}_\text{2D,I}=\left\{ z_x,z_y \left|
z_x^{(\alpha+\beta)a+\alpha b}=\frac{t_2^{b}t_3^a}{t_1^{a+b}}\frac{z_1^{(\alpha+\beta)(a+b)}}{z_2^{(a+b)b}} \ ,\,
z_y^{(\alpha+\beta)a+\alpha b}=\frac{t_1^{\beta}t_3^{\alpha}}{t_2^{\alpha+\beta}}\frac{z_2^{(\alpha+\beta)(a+b)}}{z_1^{(\alpha+\beta)\beta}}\right.\right\}\ ,
 \end{split}
\end{equation}
with $z_1,z_2$ functions of $k_1,k_2$, as defined in Eq.\eqref{EqS212}. Some comments on the relationship between $z_x,z_y$ and $z_1,z_2$ are in order. As defined by the ansatz $|\psi(x,y)\rangle\propto \sum_{m,n}z_x^mz_y^n|m,n\rangle$, $z_x,z_y$ control the spatial growth and decay of the wavefunction. However, unlike in an ``unentangled'' case like $H_\text{2D,2}$, both $z_x$ and $z_y$ depend on the GBZ spanning parameters $k_1,k_2$ in complicated manners given by Eq.~\ref{EqS212}. In other words, Eq.~\ref{EqS212} dictates how the ``non-Bloch'' scaling factors $z_x,z_y$ are related to the GBZ coordinates $k_1,k_2$ through intermediate quantities $z_1,z_2$, which are related to the effective 1D chain projections of the Hamiltonian.

\begin{figure}[h]
\centering
\includegraphics[width=.99\linewidth]{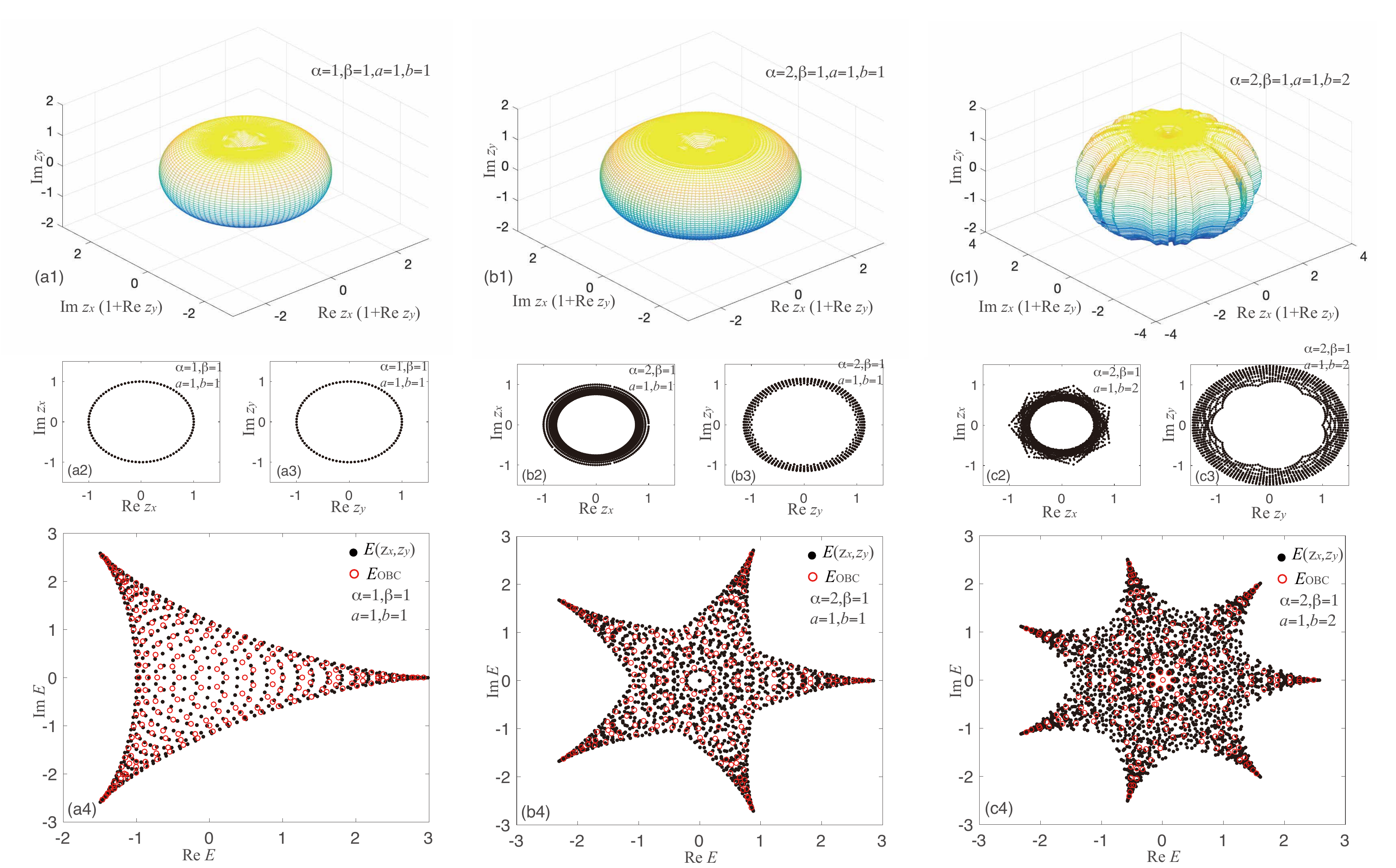}
\caption{Plots of the GBZ torus (Upper row a1-c1) , GBZs $z_x$ (Middle row a2-c2) and $z_y$ (Middle row a3-c3), as well as the energy spectra (Lower row a4-c4) of illustrative type-I lattices ($H_\text{2D,I}$ from Eq.~\eqref{eqS210}) with hoppings given by parameters $\alpha,\beta,a,b$. The upper row plots (a1-c1) are parametrized such that a torus is traced out in the trivial case without NHSE ($|z_x|=|z_y|=1$ for all $k_1,k_2$); departures from a toroidal shape depict the extent of 2D NHSE. Belonging to a 2D model, each of $z_x,z_y$ (a1-c1,a2-c2,a3-c3,) traces out a 2D region parametrized by $k_1,k_2$ (Eq.~\ref{eqS214}) as its GBZ, even though it trivially collapses into 1D loops for case (a). Perfect agreement of GBZ spectra $E(z_x,z_y)$ from Eq.~\eqref{eqS213} with OBC spectra $E_{\text{OBC}}$ is demonstrated for all cases, which for this model fills the interior of a $[(\alpha+\beta)a+\alpha b]$-sided figure (a4-c4).
Parameters are $L_x=L_y=15$, $t_1=t_1=t_3=1$, and the GBZ predictions are generated with a mesh defined by $k_1=-\frac{\pi}{\alpha+\beta}:\frac{\pi}{30}:\frac{\pi}{\alpha+\beta}$,  $k_2=-\frac{\pi}{a+b}:\frac{\pi}{30}:\frac{\pi}{a+b}$.}\label{fig:S23}
\end{figure}

From this explicit expression Eq.~\ref{eqS214}, we can already deduce that the locus of $E$ is a star-like set of branches, since $E$ factorizes into a product of terms involving the phase factors of $k_1$ and $k_2$ separately. By considering these phase factors, we find that the number of branches for a type-I hopping lattice is given by
\begin{equation}
\text{LCM}\left[\displaystyle\frac{(\alpha+\beta)a+\alpha b}{\text{GCD}((\alpha+\beta)a+\alpha b,(\alpha+\beta)a\beta)},\displaystyle\frac{(\alpha+\beta)a+\alpha b}{\text{GCD}((\alpha+\beta)a+\alpha b,(a+b)\alpha b)}\right].
\end{equation}
Their agreement with numerical OBC results is given in FIG.~\ref{fig:S23}. The phase factor is given by $\exp(\frac{2\pi i( (\alpha+\beta)a\beta k_1+(a+b)\alpha bk_2}{(\alpha+\beta)a+\alpha b})$, which evaluates to $\exp(\frac{2\pi i(  2k_1+2k_2)}{3}) \Leftrightarrow 3$ branches for (a4); $\exp(\frac{2\pi i (  3k_1+4k_2)}{5}) \Leftrightarrow 5$ branches for (b4); and $\exp(\frac{2\pi( 3 k_1+12 k_2)}{7}) \Leftrightarrow 7$ branches for (c4). 

\clearpage
\subsubsection{Type II: 2D lattice model with 3 hopping terms, two with common $\beta$ displacement} 
\begin{figure}[h]
\centering
\includegraphics[width=.99\linewidth]{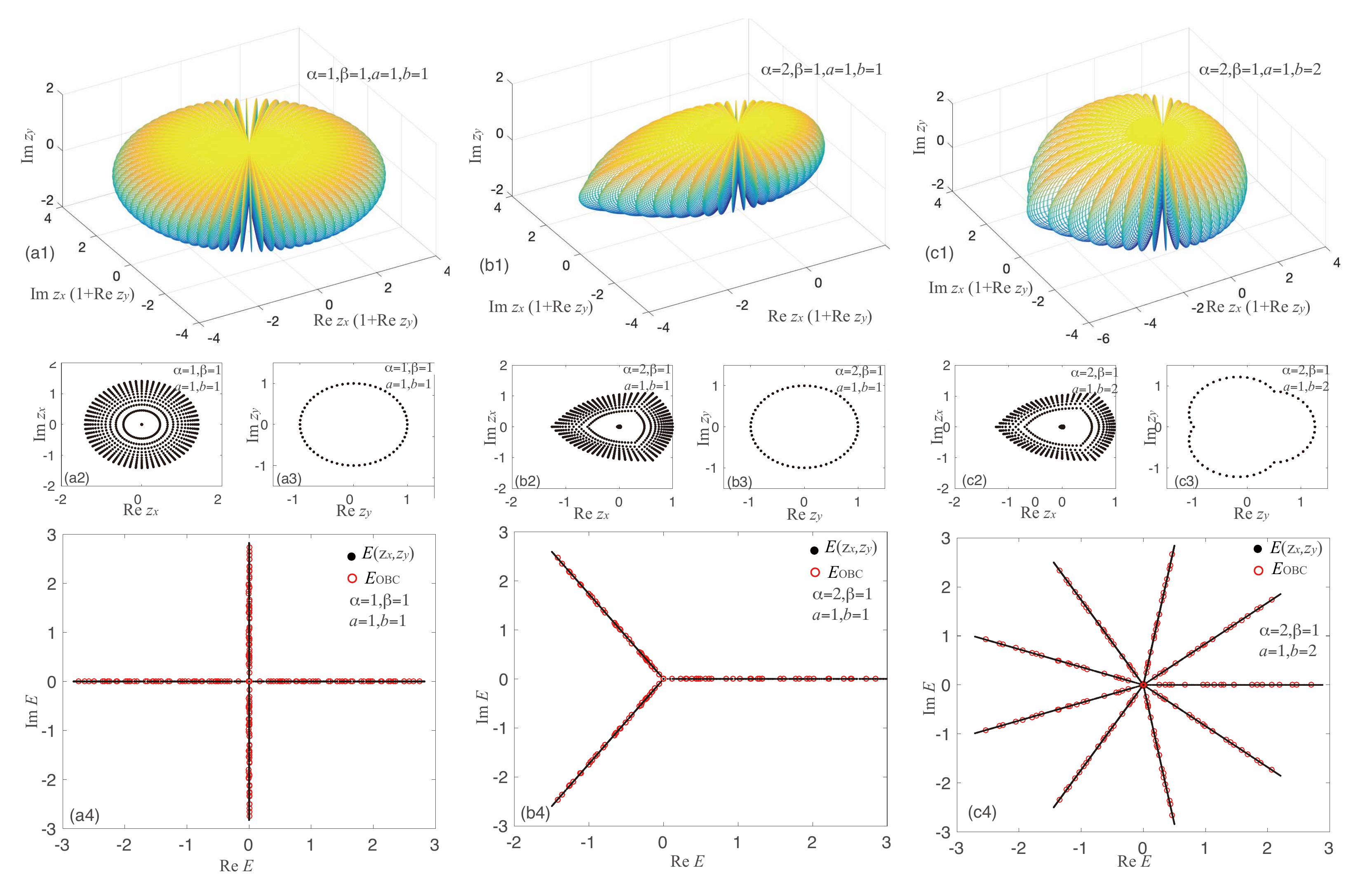}
\caption{
Plots of the  GBZ torus (Upper row a1-c1) , GBZs $z_x$ (Middle row a2-c2) and $z_y$ (Middle row a3-c3), as well as the energy spectra (Lower row a3-c3)of illustrative type-II lattices ($H_\text{2D,II}$ from Eq.~\eqref{eqS215}) with hoppings given by parameters $\alpha,\beta,a,b$. (a1-c1,a2-c2,a3-c3) While $z_x$ traces out a 2D region parametrized by the two ``momenta'' $k_1$ and $k_2$, $z_y$ only depends on one such momentum parameter, as given by Eq.~\eqref{eqS216}), thereby tracing out a 1D GBZ loop (a3-c3). (a4-c4) Perfect agreement of GBZ spectra $E(z_x,z_y)$ from Eq.~\eqref{eqS216e} with OBC spectra $E_{\text{OBC}}$ is demonstrated for all cases, with their star-like spectra consistent with the 1D nature of their effective GBZ description. The $|z_x|=0$ central dot corresponds to $k_2=\pm\pi/(a+b)$ where the factor $\left(z_2^a+z_2^{-b}\right)$ in $z_x$ is equal to 0. Parameters are $L_x=L_y=15$, $t_1=t_1=t_3=1$, and the GBZ predictions are generated with a mesh defined by $k_1=-\frac{\pi}{\alpha+\beta}:\frac{\pi}{30}:\frac{\pi}{\alpha+\beta}$, $k_2=-\frac{\pi}{a+b}:\frac{\pi}{30}:\frac{\pi}{a+b}$. 
}
\label{fig:S24}
\end{figure}  

We next consider a slightly more complicated 2D lattice model (type II). None of the hoppings are orthogonal to each other, so all 3 hoppings are ``entangled''. However, it has the simplifying property that the $t_2$ and $t_3$ hoppings are equidistant in the direction parallel to the $t_1$ hopping, such that if we set the $t_1$ hopping  normal against the horizontal $x$ open boundary, there are only two unique hopping distances in this direction. The Hamiltonian is given by
  \begin{equation}
H_\text{2D,II}= \sum_{m,n}t_1|m,n\rangle\langle m+\alpha,n|+ t_2|m,n\rangle\langle m-\beta,n+a|+t_3|m,n\rangle\langle m-\beta,n-b|\ .
\label{eqS215}
\end{equation}
Taking the ansatz $|\psi\rangle\propto\sum_{m,n}\psi_{mn}|m,n\rangle$ with $\psi_{m,n}=z_x^mz_y^n$ into the bulk equations, we obtain the energy $E(z_x,z_y)$ 
  \begin{equation}\label{eqS216e}
E(z_x,z_y)=t_1 z_x^{\alpha}+t_2z_x^{-\beta}z_y^a+t_3z_x^{-\beta}z_y^{-b}\ .
\end{equation}
Considering the boundary conditions in the same way as before, we have two relations between $z_x$ and $z_y$ from which the GBZ and OBC energy $E(z_x,z_y)$ can be obtained:
 \begin{equation}\label{eqS216}
 \begin{split}
 \text{GBZ}_\text{2D,II}=\left\{z_x,z_y\left|z_x^{\alpha+\beta}=t_1^{-1}t_2^{\frac{b}{a+b}}t_3^{\frac{a}{a+b}}\left(z_2^a+z_2^{-b}\right)z_1^{\alpha+\beta}, z_y^{a+b}=\frac{t_3}{t_2} z_2^{a+b}\right.\right\}\ ,
 \end{split}
\end{equation}
\begin{equation}\label{eqS217}
\bar E(k_1,k_2)=E(z_x,z_y)=t_1^{\frac{\beta}{\alpha+\beta}} \left(t_2^{\frac{b}{a+b}}t_3^{\frac{a}{a+b}}\left(z_2^a+z_2^{-b}\right)\right)^{\frac{\alpha}{\alpha+\beta}}\left(z_1^{\alpha}+z_1^{-\beta}\right),
\end{equation}
the forms of $z_1,z_2$ given in Eq.\eqref{EqS212} as before. 

%\subsubsection{Number of OBC spectral branches for type II hopping lattices}

From the explicit expression Eq.~\ref{eqS217}, we can similarly deduce that the locus of $E$ is a star-like set of branches. We find that the number of branches for type-II lattice hoppings is given by
%\begin{equation}
%\text{max}\,[\text{GCD}((a+b)(\alpha+\beta),a\alpha),\text{GCD}((a+b)(\alpha+\beta),a\alpha)].
%\end{equation}
\begin{equation}
\text{LCM}\left[\displaystyle\frac{(a+b)(\alpha+\beta)}{\text{GCD}((a+b)(\alpha+\beta),a\alpha)},\frac{\alpha+\beta}{\text{GCD}(\alpha+\beta,\alpha)}\right].
\end{equation}
We verify this formula with the three examples in FIG.~\ref{fig:S24}, for which there exist excellent agreement between the numerical OBC results and the above GBZ expression for $E$. The phase factor is given by$\exp(\frac{2\pi( a\alpha k_1+\alpha(a+b))k_2}{(a+b)(\alpha+\beta)}$, which evaluates to $\exp(\frac{2\pi i(  k_1+2k_2)}{4}) \Leftrightarrow 4$ branches for (a4); $\exp(\frac{2\pi i (  k_1+k_2)}{6/2}) \Leftrightarrow 3$ branches for (b4); and $\exp(\frac{2\pi( 2 k_1+6k_2)}{9}) \Leftrightarrow 9$ branches for (c4). It is not surprising that these type II 2D lattice models considered above have star-like spectra that resemble that of 1D NHSE models, since they after quickly reduce to a simple 1D effective model with x-OBCs with 2 effective hoppings.

\begin{comment}
%Section removed because example (a) is already going to be presented in the main text Fig 3b, and (b) does not agree well numerically
\subsection{Failure of successively taking GBZs in different directions}
\CH{CH: Will be good to have another figure showing additional examples where our approach gives excellent agreement with the numerical spectrum, but which taking x and y OBCs sequentially doesnt. Maybe plot something analogous to the spectral plots in the above figures, but with points of another color representing the sequential approach that doesnt agree. No need to describe these models in the text (I will intro them somewhere) We can put this subsection either here, or in Sect III.}
\begin{figure}[h]
\centering
\includegraphics[width=5.0in]{gFIGS25.pdf}
\caption{ {JH: I think the type I model with $\alpha=\beta=a=b=1$ is the best example to show that taking x and y OBCs sequentially doesn't give excellent agreement with the numerical spectrum. I also take the type I model with $\alpha=a=b=1,\beta=2$ in (b) as another example, it not as good as (a) because the size of system. And for type II model , taking x and y OBCs sequentially have the same results with our  our approach. If we consider the model without type I II,  it seems  that the connection with the context is not so strong. And there have same results with different x and y OBC sequence. }
}
\label{fig:S25}
\end{figure} 
\end{comment}

\newpage
\section{III. 2D lattices with dimensionally-reduced GBZs}\label{sec3}

The previous section gives the approach to obtain the GBZ of 2D lattice models. However, as explained in the main text, there are some classes of models where the correct GBZ is not even of the same dimensionality as the lattice. In this section, we shall provide a detailed account of the additional steps and analysis required to dimensionally reduce the GBZ to the correct one. 

\subsection{Model description and setup}

We consider 2D lattice models of the form (type III in Fig.~\ref{fig:S221})
  \begin{equation}\label{eqS31}
H_\text{2D,III}=\sum_{m,n}t_1|m,n\rangle\langle m+\alpha,n+a|+t_2|m,n\rangle\langle m-\beta,n+a|+t_3|m,n\rangle\langle m-\beta,n-b|\ ,
\end{equation}
with oblique hoppings of amplitudes $t_1,t_2,t_3$. They constitute minimal models that require dimensional reduction of the GBZ, and are still simple enough such that they can be analyzed completely analytically.  Using the ansatz $|\psi\rangle\propto\sum_{m,n}\psi_{mn}|m,n\rangle$ with $\psi_{m,n}=z_x^mz_y^n$ as before, the bulk relations give $t_1\psi_{m+\alpha,n+a}+t_2\psi_{m-\beta,n+a}+t_3\psi_{m-\beta,n-b}=E(z_x,z_y)\psi_{m,n}$, which yields
\begin{equation}\label{eqS31.5}
\begin{split}
E(z_x,z_y)=\,t_1z_x^{\alpha}z_y^{a}+t_2z_x^{-\beta}z_y^{a}+t_3z_x^{-\beta}z_y^{-b}\ .%({delete:=E(z_x,z_y)})\ .
\end{split}
\end{equation}

By treating the system as a quasi-1D system in $x$ or $y$ as before i.e. by expressing it in the form of Eq.~\ref{eqS1} with GBZ and energies given by Eqs.~\ref{eqS3} and~\ref{eqS4}, we obtain
\begin{equation}\label{eqS32}
\begin{split}
z_x^{\alpha+\beta}=\frac{t_2+t_3z_y^{-a-b}} {t_1}z_1^{\alpha+\beta}, \quad z_1^{\alpha+\beta}=\frac{\sin \beta k_1}{\sin \alpha k_1}\text{e}^{i(\alpha+\beta)k_1},\ k_1\in  \left(-\frac{\pi}{\alpha+\beta},\ \frac{\pi}{\alpha+\beta}\right]\ , 
\end{split}
\end{equation}
and
\begin{equation}\label{eqS33}
\begin{split}
z_y^{a+b}=\frac{t_3}{t_1z_x^{\alpha+\beta}+t_2} z_2^{a+b}, \quad z_2^{a+b}=\frac{\sin bk_2}{\sin ak_2}\text{e}^{i(a+b)k_2},\ \  k_2\in  \left(-\frac{\pi}{a+b},\ \frac{\pi}{a+b}\right].
\end{split}
\end{equation}
That is, the boundary conditions along the two different direction give us two relations on $z_x$ and $z_y$ respectively. In order to get $z_x,z_y$ which satisfy all the boundary conditions, we solve the above to obtain 
\begin{equation}\label{eqS34}
\begin{split}
z_x^{\alpha+\beta}=\frac{t_2}{t_1}\frac{1+z_2^{a+b}}{z_2^{a+b}-z_1^{\alpha+\beta}}z_1^{\alpha+\beta},\quad z_y^{a+b}=\frac{t_3}{t_2}\frac{z_2^{a+b}-z_1^{\alpha+\beta}}{1+z_1^{\alpha+\beta}}\ .
\end{split}
\end{equation}
Note that now, $z_x$ and $z_y$ are not even proportional to a single phase factor, and thus no longer take any conventional ``non-Bloch'' form. Substituting $z_x,z_y $ Eq.\eqref{eqS34} into the energy $E(z_x,z_y)$ Eq.\eqref{eqS31.5}, we furthermore obtain
\begin{equation}\label{eqS35}
\begin{split}
\bar E(k_1,k_2)=E(z_x,z_y)&=\frac{\left(t_1^{\beta}(1+z_2^{a+b})^{\alpha}\right)^{\frac{1}{\alpha+\beta}}\cdot\left(t_3^{a}(1+z_1^{\alpha+\beta})^{b}\right)^{\frac{1}{a+b}}}{((z_2^{a+b}-z_1^{\alpha+\beta})/t_2)^{\frac{\alpha b-\beta a}{(\alpha+\beta)(a+b)}}}z_1^{-\beta} \ ,%\text{e}^{\frac{2i\pi a v} {a+b}}\ ,
\end{split}
\end{equation}
with $v=1,2,...,a+b$, the forms of $z_1,z_2$ given by Eq.(\ref{eqS32} and~\ref{eqS33}). 

\begin{figure}[htb]
\centering
\includegraphics[width=5.5in]{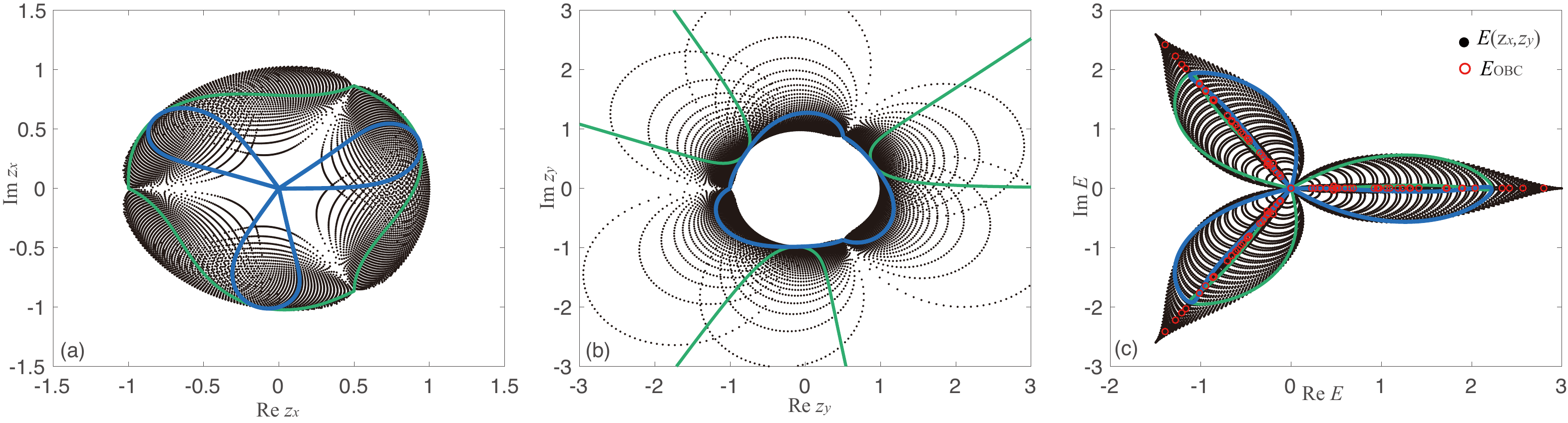}
\caption{
(a,b) Closed paths from the unconstrained GBZ from Eq.~\ref{eqS32} to~\ref{eqS34}  that result in nontrivial spectral winding, in contradiction to the known fact that OBC spectra should not enclose any nonzero area (nontrivial winding). The unconstrained GBZ $z_x$ (a) and $z_y$ (b) of $H_\text{2D,III}$ with $t_1=t_2=t_3=1$ are plotted in black for $k_1,k_2$ ranging over all values in steps of $\pi/100$, with illustrative green and blue closed paths enclosing nonzero area given by $k_1=(-\pi/3,\pi/3), k_2=\pi/6$ and $k_2=(-\pi/3,\pi/3), k_1=\pi/6$ respectively. (c) The corresponding spectrum (Eq.~\ref{eqS35}) from the unconstrained GBZ (black) and the corresponding green and blue spectral loops that enclose non-vanishing areas, which they are not supposed to do. In particular, they do not agree with the correct, numerically obtained OBC spectrum (red), which lies on a subset of $E(z_x,z_y)$ points, but which does not exhibit any nontrivial winding.
 }\label{fig:S31}
\end{figure}

Importantly, this 2D GBZ $(z_x,z_y)$ as it is currently defined does \emph{not} form a valid GBZ because there exists certain paths on it where the spectral winding 
 \begin{equation}\label{eqS36}
   \begin{split}
   \omega_{i}(E)&=\oint\text{d} z_i \partial_{z_i} \log(E(\bm k)-E_b)\ ,\\
   \end{split}
\end{equation}
$i=1,2 $ is nonzero for some arbitrary reference energies $E_b$. In other words, there exist some closed paths in $(k_1,k_2)$ space (defined in Eqs.~\ref{eqS32},~\ref{eqS33}) such that the energy $E(k_1,k_2)$ loop encloses a nonzero area as we cycle over $k_1$ or $k_2$. An illustrative example is shown in Fig.~\ref{fig:S31}.

Since it is known\cite{okuma2020topological,zhang2019correspondence} that OBC spectra should never possess nonzero spectral winding, which presupposes incomplete NHSE equilibration, the above results Eqs.~\ref{eqS34} and~\ref{eqS35} cannot possibly give the correct OBC spectrum of $H_\text{2D, III}$. Indeed, this is shown in the lack of agreement between the numerically obtained $E_\text{OBC}$ spectrum and the (currently unconstrained) GBZ spectrum $E(z_x,z_y)$ in Fig.~\ref{fig:S31}. Yet, the above results can be rigorously traced to satisfy all the relations pertaining to the real-space hoppings, and cannot be incorrect. 
 {In detail, $z_x$ $z_y$  are not even proportional to a single phase factor , but $z_x=z_x(k_1,k_2)$,$z_y=z_y(k_1,k_2)$ which 2 phases are twisted with each other.
While consider the $x$-direction boundaries, we have $z_x(z_y,k_1)=z_x(z_y,k’_1)$ for same $z_y$, it’s possible that we can not found the same $z_y$ for different $k_1, k’_1$ due to  twisted momentum,  which indicates that this is not  belong to effective BZ.}
Hence the natural conclusion is that the correct GBZ must be a subset of the GBZ, as currently defined by Eqs.~\ref{eqS32} to~\ref{eqS34} with the range of $k_1,k_2$ prescribed above. In the below, we shall carefully derive the constraints that extracts the subset of $k_1,k_2$ that generates the correct GBZ for this model; at the end of this supplement, we shall generalize this correct GBZ construction to arbitrary non-Hermitian lattice models.

\subsection{Detailed analysis of constraints leading to dimensional reduction}\label{sec32}

Here, we elaborate on the possible constraints that can make the energy spectrum (Eq.~\ref{eqS35}) exhibit zero winding. Since the unconstrained GBZ is a 2D torus that is spanned by 2 homotopy generators, a sure way to remove all possible nonzero spectral windings is to remove the homotopy generator/s that lead to particular nontrivial spectral winding paths. In principle, there can be many ways to remove a homotopy generator, since a combination of generators form another generator. However, for our model $H_\text{2D,III}$, it turns out that the requisite constraints can simply take the linear form of $k_1=\gamma k_2$, with $\gamma$ taking particular values that we shall elaborate on. (Note that we cannot possibly remove both homotopy generators, since there will be no GBZ left then.)\\

\subsubsection{Analysis of different possible parametrizations}
%For the model Eq.\eqref{eqS31},  assuming the connection between $k_1$ and $k_2$ is linear,  the restriction $k_1=\gamma k_2$ make  the $E(z_x,z_y)$ Eq.\eqref{eqS35} have zero winding number.
We now justify why it suffices to consider $k_1=\gamma k_2$, such that the energy takes the 1-parameter form $\bar E(k_2)$ (or equivalently $\bar E(k_1)$), with an slight abuse of notation. First, we establish the range and offset of $k_1,k_2$. given that $z_1,z_2$ (Eqs.~\ref{eqS32},\ref{eqS33}) satisfy periodicity conditions $z_{1}(k_{1})=z_{1}(k_{1}+2m'\pi/(\alpha+\beta))$, $z_{2}(k_{2})=z_{2}(k_{2}+2n'\pi/(a+b))$ with integer $n',m'$ ($n',m' \in \mathbb{Z}$). We can hence write down a valid linear reparametrization as $k_1=\gamma k_2+{2 n' \pi \gamma}/{(a+b)}+{2 m' \pi}/{(\alpha+\beta)}$ with integer $n',m'$, such that the energy Eq.\eqref{eqS35} is transformed into $\bar E'$ as given by
\begin{equation} 
\begin{split}
   &\bar E'= E(z_x,z_y) \Big |_{k_1=\gamma k_2+\frac{2 n' \pi \gamma}{a+b}+\frac{2 m' \pi}{\alpha+\beta}}= E\left(z_1(k_1),z_2(k_2)\right) \Big |_{k_1=\gamma k_2+\frac{2 n' \pi \gamma}{a+b}+\frac{2 m' \pi}{\alpha+\beta}}\\
&\quad=E(z_1(\gamma k_2+2n'\pi\gamma/(a+b)),z_2(k_2))\\
&\quad= \bar E\left(k_2+\frac{2n'\pi}{a+b}\right)\ ,
\end{split}
\end{equation}
such as $E'\in \{E(k_2)\}$. Since this hence just a translation of a single coordinate, we can hence just consider constraints of the form $k_1=\gamma k_2$ without any constant offset. With that, 
\begin{equation}
\bar E(k_2)= E(z_x,z_y) \Big |_{k_1=\gamma k_2}= E\left(z_1(k_1),z_2(k_2)\right) \Big |_{k_1=\gamma k_2}=E(z_1(\gamma k_2),z_2(k_2))\ .
\end{equation}\\

Next, we shall show that there exists $3$ possible values of $\gamma$ that guarantee zero spectral winding number while keeping $z_x,z_y$ periodic, namely $\gamma=b/\beta,\ a/\alpha,\  (a+b)/(\alpha+\beta)$. 
%Due to the linear constraint, the $z_x$ and $z_y$ of the system may no longer be connected end to end with 'momentum'. That is to say, when the system find the zero winding number energy, we must ensure that the reasonable $z_x$ and $z_y$ distribution is end to end with 'momentum'. 
Below are the detailed justifications for the possible choices of $\gamma$ (we call them $\gamma_1,\gamma_2$ and $\gamma_3$):%Concrete analysis is as follows:
\begin{itemize}
\item The case with $k_1=\gamma_1 k_2$, $\gamma_1=b/\beta$: Here $z_x,z_y$~\eqref{eqS34} and energy~\eqref{eqS35} with $k_1=b k, k_2=\beta k$ are given by
\begin{equation}\label{eqS37}
\begin{split}
 &z_{x,1}^{\alpha+\beta} =\left. z_x^{\alpha+\beta}\right|_{k_1=\gamma_1 k_2}=\frac{t_2\sin((a+b) \beta k)}{t_1\sin((\alpha b-a\beta )k)}\text{e}^{i(\alpha b+\beta b)k}\ , \\
& \left.  z_{y,1}^{a+b} = z_y^{a+b}\right|_{k_1=\gamma_1 k_2}=\frac{t_3\sin(\beta b k)\sin((\alpha b- a \beta)k)}{t_2\sin( a \beta k)\sin((\alpha b+\beta b)k)}\ ,
 \end{split}
\end{equation}
\begin{equation}\label{eqS38}\begin{split}
\bar E_1(k)&=\left. E(z_x,z_y)\right|_{k_1=\gamma_1 k_2=bk}\ ,\\
 &=t_1^{\frac{\beta}{\alpha+\beta}}t_2^{\frac{\alpha b-\beta a}{(\alpha+\beta)(a+b)}}t_3^{\frac{a}{a+b}} \text{e}^{\frac{i2\pi av}{a+b}+\frac{2\pi\alpha v'}{\alpha+\beta}} \left(\frac{\sin(\alpha b+\beta b) k}{\sin \alpha bk}\right)^{\frac{b}{a+b}}\\
&\qquad \times\left(\frac{\sin(a\beta+\beta b)k}{\sin a\beta k}\right)^{\frac{\alpha}{\alpha+\beta}}\left(\frac{\sin\beta  bk}{\sin \alpha bk}\right)^{-\frac{\beta}{\alpha+\beta}} \left(\frac{\sin(a+b)\beta k}{\sin a\beta k}-\frac{\sin(\alpha b+\beta b)k}{\sin \alpha b k}\right)^{\frac{\beta a-\alpha b}{(\alpha+\beta)(a+b)}}\ ,\\
\end{split}
\end{equation}
where $v=1,2,..,a+b$ ,  $v'=1,2,..,\alpha+\beta$. The range of $k$ has to be dependent on the ratio between $\alpha/\beta$ and $a/b$ as follows:

\begin{itemize}
    \item[-]  $\alpha/\beta>a/b$: $k$ can take values in $\left(-\frac{\pi}{(\alpha+\beta) b},\frac{\pi}{(\alpha+\beta) b}\right]$.  Both $z_{x,1}$ and $z_{y,1}$ are connected end to end with `momentum' $k$, and the path of $z_{x,1}$ in the complex plane forms a closed loop.  Hence, the results of $z_{x,1},z_{y,1}$ (Eq.\eqref{eqS37}) are legitimate, and zero winding in the energy $\bar E_1$ (Eq.\eqref{eqS38}) is respected. 
    \item[-] $\alpha/\beta=a/b$:  $k$ can take values in $\left(-\frac{\pi}{(a+b) \beta},\frac{\pi}{(a+b) \beta}\right]$. And $z_{x,1}^{\alpha+\beta}\rightarrow \infty$, \ $z_{y,1}^{a+b}\rightarrow 0$, but the path of $ z_{x,1}^{\alpha+\beta}z_{y,1}^{a+b}=\frac{\sin(\beta bk)}{\sin(a\beta k)}\text{e}^{i(a+ b)\beta k}$ in the complex plane  forms a closed loop with `momentum'  $k$.  Hence, the results of $z_{x,1},z_{y,1}$ Eq.\eqref{eqS37} are legitimate,  and zero winding in the  energy $\bar E_1$ Eq.\eqref{eqS38} is respected.
   \item[-] $\alpha/\beta<a/b$:  $k$ can  take values in  $\left(-\frac{\pi}{(a+b) \beta},\frac{\pi}{(a+b) \beta}\right]$. None of   $z_{x,1}^{\alpha+\beta}$, $z_{y,1}^{a+b}$, $z_{x,1}^{\alpha+\beta}z_{y,1}^{a+b}$ can form  closed loop parametrized by  `momentum' $k$. The results of $z_{x,1},z_{y,1}$ and the zero winding requirement of the energy $\bar E_1$ Eq.\eqref{eqS38} are inconsistent. It is worth noting that when $z^{\alpha+\beta}_1=z_2^{a+b}$ in Eq.\eqref{eqS34}, the $z_{x}^{\alpha+\beta}\rightarrow \infty,z_{y}^{a+b}\rightarrow 0$ which can ignore the end-to-end condition. And in this case, the energy $\bar E=0$.

\end{itemize}
\item The case with $k_1=\gamma_2 k_2$, $\gamma_2=a/\alpha$: $z_x,z_y$~\eqref{eqS34} and the energy~\eqref{eqS35} with $k_1=a k', k_2=\alpha k'$ are given by 
\begin{equation}\label{eqS39}
\begin{split}
& z_{x,2}^{\alpha+\beta} =\left. z_x^{\alpha+\beta}\right|_{k_1=\gamma_2 k_2}=\frac{t_2\sin(\beta a k')\sin((\alpha a+ \alpha b)k')}{t_1\sin( a \alpha k')\sin((\alpha b-\beta a)k')}\ , \\
 &\left.  z_{y,2}^{a+b} = z_y^{a+b}\right|_{k_1=\gamma_2 k_2}=\frac{t_3\sin((\alpha b-a\beta )k'))}{t_2\sin((\alpha a+a\beta )k')}\text{e}^{i(\alpha a+\alpha b)k'}\ ,
 \end{split}
\end{equation}
\begin{equation}\label{eqS310}\begin{split}
\bar E_2(k')&=\left. E(z_x,z_y)\right|_{k_1=\gamma_2 k_2=ak'}\ ,\\
 &=t_1^{\frac{\beta}{\alpha+\beta}}t_2^{\frac{\alpha b-\beta a}{(\alpha+\beta)(a+b)}}t_3^{\frac{a}{a+b}} \text{e}^{\frac{i2\pi av}{a+b}+\frac{2\pi\alpha v'}{\alpha+\beta}} \left(\frac{\sin(\alpha a+\beta a) k'}{\sin \alpha ak'}\right)^{\frac{b}{a+b}}\\
&\qquad \times\left(\frac{\sin(a\alpha+\alpha b)k'}{\sin a\alpha k'}\right)^{\frac{\alpha}{\alpha+\beta}}\left(\frac{\sin\beta ak'}{\sin \alpha ak'}\right)^{-\frac{\beta}{\alpha+\beta}} \left(\frac{\sin(\alpha b-a\beta ) k'}{\sin a\alpha k'}\right)^{\frac{\beta a-\alpha b}{(\alpha+\beta)(a+b)}}\ ,\\
\end{split}
\end{equation}
where $v=1,2,..,a+b$ , $v'=1,2,..,\alpha+\beta$. The range of $k'$ similarly has to depend on the ratio between $\alpha/\beta$ and $a/b$:
\begin{itemize}
    \item[-]  $\alpha/\beta>a/b$,  $k'$ can take values in $\left(-\frac{\pi}{(a+b) \alpha},\frac{\pi}{(a+b) \alpha}\right]$.  Both $z_{x,2}$ and $z_{y,2}$ are connected end to end with `momentum' $k$, and the path of $z_{y,2}$ in complex plane forms a closed loop. Hence, the results of $z_{x,2},z_{y,2}$ (Eq.\eqref{eqS39}) are legitimate, and  zero winding in the energy $\bar E_2$ Eq.\eqref{eqS310} is repected. 
    \item[-] $\alpha/\beta=a/b$,  $k'$ can take values in  $\left(-\frac{\pi}{(a+b) \alpha},\frac{\pi}{(a+b) \alpha}\right]$. And  $z_{x,2}^{\alpha+\beta}\rightarrow \infty$, \ $z_{y,2}^{a+b}\rightarrow 0$, but the path of $ z_{x,2}^{\alpha+\beta}z_{y,2}^{a+b}=\frac{\sin(\alpha b k')}{\sin(\alpha a k')}\text{e}^{i( a+ b)\alpha k'}$  in the complex plane  forms a closed loop with `momentum' $k$. Hence, the results of $z_{x,2},z_{y,2}$ Eq.\eqref{eqS39} are legitimate, and  zero winding in the energy $\bar E_2$ Eq.\eqref{eqS310} is repected.
   \item[-] $\alpha/\beta<a/b$,  $k'$ can take values in  $\left(-\frac{\pi}{(\alpha+\beta) a},\frac{\pi}{(\alpha+\beta) a}\right]$. None of   $z_{x,2}^{\alpha+\beta}$, $z_{y,2}^{a+b}$, $z_{x,2}^{\alpha+\beta}z_{y,2}^{a+b}$ can form closed loop parametrized by `momentum' $k'$. The results of $z_{x,2},z_{y,2}$ and the zero winding requirement of the energyy $\bar E_2$ Eq.\eqref{eqS310} seem unreasonable. It is worth noting that when $z^{\alpha+\beta}_1=z_2^{a+b}$ in Eq.\eqref{eqS34}, the $z_{x}^{\alpha+\beta}\rightarrow \infty,z_{y}^{a+b}\rightarrow 0$ which can ignore the end-to-end condition. And in this case, the energy $\bar E=0$.

\end{itemize}

\item The case with $k_1=\gamma_3 k_2$, $\gamma_3=(a+b)/(\alpha+\beta)$: $z_x,z_y$~\eqref{eqS34} and energy~\eqref{eqS35}  with  $k_1=(a+b) k'', k_2=(\alpha+\beta) k''$  take the forms 
\begin{equation}\label{eqS311}
\begin{split}
& z_{x,3}^{\alpha+\beta} =\left. z_x^{\alpha+\beta}\right|_{k_1=\gamma_3 k_2}=\frac{t_2\sin (\beta (a+b)k'')}{t_1\sin(\alpha b-\beta a)k'')}\text{e}^{ib(\alpha+\beta)k''}\ , \\
 &\left.  z_{y,3}^{a+b} = z_y^{a+b}\right|_{k_1=\gamma_3 k_2}=\frac{t_3\sin(\alpha b-\beta a)k'')}{t_2\sin(a(\alpha+\beta)k'')}\text{e}^{i\alpha (a+b)k''}\ ,
 \end{split}
\end{equation}
\begin{equation}\label{eqS312}\begin{split}
\bar E_3(k'')&=\left. E(z_x,z_y)\right|_{k_1=\gamma_3 k_2=(a+b)k''}\ ,\\
 &=t_1^{\frac{\beta}{\alpha+\beta}}t_2^{\frac{\alpha b-\beta a}{(\alpha+\beta)(a+b)}}t_3^{\frac{a}{a+b}} \text{e}^{\frac{i3\pi av}{a+b}+\frac{3\pi\alpha v'}{\alpha+\beta}}  \left(\frac{\sin(\alpha+\beta)(a+b)k''}{\sin \alpha (a+b)k''}\right)^{\frac{b}{a+b}}\left(\frac{\sin(a+b)(\alpha+\beta)k''}{\sin a(\alpha+\beta)k''}\right)^{\frac{\alpha}{\alpha+\beta}}\\
&\qquad \times\left(\frac{\sin \beta (a+b)k''}{\sin \alpha (a+b)k''}\right)^{-\frac{\beta}{\alpha+\beta}}\left(\frac{\sin b (\alpha+\beta)k''}{\sin a(\alpha+\beta)k''}-\frac{\sin \beta (a+b)k''}{\sin \alpha (a+b)k''}\right)^{\frac{\beta a-\alpha b}{(\alpha+\beta)(a+b)}}\ ,\\
\end{split}
\end{equation}
where $v=1,2,..,a+b$ , $v'=1,2,..,\alpha+\beta$. Here, the range of $k''$ is always $\left(-\frac{\pi}{(\alpha+\beta)(a+b)},\frac{\pi}{(\alpha+\beta)(a+b)}\right]$.
 \begin{itemize}
    \item[-]  $\alpha/\beta\neq a/b$: none of $z_{x,3}^{\alpha+\beta}$, $z_{y,3}^{\alpha+\beta}$, $z_{x,3}^{\alpha+\beta}z_{y,3}^{\alpha+\beta}$ can form closed loop parametrized by momentum $k''$. Hence the results of $z_{x,3},z_{y,3}$ (Eq.\eqref{eqS311}) are inadmissible, since they are inconsistent with the zero winding requirement of the energy $E_3$ Eq.\eqref{eqS312}.
        \item[-] $\alpha/\beta=a/b$: this case is equivalent to the case $k_1=\gamma_{1} k_2$ or $k_1=\gamma_{2} k_2$ with $\alpha/\beta=a/b$.
\end{itemize}
Hence, the case with $k_1=\gamma_3 k_2$, $\gamma_3=(a+b)/(\alpha+\beta)$ can  incorporate the cases with $k_1=\gamma_{1,2} k_2$ ($\gamma_1=b/\beta,\gamma_2=a/\alpha$).
\end{itemize}

\subsubsection{Summary}
\begin{table*}[h!]
\renewcommand\arraystretch{1.5}
  \centering
  \caption{Brief summary of admissible GBZs for our type III model (Eq.~\ref{eqS31})}
  {(Shaded boxes represent valid contributing GBZ sectors)}\label{tab:sum}
  \begin{tabular}{c|c|c|c}
\hline
\hline
& \bm{ $k_1=\gamma_1k_2$} {\small$(\gamma_1=\frac{b}{\beta})$} & \bm{$k_1=\gamma_2k_2$} {\small $(\gamma_2=\frac{a}{\alpha})$}& \bm{ $k_1=\gamma_3k_2$}{\small$(\gamma_3=\frac{a+b}{\alpha+\beta})$}\\
%& \quad {\small$(\gamma_1=\frac{b}{\beta})$} &  \quad  {\small $(\gamma_2=\frac{a}{\alpha})$} &   \quad {\small$(\gamma_3=\frac{a+b}{\alpha+\beta})$}\\
&$z_{x(y),1}=\left. z_{x(y)}\right|_{k_1=\gamma_1 k_2}$ &$z_{x(y),2}=\left. z_{x(y)}\right|_{k_1=\gamma_2 k_2}$&$z_{x(y),3}=\left. z_{x(y)}\right|_{k_1=\gamma_3 k_2}$\\
\hline
\multirow {3}{*}{${\alpha/\beta>a/b}$}&\multicolumn{1}{>{\columncolor{mygray}}c|}{\bm{$z_{x,1}$}{\bf: closed loop}}  & \multicolumn{1}{>{\columncolor{mygray}}c|}{{$z_{x,2}$}{: real}}  &  \multirow {2}{*}{$z_{x,3}$, $z_{y,3}$, $z_{x,3}^{\alpha+\beta}z_{y,3}^{\alpha+\beta}$:  open loop}\\
&\multicolumn{1}{>{\columncolor{mygray}}c|}{{$z_{y,1}$}{: real}} &\multicolumn{1}{>{\columncolor{mygray}}c|}{\bm{$z_{y,2}$}{\bf: closed loop}} &\\
\cline{2-4}
& \multicolumn{3}{c}{ Hence {GBZ=$\text{GBZ}_1\cup\text{GBZ}_2$}\quad(with $\text{GBZ}_1=\left\{z_{x,1},z_{y,1}\right\}$, $\text{GBZ}_2=\left\{z_{x,2},z_{y,2}\right\}$)}\\
\hline
\multirow {3}{*}{${\alpha/\beta=a/b}$}&\multicolumn{3}{>{\columncolor{mygray}}c}{$z_{i,1}=z_{i,2}=z_{i,3}$\quad ($i=x,y$) and $z_{x,1}\rightarrow \infty,\ z_{y,1}\rightarrow 0$}  \\
&\multicolumn{3}{>{\columncolor{mygray}}c}{\bm{$z^{\alpha+\beta}_{x,1} z^{a+b}_{y,1}$}{\bf: closed loop}} \\
\cline{2-4}
& \multicolumn{3}{c}{Hence $\text{GBZ}=\left\{z^{\alpha+\beta}_{x,1}z^{a+b}_{y,1}\right\}$ }\\
\hline
\multirow {2}{*}{${\alpha/\beta<a/b}$}&\multicolumn{3}{c}{$z_{x,j}\rightarrow \infty, \ z_{y,j}\rightarrow 0 $ \quad ($j=1,2,3$)}  \\
&\multicolumn{3}{c}{ $\bar E=0$}  \\
\hline
\hline
\end{tabular}
\end{table*}

\begin{figure}[h]
\centering
\includegraphics[width=4.3in]{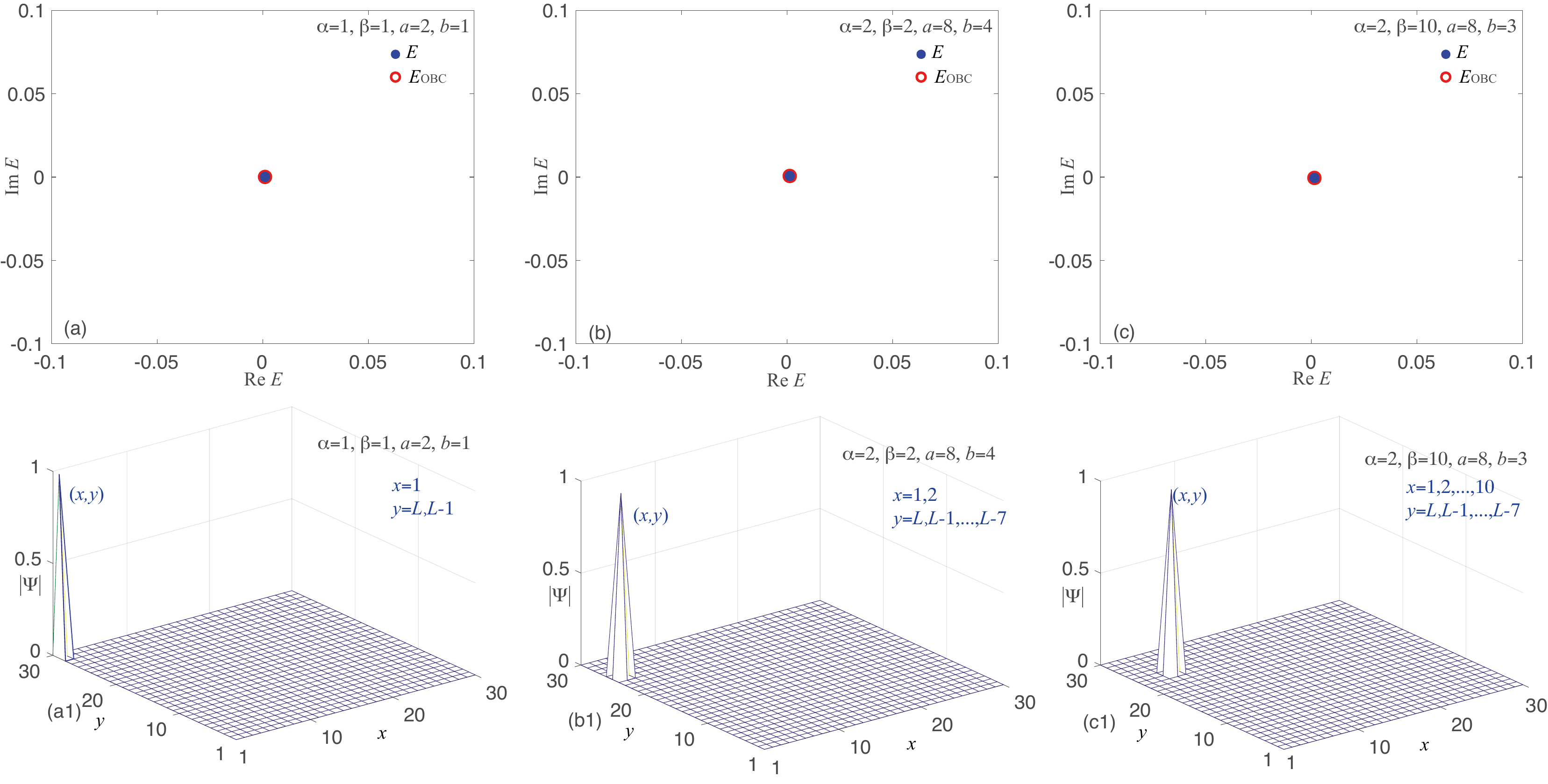}
\caption{(Top) Identically zero energies for the case of $\alpha\beta<a/b$ of the type III system Eq.\eqref{eqS31} ($ \bar E=0$ are from theoretical analysis , $E_{\text{OBC}}=0$ are numerical OBC eigenvalues). (Bottom) corresponding corner-localized distribution of eigenstates. The system parameters are $t_1=t_1=t_3=1$, with $\alpha,\beta,a,b$ indicated in the figures.}
\label{fig:S34}
\end{figure}

To summarize, the (correct) constrained GBZ of $H_\text{2D, III}$ of Eq.\eqref{eqS31} is not 2D, but is made up of one or more disconnected 1D loops as described above, depending on the hopping lengths $\alpha,\beta, a,b$. We call the effective Hamiltonian constrained to this reduced GBZ $H_\text{2D-red}$:
 \begin{itemize}
    \item[$\blacktriangleright$]   When  $\alpha/\beta>a/b$, the GBZ and spectrum consists of the union of the two sectors
    \begin{equation}\label{eqS313}
\begin{split}
&\ \quad\qquad\qquad\text{GBZ}=\text{GBZ}_1\cup\text{GBZ}_2,\\
&\text{GBZ}_1=\left\{z_{x,1},z_{y,1}\left|k\in\left(-\frac{\pi}{(
\alpha+\beta) b},\frac{\pi}{(\alpha+\beta) b}\right]\right.\right\}, \\
&\text{GBZ}_2=\left\{z_{x,2},z_{y,2}\left|k'\in\left(-\frac{\pi}{(a+b) \alpha},\frac{\pi}{(a+b) \alpha}\right]\right.\right\},
 \end{split}
\end{equation}
\begin{equation}\label{eqS314}
\begin{split}
\{\bar E\}=\{\bar E_1\}\cup\{\bar E_2\},\\
 \end{split}
\end{equation}
where the forms of $z_{x(y),1(2)}$ can be found in Eq.(\ref{eqS37},\ref{eqS39}) and that of $E_{1,2}$ can be found in Eq.(\ref{eqS38},\ref{eqS310}). Unlike with the unconstrained GBZ, they agree excellently with numerics, as shown in the example in FIG.~\ref{fig:S32}. In other words, the effective surrogate OBC Hamiltonian of the case $\alpha/\beta>a/b$ takes the form
\begin{equation}\label{eqS314.5}
\bar H_\text{2D-red}=\bar H_\text{2D,III}=\bar H_1\oplus\bar H_2 \ ,
\end{equation}
where $\bar H_1$ and $\bar H_2$ are the Hamiltonian operators with corresponding spectra $\bar E_1$ and $\bar E_2$ (Eqs.~\ref{eqS38} and~\ref{eqS310}). 

\begin{figure}[htb]
\centering
\includegraphics[width=5in]{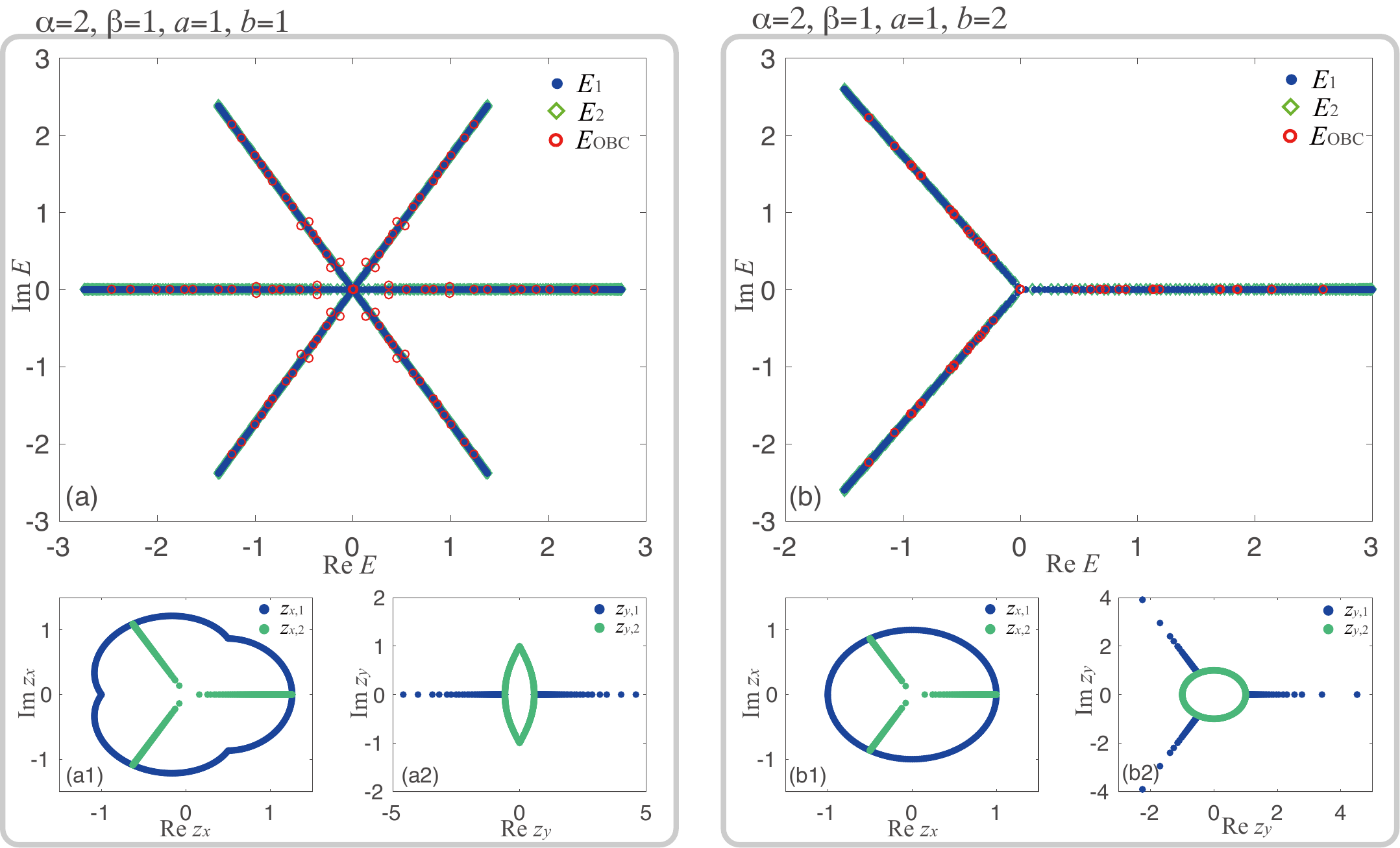}
\caption{The constrained GBZs and corresponding spectra $\bar E$ of two illustrative cases of $H_\text{2D, III}$ with hopping lengths $\alpha/\beta > a/b$, as indicated on top of the panels. Top) Coincident spectral contributions $\bar E_1,\bar E_2$ from 1D GBZ sectors $1$ and $2$ of these 2D lattices, as defined in Eq.(\ref{eqS38},\ref{eqS310}). They agree well with the numerical $E_\text{OBC}$ which is here computed with $10\times 10$ lattice sites, with density of states comparison presented later.  
Bottom) 1D GBZs $z_x,z_y$ Eq.(\ref{eqS37},\ref{eqS39}) for sectors $1$ and $2$, which traces loops or flattened loops in the complex plane, consistent with the conclusions from Table.~\ref{tab:sum}. The two GBZ sectors may look totally different from each other, but they combine to describe the system as a coherent whole. Parameters are  $t_1=t_1=t_3=1$ and the GBZs are plotted with $k,k'$ points at intervals of $\pi/300$.
}\label{fig:S32}
\end{figure}
		
 \item[$\blacktriangleright$] When  $\alpha/\beta=a/b$, the GBZ and spectrum are given by
 \begin{equation}\label{eqS315}
\begin{split}
\text{GBZ}=\left\{z_{x,1}^{\alpha+\beta}z_{y,1}^{a+b}=\frac{\sin(\beta bk)}{\sin(a\beta k)}\text{e}^{i(a\beta+\beta b)k}  \left|k\in\left(-\frac{\pi}{(a+b) \beta},\frac{\pi}{(a+b) \beta}\right]\right.\right\},
 \end{split}
\end{equation}
\begin{equation}\label{eqS316}
\begin{split}
\bar E(k)=t_1^{\frac{b}{a+b}}t_3^{\frac{a}{a+b}}\left( \frac{\sin (a+b)\beta k}{\sin a \beta k}\right)\left( \frac{\sin \beta b  k}{\sin a \beta k}\right)^{-\frac{b}{a+b}}\text{e}^{\frac{2i\pi a v'} {a+b}}\ ,\\
 \end{split}
\end{equation}
with excellent numerical agreement as shown in FIG.~\ref{fig:S33}.

\begin{figure}[htb]
\centering
\includegraphics[width=3.4in]{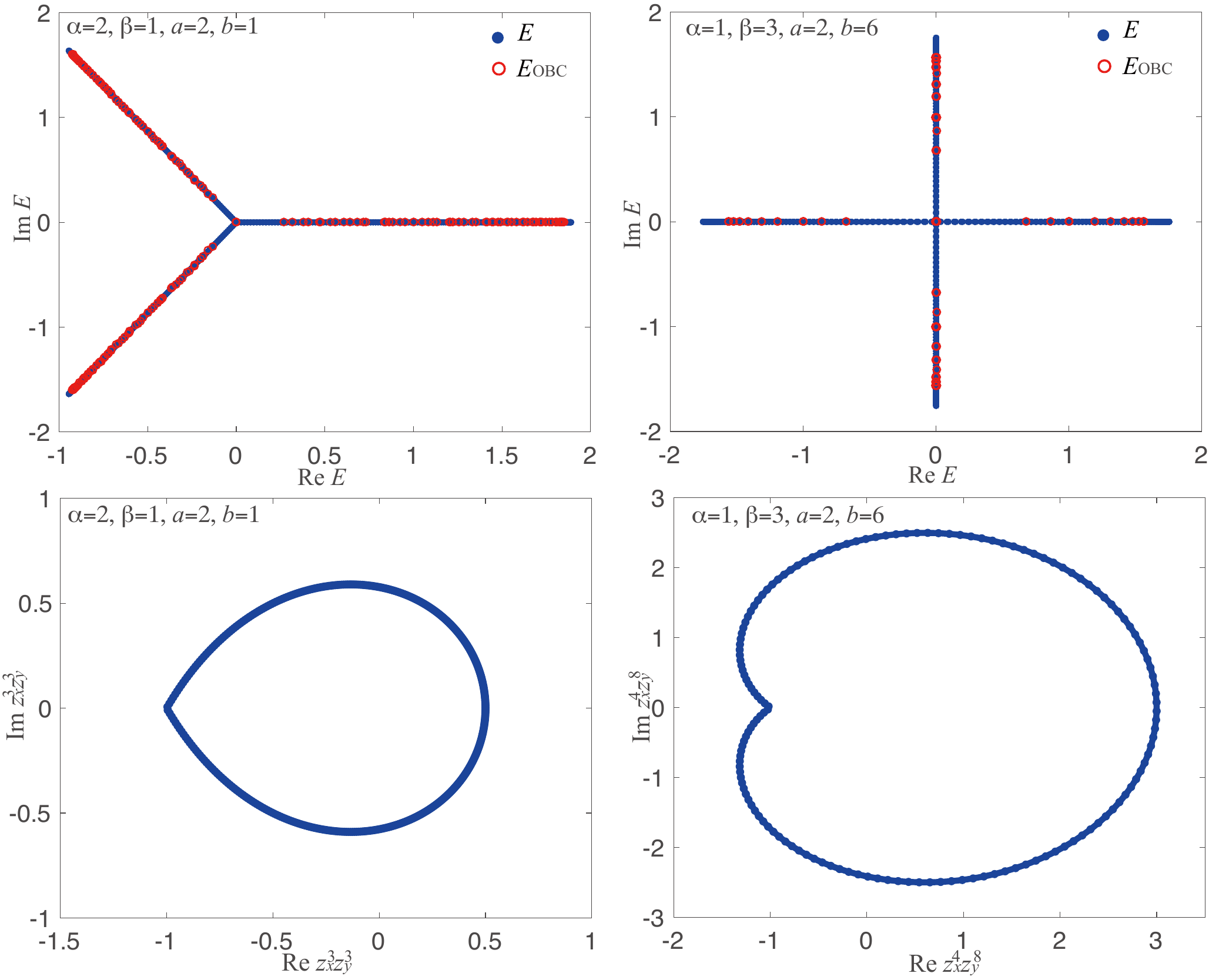}
\caption{
The constrained GBZs and corresponding spectra $\bar E$ of $H_\text{2D, III}$ of two illustrative cases with hopping lengths $\alpha/\beta=a/b$ as indicated in the panels. Top) Spectrum according to the GBZ (Eq.\eqref{eqS316}), which agree well with the numerical $E_\text{OBC}$ spectrum which is here computed with $20\times 20$ lattice sites.  
Bottom) Corresponding constrained 1D GBZs $z_x,z_y$ (Eq.~\eqref{eqS315}) of these 2D lattices, with cusps that are not present in the GBZs of 1D non-Hermitian lattices. Parameters are  $t_1=t_1=t_3=1$ and the GBZ is plotted with $k$ points at intervals of $\pi/600$.
}\label{fig:S33}
\end{figure} 

\item[$\blacktriangleright$]   When  $\alpha/\beta<a/b$, the energy is always zero and $z_{x}^{\alpha+\beta}\rightarrow \infty$, \ $z_{y}^{a+b}\rightarrow 0$, which means that the eigenstates always corner-localize, as illustrated in FIG.~\ref{fig:S34}. This is due to non-Bloch collapse, since hoppings $t_1$ and $t_3$ do not contain any net component that in the inverse direction of hopping $t_2$. 
\end{itemize}

% May confuse the reader from the main message
%The constraint of zero winding number energy makes  dimension of the system more than the  dimension of GBZ Hamiltonian, that can be called dimensional reduction. And  dimensional reduction does not only happen in 2 or higher dimensional system, there also have dimensional reduction in 1D system, like the dimension of GBZ Hamiltonian is 0 when the 1D system only have unidirectional hopping term.

\subsubsection{Number of OBC spectral branches for type III hopping lattices}

%$H_\text{2D,I},H_\text{2D,II}$ and 
To find the number of OBC spectral branches for $H_\text{2D,III}$ satisfies the symmetry 
\begin{equation}\label{mastermodelsymmetry}
\begin{split}
&U_{v,v'}^{-1}H_\text{2D,III}U_{v,v'}=\exp\left(i\frac{2 \alpha \pi}{\alpha+\beta}v+i\frac{2a\pi}{a+b}v'\right)  H_\text{2D,III}\ , \\
&U_{v,v'}=\exp\left(i\frac{2 m\pi}{\alpha+\beta}v+i\frac{2n\pi}{a+b}v'\right)|m,n\rangle\langle m,n|\ ,
\end{split}
\end{equation}
with integer indices $v,v'$ ($v=1,2,..,\alpha+\beta,\,v'=1,2,...,a+b$), where $|m,n\rangle$ are real space basis orbitals. $U_{v,v'}$ is a unitary operator which satisfies 
$U_{v,v'}U_{v,v'}^{\dagger}=\text{I}$ , (`I' represents the unit operator).  Assuming that the wave function $|\Psi^{n}\rangle $ is an eigenstate of the model Eq.\eqref{eqS31}   with eigenvalue $E_n$, the other eigenstates $U_{v,v'}|\Psi^{n}\rangle$ can be found via symmetry Eq.\eqref{mastermodelsymmetry} with eigenvalue $\exp\left(i\frac{2 \alpha \pi}{\alpha+\beta}v+i\frac{2a\pi}{a+b}v'\right) E_n$, that is, 
  \begin{equation}
\begin{split}
H_\text{2D,III}|\Psi^{n}\rangle&=E_n|\Psi^{n}\rangle\ , \\
\exp\left(i\frac{2 \alpha \pi}{\alpha+\beta}v+i\frac{2a\pi}{a+b}v'\right)H_\text{2D,III}|\Psi^{n}\rangle&=\exp\left(i\frac{2 \alpha \pi}{\alpha+\beta}v+i\frac{2a\pi}{a+b}v'\right)E_n|\Psi^{n}\rangle\ ,\\
U_{v,v'}^{-1}H_\text{2D,III}U_{v,v'}|\Psi^{n}\rangle&=\exp\left(i\frac{2 \alpha \pi}{\alpha+\beta}v+i\frac{2a\pi}{a+b}v'\right)E_n|\Psi^{n}\rangle\ ,\\
H_\text{2D,III}U_{v,v'}|\Psi^{n}\rangle&=\exp\left(i\frac{2 \alpha \pi}{\alpha+\beta}v+i\frac{2a\pi}{a+b}v'\right)E_nU_{v,v'}|\Psi^{n}\rangle\ .\\
\end{split}
\end{equation}
Hence we see that the energy spectrum of $H_\text{2D,III}$ is parametrized by a 1-parameter family of states, with indices $v,v'$ taking $\alpha+\beta$,$a+b$ possible values. As such, there are 
\begin{equation}
\text{LCM}\left[\displaystyle\frac{a+b}{\text{GCD}(a+b,a)},\frac{\alpha+\beta}{\text{GCD}(\alpha+\beta,\alpha)}\right]\ ,
\end{equation} 
OBC spectral branches for type III lattices.

\clearpage
\subsubsection{Flowchart of our approach}
All in all, the procedure of obtaining the effective BZ and spectrum is illustrated in the following flowchart for generic 2D models. Higher-dimensional models can be dealt with analogously, and will be discussed later.

\begin{figure}[h]
\centering
\includegraphics[width=3in]{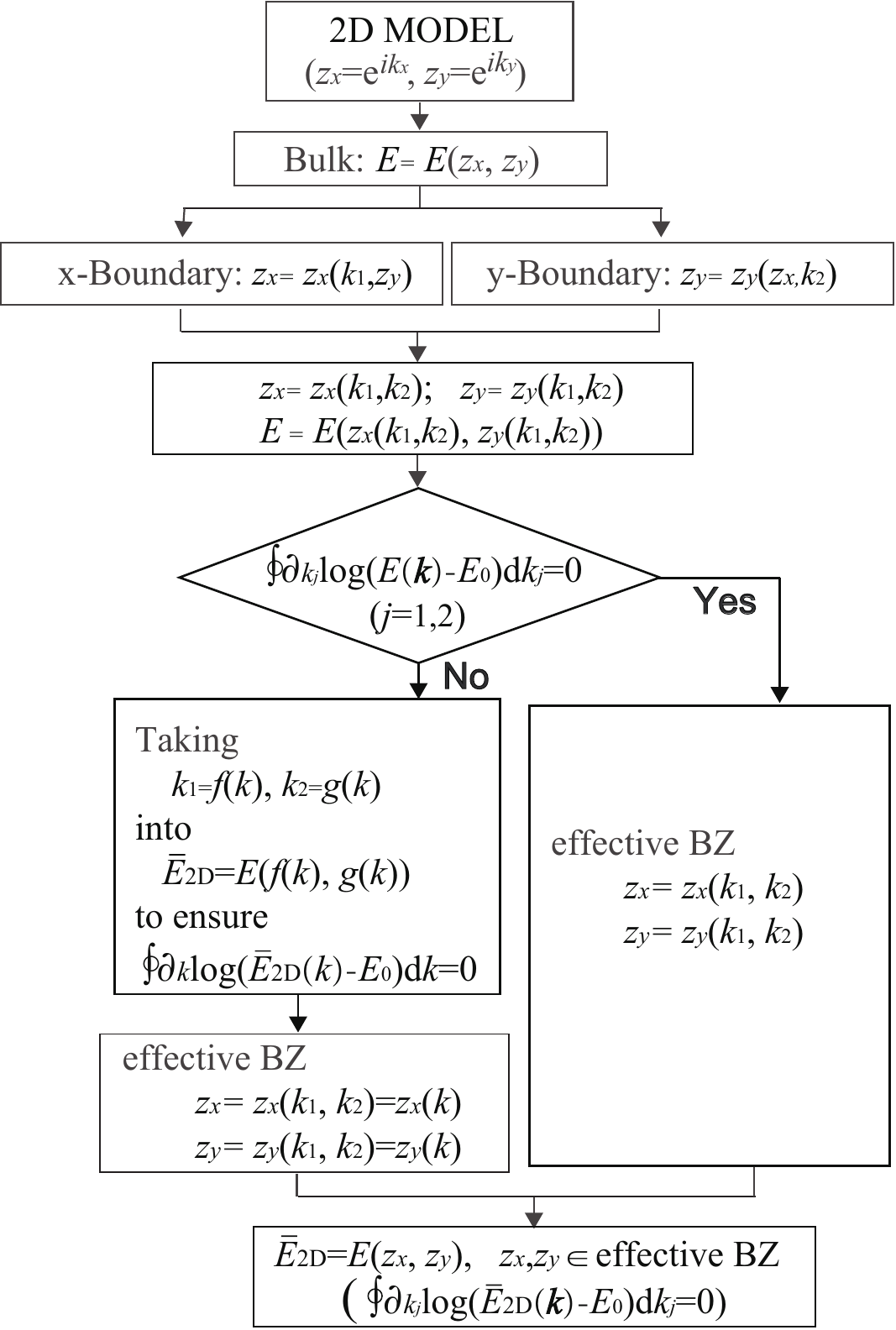}
\caption{Flowchart of our approach for generic 2D models. Starting with the original model which can be expressed as $E(z_x,z_y)$ by identifying $z_x=e^{ik_x}$, $z_y=e^{ik_y}$, we find how $z_x,z_y$ depend on $k_1,k_2$ by implementing x and y-OBCs separately. If the resulting spectrum does not contain nontrivial spectral winding, the effective BZ is still 2D and we are done. If not, we find 1D parametrizations $k$ of a submanifold of the $(k_1,k_2)$ torus such that the spectral windings are trivial. The effective 1D BZ comprises all such 1D paths. } 
\label{fig:S35}
\end{figure}  

\clearpage
\subsection{Density of states in the dimensionally-reduced effective system}
\label{sec33}

In the subsection above, we have shown that in order for the GBZ spectrum of $H_\text{2D,III}$ (Eq.~\ref{eqS31}) to exhibit nonzero winding for any closed path, there is a need to constrain the valid GBZ to a 1D subspace of the original naively obtained 2D GBZ. Here, we show that the numerically obtained OBC spectrum indeed correspond to equally spaced states in the 1D constrained BZ, just like how the momentum eigenstates are distributed in 1D BZ corresponding to a physically 1D system.
%In the above subsections, the approach give the GBZ results of the model Eq.\eqref{eqS31}, and find  the dimension of GBZ Hamiltonian is less than the system-----dimensional reduction.
%This subsection give special example to see the properties of these model.\\

%is rotationally symmetric. In other words,  the overall information can be known though studying  one branch of the energy spectrum.

To verify that the spectrum of $H_\text{2D-red}$ indeed comprises equally-spaced ``momentum'' eigenstates, we examine the density of states (DOS) of the OBC spectrum. The DOS is given by
% \begin{equation}
%\begin{split}
%\rho (E)=\sum_{i}\rho_i (E) =\sum_{i}\frac{\delta n_i}{\delta E_i}\ ,
%\end{split}
%\end{equation}
\begin{equation}\label{DOS}
\begin{split}
\rho (E)=\frac1{N}\frac{\delta n}{\delta E}\ ,
\end{split}
\end{equation}
which means that there are $\delta n$  eigenstates in the energy range $[E,E+\delta E)$ and  normalization constant $N$ is equal to the length of the energy to be measured. What we would like to check is the quantity $\sigma(k)$, which is the DOS in momentum space.  
That is,  the momentum range $[k_i,k_i+\delta k_i)$ $(i=1,2) $ which corresponds to energy range $[E,E+\delta E)$ need to be considered, where $i=1,2$ refers to the possible sectors for the GBZs.
 If the states are indeed uniformly labeled by momentum, $\sigma_i(k)\propto \rho_i(E){\delta E_i}/{\delta k}$ should be uniform in each sector $i$, and $\sigma(k)=\sum_i\sigma_i(k)$ should only be proportional to the number of sectors corresponding to a particular value of $k$. For OBC states predicted by the GBZ sector energies $\bar E_i$, we have

%where $i=1,2$ refers to the possible sectors for the GBZs and $\mathscr{N}$ . For each $i$, there are $\delta n_i$ ($i=1,2$) eigenstates in the energy range $[E_i,E_i+\delta E_i)$. What we would like to check is the quantity $\sigma(k)$, which is the DOS in momentum space. If the states are indeed uniformly labeled by momentum, $\sigma_i(k)=\rho_i(E)\frac{\delta E_i}{\delta k}$ should be uniform in each sector $i$, and $\sigma(k)=\sum_i\sigma_i(k)$ should only be proportional to the number of sectors corresponding to a particular value of $k$. We have
%Therefore, the physical quantity $\rho (k) $ according to the DOS can be defined by

\begin{equation}\label{eqS322}
\begin{split}
\sigma (k) &=\sum _i\sigma_i (k)=\frac{2\pi}{\mathcal{N}}\sum _i\frac{\delta n_i(k)}{\delta E_i(k)}\frac{\delta  E_i(k)}{\delta k} =\sum_i\delta n_i(k)\ ,\\
\end{split}
\end{equation}
with $\delta k=2\pi/ \mathcal{N}$. %$ E_{\text{GBZ}}(k) $  connects momentum $k$ with approximate eigenvalues of $E_{\text{OBC}}$.
For each energy interval $\delta E(k)= \bar E_i(k+\delta k)-\bar E_i(k)$, 
one can obtain the number $\delta n(k)$ of occupied states within the interval.  With the knowledge of the
dependence between $\bar E_i$ of the GBZ and momentum $k$ , it's easy to get the DOS in momentum space $\sigma(k)$. This is shown in FIG.\ref{fig:S35}(a,b) for two illustrative sets of parameters. 
%That is,  for  momentum range $k\sim k+\delta k$ (left vertical axis, FIG.\ref{fig:S35}(a,b)),  using $E_{\text{GBZ}}(k)$) (Re$E-k/\pi$ plane,FIG.\ref{fig:S35}) as stopover to describe number $ \delta n(k)$ of states ($E_{\text{OBC}}$, red part , FIG.\ref{fig:S35}(a,b)) that are to be occupied by the system at each momentum. And  the subscript $i$ ($i=1,2$) means the sector of GBZ  ($=\text{GBZ}_1\cup\text{GBZ}_2$).
%Consider the system have LCM$(a + b, \alpha+\beta)$ spectral branches, $ \sigma (k)$ must be LCM$(a + b, \alpha+\beta)\times n$ with integer $n$. 

\begin{comment}
 {For new function of $\sigma (k)$, (1) $\sigma_i(k)$ is  an integer,
$2\pi$ is equal to the length of the momentum to be measured $( \delta k=2\pi/\mathcal{N})$. $\sum_k \sigma (k) = \sum_k \sum _i\sigma_i (k)= \sum_k \sum _i n_i (k)$ is several times of  total no. of states, like FIG.\ref{fig:S35}(a), $E_1=E_2$,$\sum_k \sigma (k) =2\times$ total no. of states.\\
(2) we can not tell which energy $E_{\text{OBC}}$ belongs to $E_1$ or $E_{2}$.  $\sigma_1$,$\sigma_2$ can be seen as 2  independent channels.
For same momentum space $k-k+\delta k$,  there have energy $E_{\text{OBC}}$ both belong to $\delta E_1, \delta E_2$. We just add $\delta n_1$ and $\delta n_2$ to get $\sigma(k)$.\\
(3)the system have LCM$(a + b, \alpha+\beta)$ spectral branches, $ \sigma (k)$ must be LCM$(a + b, \alpha+\beta)\times n$ with integer $n$.}
\end{comment}

\subsubsection{Model with the GBZ sectors possessing the same energies}

For definiteness, we first consider $H_\text{2D,III}$ Eq.\eqref{eqS31} with $\alpha=2,\beta=1,a=1,b=2$, such that the energy takes the form
\begin{equation}\label{eqS318}
\begin{split}
E(z_x,z_y)=t_1z_x^{2}z_y+t_2z_x^{-1}z_y+t_3z_x^{-1}z_y^{-2},
\end{split}\end{equation}
where each of $z_x,z_y$ belongs to the 1D GBZ sectors $\text{GBZ}_1$ and $\text{GBZ}_2$, which from Sec.~\ref{sec32} Eq.(\ref{eqS313},~\ref{eqS314}) simplify to 
\begin{equation}\label{eqS319}
\begin{split}
\text{GBZ}_1&=\left\{\left.z_{x,1}^3=\frac{t_2}{t_1}\text{e}^{ik}, z_{y,1}^3=\frac{t_3}{t_2}\frac{1}{2\cos \left(k/3\right)-1}\right|k\in[-\pi,\pi)\right\}\ ,\\
\text{GBZ}_2=&\left\{\left.z_{x,2}^3=\frac{t_2}{t_1}\left(2\cos \left(k'/3\right)-1\right),z_{y,2}^3=\frac{t_3}{t_2}\text{e}^{-ik'},\right|k'\in[-\pi,\pi)\right\}\ ,\\
&\qquad \qquad \text{GBZ}=\text{GBZ}_1\cup \text{GBZ}_2\ .
\end{split}
    \end{equation}
In other words, the set of GBZ eigenstates consists of eigenstates from both GBZ sectors. The OBC eigenenergies are approximated by GBZ energies, which are given either by $E(z_{x,1},z_{y,1})=\bar{E}_1(k)$ or   $E(z_{x,2},z_{y,2})=\bar{E}_2(k')$, where $k$ and $k'$ parametrize their respective GBZs. The GBZ Hamiltonian is hence given by
\begin{equation}\label{eqS320}
\begin{split}
H_\text{2D-red}&=\bar H_\text{2D,III}={H}_1\oplus {H}_2\ , \\
\bar{E}_1(k)=\bar{E}_2(k)&= t_1^{1/3}t_2^{1/3}t_3^{1/3} (2\cos (k/3)+1)^{1/3}(2\cos (2k/3)+1)^{2/3}\text{e}^{2i\pi v/3}\ ,
\end{split}
\end{equation}
with $k,k'\in(-\pi,\pi]$, $v=1,2,3$, see FIG.~\ref{fig:S35}(a). $ \bar E_1$ and $\bar E_2$ are the corresponding eigenenergies of $ H_1$ and $ H_2$; in this case, after substituting the sector-dependent forms of $z_{x,\nu}$ and $z_{y,\nu}$, they happen to take identical functional forms.

To justify the integrity of this GBZ construction, we first note that the spectral winding number is zero, since  Eq.\eqref{eqS320} is obviously real. Also, both the GBZ$_1$ and GBZ$_2$ are period in $k,k'$. The correctness of Eq.(\ref{eqS319},\ref{eqS320}) is demonstrated in FIG.~\ref{fig:S32}(b), which not only shows that the results of the GBZ and GBZ Hamiltonian can be trusted, but also that the relation between OBC energies and momentum $k$ are as expected.
\begin{figure}[htb!]
\centering
\includegraphics[width=4.5in]{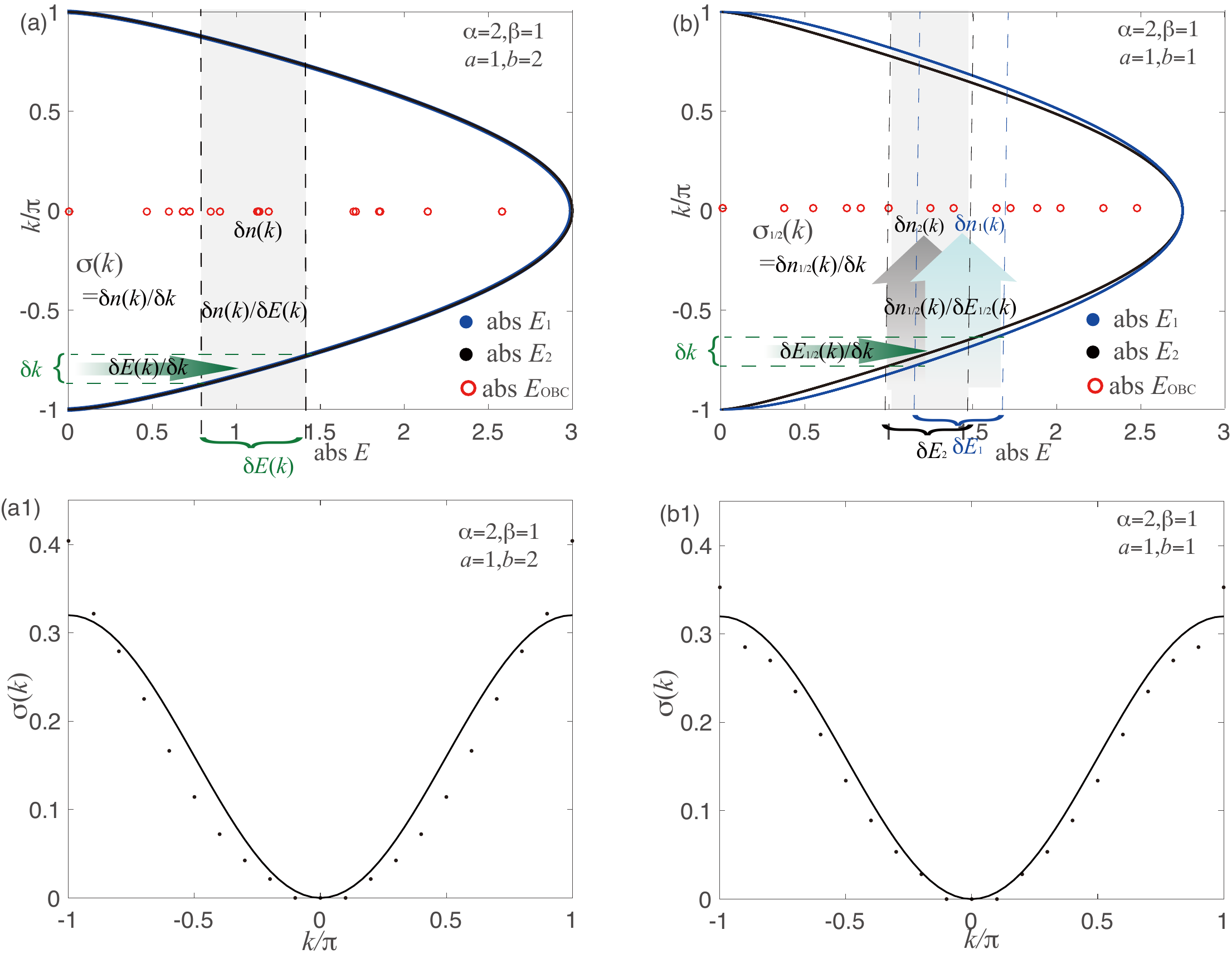}
\caption{Density of states (DOS) in momentum space. (a,b) The solid curves denote the dependence of effective 1D GBZ sector energies $E_1,E_2$ with momentum $k$ for two possible models Eq.\eqref{eqS31.5}. Numerically, we can obtain $k\approx k+\delta k$ from each $\delta E(k)$ point from $E_\text{OBC}$ data (red circles, shown on the central horizontal), and obtain the momentum space DOS $\sigma(k)$. \\ 
(a1,b1) The momentum space DOS profile $\sigma(k)$ in our model Eq.\eqref{eqS31.5}, with $\delta k=\pi/8000$. For both models, the DOS (black dots) can be approximately fitted to a sinusoidal curve (black curve). Parameters are  $t_1=t_1=t_3=1$, with a OBC lattice of $L^2=10^2=100$ sites.
}\label{fig:S35}
\end{figure}  

\subsubsection{Model with the GBZ sectors possessing different energies}

\indent  In the above, we considered $H_\text{2D,III}$ (Eq.\eqref{eqS31}) with  $\alpha=b$ and $\beta =a$, such that the two 1D GBZ sectors correspond to eigenenergies $\bar E_1$ and $\bar E_2$ Eq.\eqref{eqS320} with equivalent forms. That, however, is not the case for more generic models. Here, we consider an illustrative model with  $\alpha=2 $,  $\beta =1, a=1,b=1$ with dissimilar GBZ eigenenergies for its two sectors. We first write it in terms of $z_x,z_y$: 
\begin{equation}\label{eqS3181}
\begin{split}
E(z_x,z_y)=t_1z_x^{2}z_y+t_2z_x^{-1}z_y+t_3z_x^{-1}z_y^{-1}\ ,
\end{split}\end{equation}
where each of $z_x,z_y$ belongs to either of the two GBZ sectors 
\begin{equation}\label{eqS3191}
\begin{split}
\text{GBZ}_1&=\left\{\left.z_{x,1}^3=\frac{t_2}{t_1}2\cos(k/3)\text{e}^{ik}, z_{y,1}^3=\frac{t_3}{t_2}\frac{1}{2\cos \left(2k/3\right)+1}\right|k\in[-\pi,\pi)\right\}\ ,\\
\text{GBZ}_2=&\left\{\left.z_{x,2}^3=\frac{t_2}{t_1}\frac{\sin(3k'/4)}{\sin(k'/2)},z_{y,2}^3=\frac{t_3}{t_2}\frac{1}{2\cos \left(k/2\right)+1}\text{e}^{-ik'},\right|k'\in[-\pi,\pi)\right\}\ ,\\
&\qquad \qquad \text{GBZ}=\text{GBZ}_1\cup \text{GBZ}_2\ ,
\end{split}
    \end{equation}
as introduced in Sec.~\ref{sec32} Eq.(\ref{eqS313},~\ref{eqS314}).
GBZ eigenenergies $\bar E$ take the form of either $E(z_{x,1},z_{y,1})=\bar E_1(k)$ or $E(z_{x,2},z_{y,2})=\bar E_2(k')$, with their GBZ Hamiltonian and explicit eigenenergies
\begin{equation}\label{eqS3201}
\begin{split}
&\bar H_\text{2D-red}=\bar{H}_\text{2D,III}= \bar H_1\oplus \bar H_2 ,\\
\bar{E}_1(k)=\pm t_1^{1/3}t_2^{1/6}t_3^{1/2} &(2\cos (2k/3)+1)^{1/2}(2\cos (k/3))^{2/3}\text{e}^{2i\pi v/3}\ ,\\
\bar {E}_2(k')=\pm t_1^{1/3}t_2^{1/6}t_3^{1/2} &(2\cos (k'/2)+1)^{1/2}(2\cos (k'/2))^{2/3}(\sin(k'/4))^{2/3}\text{e}^{2i\pi v/3}\ ,
\end{split}
\end{equation}

with $k,k'\in(-\pi,\pi]$, $v=1,2,3,$ for the model Eq.\eqref{eqS3181}. As in the previous example, the spectral winding in Eq.\eqref{eqS3201} are both zero, and both GBZ$_1$ and GBZ$_2$ are periodic. However, here $\bar E_1(k)$ and $\bar E_2(k)$ manifestly take different functional forms. As presented in  FIG.~\ref{fig:S32}(a), the numerical results demonstrate the correctness of Eq.(\ref{eqS3191},\ref{eqS3201}), with both $\bar E_1(k)$ and $\bar E_2(k)$ agreeing with the distribution of numerically obtained $E_\text{OBC}$. Note that strictly, the above GBZ results are exact only in the continuum limit ($L\rightarrow\infty$) where the set of momenta $k$ tends towards an equally spaced set of points with $\sim L^{-1}$ separation. %In finite size cases, however, there can some special momentum points with different density of states, as in the two ``spikes'' in $\sigma(k)$ in FIG.~\ref{fig:S35} due to $L=10$ not being a multiple of $a+b=3$ or $\alpha+\beta=3$. 

%Hereupon,  the results of GBZ and GBZ Hamiltonian is trustworthy. And the relation between energy and momentum $k$ could confirm.   In  FIG.~\ref{fig:S35}(b1), $E_1(k)=E_2(k)$, $\sigma (k)$ =$2*3n$  with integer $n$, and in  FIG.~\ref{fig:S35}(b2), $E_1(k)\neq E_2(k)$, $\sigma (k)$ =$6n$ with integer $n$ which agree with $\sigma (k)$ = LCM$(a + b, \alpha+\beta) \times n$ values with integer $n$.  And the other values near $ k=\pm \pi$ is caused by Mod$(L^2,\text{LCM}(a + b, \alpha+\beta))\neq 0$.

\subsection{Circuit simulation of non-Hermitian lattices } 

\begin{figure*}[htbp]
    \begin{centering}
    \includegraphics[width=0.8\linewidth]{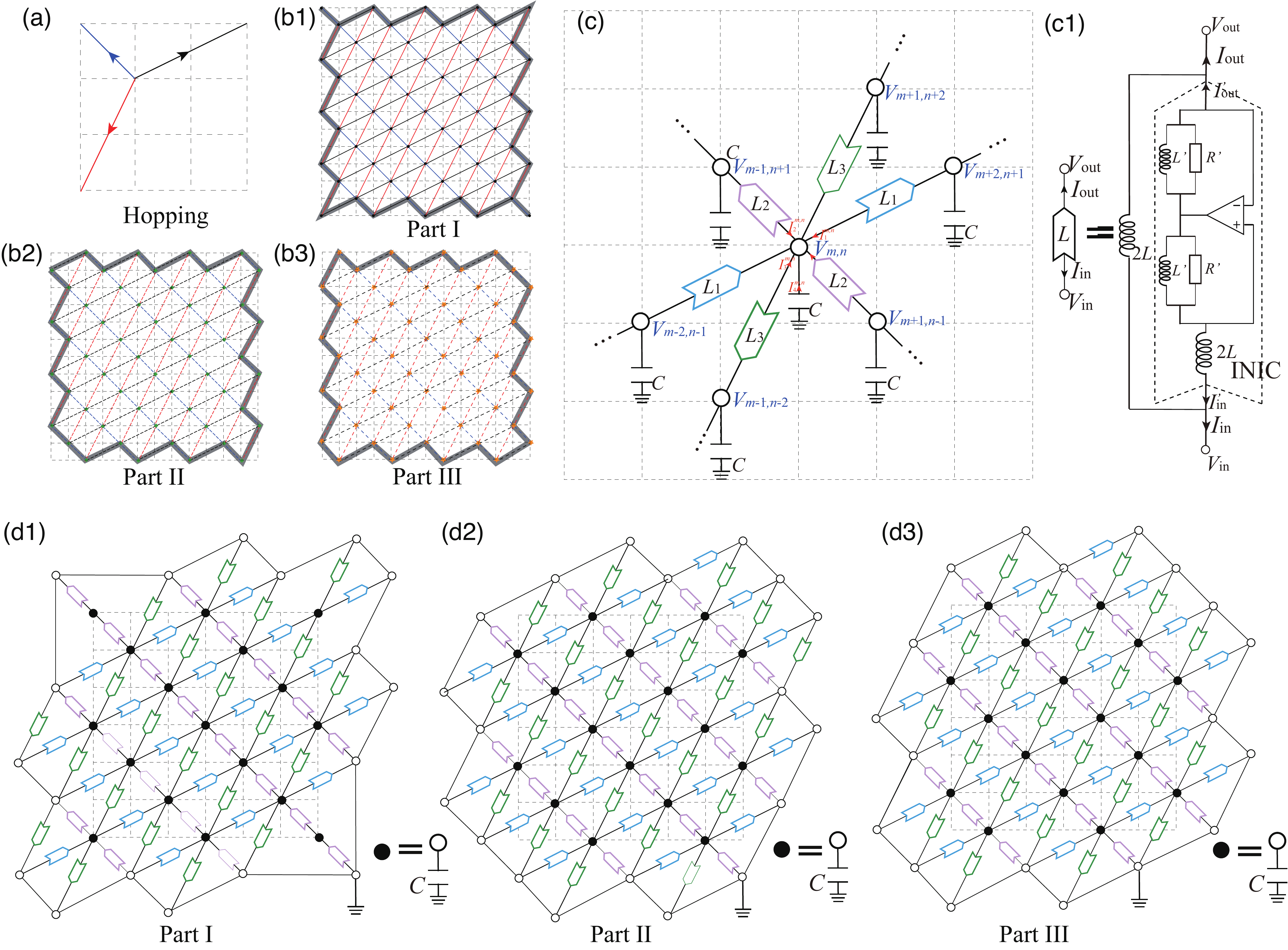}
    \par\end{centering}
    \protect\caption{\label{fig:circuit3}
(a) The three hoppings (black,blue,red) that define the model of Eq.\eqref{cceq2}. 
(b1-b3) Different double OBC boundary orientations on this model lattice gives rise to effective models with hoppings that are oriented differently (labeled as Parts I, II and III). 
(c) To realize this model with an electrical circuit, each asymmetric hopping can be realized with an INIC $L_i$, $i=1,2,3$. (c1) gives the internal make-up of an INIC, with the operation amplifier (triangle) giving rise to the hopping asymmetry. (d1-d3) Circuit realizations of the model variations (b1) to (b3).
}
\end{figure*}

\indent Any electrical circuit network can be represented by a graph whose nodes and edges correspond to the circuit junctions and connecting wires/elements. The circuit
 behavior is fundamentally described by Kirchhoff’s law. Non-Hermiticity can be introduced in a RLC electrical circuit by means of negative impedance converters with current inversion (INICs)\cite{ningyuan2015time,lee2018topolectrical,hofmann2019chiral,stegmaier2021topological,circuitliu2021non,hofmann2020reciprocal,helbig2020generalized,ezawa2019non} (Fig.~\ref{fig:circuit3}(c1)). By means of Kirchhoff's law, it can be shown that~\cite{ningyuan2015time,lee2018topolectrical,hofmann2019chiral,stegmaier2021topological,imhof2018topolectrical,helbig2019band,circuitPhysRevB.100.184202,helbig2020generalized,lee2019imaging,circuitPhysRevLett.124.046401,hofmann2020reciprocal,lenggenhager2021electric,circuitliu2021non,zou2021observation,circuitliu202001,hohmann2022observation,zhang2022anomalous,bergholtz2021exceptional,ezawa2019non} each INIC possess the reduced 2-node Laplacian

\begin{equation}
\begin{split}
  \left(
    \begin{array}{c}
      I'_{\text{in}} \\
      I'_{\text{out}} \\
    \end{array}
  \right)
  =\frac{1}{2i\omega L}
  \left(
            \begin{array}{cc}
              1& -1 \\
              1 & -1 \\
            \end{array}
          \right)
          \left(
    \begin{array}{c}
      V_{\text{in}} \\
      V_{\text{out}} \\
    \end{array}
  \right).
  \end{split}
\end{equation}
If we further connect an INIC with an inductor of inductance $2L$ in parallel, we obtain
\begin{equation}
\begin{split}
  \left(
    \begin{array}{c}
      I_{\text{in}} \\
      I_{\text{out}} \\
    \end{array}
  \right)
  =\frac{1}{i\omega L}
  \left(
            \begin{array}{cc}
              1& -1 \\
              0 & 0 \\
            \end{array}
          \right)
          \left(
    \begin{array}{c}
      V_{\text{in}} \\
      V_{\text{out}} \\
    \end{array}
  \right). 
  \end{split}
\end{equation}
for this parallel configuration pair. This INIC-inductor pair is very versatile, and can be used as the building block of arbitrary non-Hermitian circuit Laplacians. In general, circuit Laplacians $J$ are obtained via Kirchhoff rule $\bm{I} = J\bm{V}$ , where $I$ denotes the current input and the voltage $\bm{V}$ measures against ground at each node. As an initial step towards identifying circuits with tight-binding lattice models, we can get $\bm{I} = J\bm{V}$ in compact matrix form (Fig.~\ref{fig:circuit3}(c))
\begin{equation}
\begin{split}
  \left(
    \begin{array}{c}
      I_{1}^{m,n} \\
     I_{2}^{m,n} \\
     I_{3}^{m,n} \\
     I_{4}^{m,n} \\
    \end{array}
  \right)
  =
  \left(
            \begin{array}{cccc}
              -{1}/{(i\omega L_1)}& 0&0& {1}/{(i\omega L_1)}\\
              0 & -{1}/{(i\omega L_2)}& 0&{1}/{(i\omega L_2)} \\
            0&0  & -{1}/{(i\omega L_3)}&{1}/{(i\omega L_3)}\\
            0&0&0&i\omega C 
            \end{array}
          \right)
          \left(
    \begin{array}{c}
      V_{m+2,n+1}\\
      V_{m-1,n+1} \\
      V_{m-1,n-2}\\
      V_{m,n}
    \end{array}
  \right). 
  \end{split}
\end{equation}
For instance, the circuit in Fig.~\ref{fig:circuit3} (c), shown for just one unit cell, is mathematically described by a Laplacian matrix, which in momentum space under PBCs, takes the form
\begin{equation}\label{cceq5}
\begin{split}
&I(\bm{k})=J(\bm{k})V(\bm{k})\ ,\\
J(\bm{k})=\mu+&t_1\text{e}^{2ik_x+ik_y}+t_2\text{e}^{-ik_x+ik_y}+t_3\text{e}^{-ik_x-2ik_y}\ ,
\end{split}
\end{equation}
with $t_j=-1/i\omega L_j$, $j=1,2,3,$ and $\mu=i\omega C-\sum_{j=1,2,3}t_j$.
This takes the same form as a tight-binding Hamiltonian of the form (FIG.~\ref{fig:circuit3}(a))
\begin{equation}\label{cceq2}
\begin{split}
&H=\sum_{m,n}t_1|m,n\rangle\langle m+2,n+1|+t_2|m,n\rangle\langle m-1,n+1+t_3|m,n\rangle\langle m-1,n-2|\ ,
\end{split}
\end{equation}
with $H=t_1\text{e}^{2ik_x+ik_y}+t_2\text{e}^{-ik_x+ik_y}+t_3\text{e}^{-ik_x-2ik_y}=J(\bm{k})-\mu$ when expressed in momentum space.\\
\\
Notably, the system is sensitive to the boundary orientation, in that the effective lattice takes different forms under different boundary orientations, as shown in FIG.~\ref{fig:circuit3} (b1-b3), which correspond to the type I to III models discussed in this work. They correspond to their respective circuits in FIG.~\ref{fig:circuit3} (d1-d3).

\section{IV. 2D model with topological zero modes from 1D topological invariant}\label{sec4}

Here we present a nontrivial implication of the effectively 1D GBZ of our 2D model of the class $H_\text{2D,III}$. Since the GBZ construction pertains to not just the spectrum, but represents a complex analytic deformation of the Hamiltonian \emph{operator} itself (to a so-called \emph{surrogate} Hamiltonian, see Ref.~\cite{lee2020unraveling}), it means that the OBC properties of such Hamiltonians are also nontrivially modified. In particular, an OBC feature of particular interest is the presence of topological zero modes. Below, we shall see how our model, which is defined in 2D, nevertheless host topological zero modes defined by a 1D topological invariant.

Noting that the GBZ construction is completely determined by the characteristic polynomial (bulk energy dispersion), we can construct a 2-band topological Hamiltonian with a 1D constrained GBZ by writing down a model with exactly the same eigenenergy dispersion as Eq.\eqref{eqS31.5} up to a constant offset $c$ that gaps the system, i.e. \begin{equation}\label{eqS41}
 H=\left(
      \begin{array}{cc}
        0 &H_{12}\\
        H_{21}& 0 \\
      \end{array}
    \right),
    \end{equation}
with eigenenergies $\mathcal{E}_{1,2}$, which are stipulated to satisfy $\left(\mathcal{E}_{1,2}\right)^2=H_{12}H_{21}=E+c=t_1z_x^{\alpha}z_y^{a}+t_2z_x^{-\beta}z_y^{a}+t_3z_x^{-\beta}z_y^{-b}+c$, and 
\begin{equation}\label{eqS42}
\begin{split}
    H_{12}=\left(E+c\right)/Z&=\left(t_1z_x^{\alpha}z_y^{a}+t_2z_x^{-\beta}z_y^{a}+t_3z_x^{-\beta}z_y^{-b}+c\right)/Z\ ,\\
    &\quad\qquad H_{21}=Z\ ,
    \end{split}
    \end{equation}
%where the effect of constant $c$ is to get gapped system,  
where we allow $Z$ to assume either of the two forms for illustrative purposes: $Z=z^{a}_y$ or  $Z=z^{-\beta}_x$. $E$ is the energy function of Eq.\ref{eqS31.5}. %With $z_x=\text{e}^{ik_x}$, $z_y=\text{e}^{ik_y}$, Eq.\eqref{eqS41} represents a 2-band model with our desired properties. 
The gap induced by $c$ does not affect the GBZ solutions since energy degeneracies are not sensitive to a constant offset. \\

First, for the model Eq.\eqref{eqS31} with OBCs, either along one or both boundary directions, the spectrum of the GBZ hamiltonian must have a net zero winding number, even though the winding number of the off-diagonal term $H_{12}$ is dependent on the $Z$ term.  \\

With $\alpha/\beta>a/b$,  its GBZ Hamiltonian takes the form 
%\begin{equation}\label{eqS43}
%\mathbb{H}=\sum_{k,k'}E_1(k)\oplus E_2(k').
%    \end{equation}
% see Sec.~\ref{sec32} for detail. Corresponding 2-dimensional equivalence model can be
\begin{equation}\label{eqS44}
 %\mathscr{ H}
\bar H=\bar H_1\oplus \bar H_2=\left(
      \begin{array}{cc}
        0 &\bar H_{12}\\
        \bar H_{21}& 0 \\
      \end{array}
    \right)=\left(
      \begin{array}{cc}
        0 &\bar H^1_{12}\\
        \bar H^1_{21}& 0 \\
      \end{array}
    \right)\oplus\left(
      \begin{array}{cc}
        0 &\bar H^2_{12}\\
        \bar H^2_{21}& 0 \\
      \end{array}
    \right)\ ,
    \end{equation}
		\begin{equation}
\bar H^i_{12}=(\bar E_i+c)/Z_i,\quad \bar H^i_{21}=Z_i\  .
\end{equation}
with $\bar E_{i}=\bar H^i_{12}\bar H^i_{21}$, ($i=1,2$) Eq.(\ref{eqS38},\ref{eqS310}).   Both $\bar H^i_{12}$, $i=1,2$  have the same forms of $z_{x,i},z_{y,i}$: 
\begin{equation}
\begin{split}
&z_{x,1}^{\alpha+\beta} =\frac{t_2\sin((a+b) \beta k)}{t_1\sin((\alpha b-a\beta )k)}\text{e}^{i(\alpha b+\beta b)k}, \quad  z_{y,1}^{a+b} =\frac{t_3\sin(\beta b k)\sin((\alpha b- a \beta)k)}{t_2\sin( a \beta k)\sin((\alpha b+\beta b)k)}, \qquad k\in\left[-\frac{\pi}{(\alpha +\beta) b},\frac{\pi}{(\alpha +\beta) b}\right],\\
& z_{x,2}^{\alpha+\beta} =\frac{t_2\sin(\beta a k')\sin((\alpha a+ \alpha b)k')}{t_1\sin( a \alpha k')\sin((\alpha b-\beta a)k')}\ , 
\  z_{y,2}^{a+b} =\frac{t_3\sin((\alpha b-a\beta )k'))}{t_2\sin((\alpha a+a\beta )k')}\text{e}^{i(\alpha a+\alpha b)k'},\  k'\in\left[-\frac{\pi}{\alpha(a +b)},\frac{\pi}{\alpha(a +b)}\right]\ ,
 \end{split}
\end{equation}
For each GBZ sector $i$, we have $Z_i=z^a_{y,i}$ or $Z_i=z^{-\beta}_{x,i}$ depending on the choice of illustrative example.

\begin{figure}[h!]
\centering
\includegraphics[width=7in]{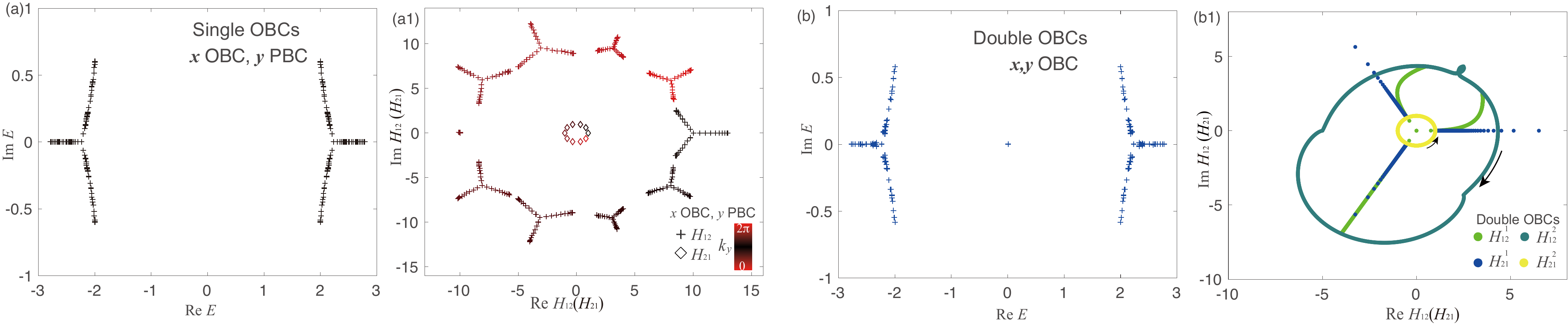}
\caption{ Energy spectrum of the 2-band model (Eq.(\ref{eqS41})) with (a) single OBCs and (b) double OBCs, showing double OBC topological modes that originate from a 1D topological invariant. We have $Z=z_y$,  $\alpha=2,\beta=1,a=1,b=2$, and the constant is set to $c=5$ to make sure that the 2-band model is gapped under PBC. In (a), there is no topological zero mode, which is corroborated by the fact that $H_{12}$ in (a1) has zero winding. In (b), the presence of a double OBC topological zero mode agrees with the fact that for each of $H^i_{12}$ and $H^i_{21}$, the sum of their windings over $i=1,2$ is nonzero. The winding of $H_{12}^2$ is $\omega(H_{12}^2)=-1=-\omega(H_{21}^2)$  and $\omega(H_{12}^1)=\omega(H_{21}^1)=0$. Note that for the single OBC (xOBC, yPBC) case, the winding is about $k_x$, not $k_y$, so the winding of $H_{12}$ and $H_{21}$ is zero. 
The plots are generated with (a) $k_y=-\pi:\pi/30:\pi$, $L=10$,  (b) with $L_x=L_y=10$. 
}\label{fig:S41}.
\end{figure}

Since we have designed this model such that it harbors nontrivial topological zero modes under double OBCs, from established results~\cite{Lee2019anatomy}, both $\bar H_{12}$ and $\bar H_{21}$ must have nonzero winding about $E=0$ (we shall suppress the GBZ sector label $i$ unless it is explicitly referred to), and these windings sum to zero. %We already know that $\bar H_{21}=Z$ has nonzero winding by its definition. 
For the off-diagonal term $\bar H_{12}$, the winding number about $E=0$
 \begin{equation}
   \begin{split}
   \omega(\bar H_{12})&=\oint\text{d}k\, \partial_k \log\bar H_{12}=\omega(\bar H^1_{12})+\omega(\bar H^2_{12})\neq 0\ ,\\
   \end{split}
\end{equation} 
is also nonzero. We emphasize that although this nonzero winding criterion for topological modes was originally formulated for a 1D system, we have now used it for the 1D effective GBZ of a \emph{physically 2D} system. Some examples of cases with topological zero modes for double OBCs but not single OBCs are given in Fig.~\ref{fig:S41}.\\%because the paths of $z_{x,1}$ and $z_{y,2}$ Eq.(\ref{eqS37},\ref{eqS39}) in the complex plane form closed loops . It is known that \cite{lee2019anatomy, yao2018edge} a non-zero winding number implies a nontrivial 1D topological invariant which protects topological zero modes - here for our OBC model in \emph{two} dimensions.\\
   
However, this model with a single OBC exhibit very different results.
If we choose $Z=z^{a}_y$,  only when the $\bm x$ PBC, $\bm y$ OBC,  the system has  topological zero modes, the winding number of  $\bar H_{12}$  is non-zero on the single OBC GBZ.  By contrast, the same system with $\bm x$ OBC, $\bm y$ PBC has no topological zero modes (Fig.~\ref{fig:S41}). With double OBCs, the 1D GBZ for the double OBCs give rise to zero winding number for $\bar H^1_{12}$, and nonzero winding number for $\bar H^2_{12}$, both summing to a nonzero total winding of $\bar H_{2}$ from Eq.~\ref{eqS44}. 

If we choose $Z=z^{-\beta}_x$ instead, the opposite is the case, the system with $\bm x$ OBC, $\bm y$ PBC have topological zero modes, the system with $\bm x$ PBC, $\bm y$ OBC does not have.
The winding number of  $\bar H^1_{12}$  is non-zero, the winding number of  $\bar H^2_{12}$  is zero in the GBZ Hamiltonian with double OBC. In an analogous way, the 2D model  with double OBC also  has  topological zero modes which is  contributed by $\bar H_{1}(k)$ from Eq.\ref{eqS44}.\\

\begin{figure}[h!]
\centering
\includegraphics[width=3.5in]{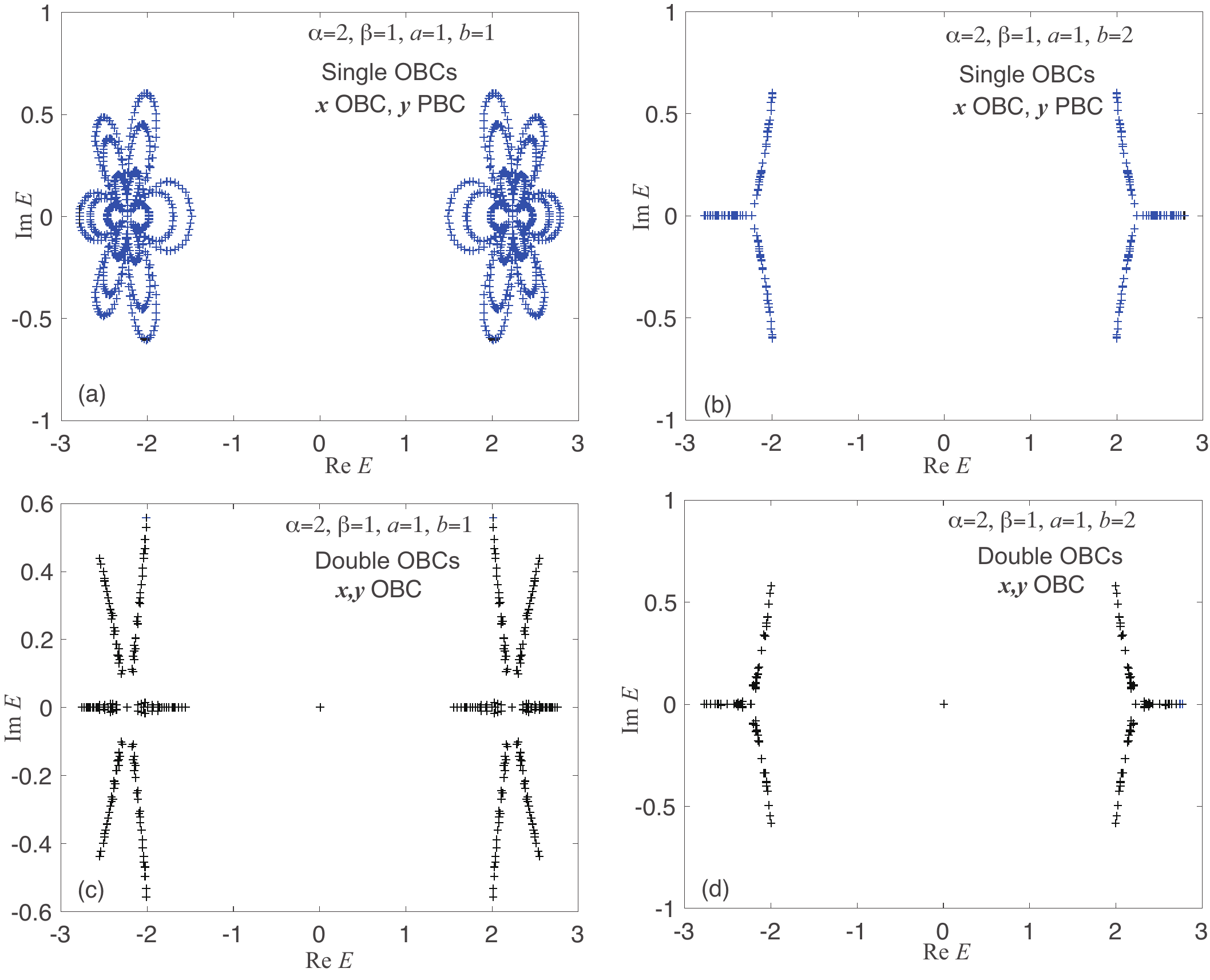}
\caption{ Energy spectrum of the 2-band model (Eq.(\ref{eqS41})) with $Z=z_y$, under single and double open boundary conditions. The Left and Right columns denote  different sets of parameters (Marked in the figure). Evidently, under double OBCs, we observe topological zero modes that originate from a 1D topological invariant, indicating that the GBZ is an effective 1D GBZ, not a 2D GBZ. The constant offset in Eq.(\ref{eqS41}) is set to $c=5$ to make sure that the 2-band model is gapped under PBCs. The system is numerically discretized such that (a,b) $k_y=-\pi:\pi/30:\pi$, $L_x=10$, (c,d) $L_x=L_y=10$.}
\label{fig:S43}
\end{figure}

We demonstrate the above arguments for two illustrative cases of $\alpha,\beta,a,b$ parameters. %, the detailed information can be found in Sec.~\ref{sec33},  
We use $Z=z_y$ and examine the energy spectrum of our 2D two-band model Eq.\ref{eqS41} with different boundary conditions.  The observed numerical results prove the correctness of the above theoretically established results, in FIG.~\ref{fig:S41},\ref{fig:S43}. Specifically, it demonstrates that the construction of the effective 1D GBZ Hamiltonian can correctly predict the topological zero mode, despite the system being physically a 2D system.

\begin{comment}\begin{figure}[h!]
%This fig is the same as the next fig(a,c)!
\centering
\includegraphics[width=4.5in]{gfigures42.pdf}
\caption{ Energy spectrum of the 2-band model (Eq.(\ref{eqS41})) with different boundary conditions and $Z=z_y$ with a different set of parameters ($\alpha=2,\beta=1,a=1,b=1$), also showing double OBC topological modes that originate from a 1D topological invariant.  The constant is set to $c=5$ to make sure that the 2-band model is gapped under PBC, 
(a) with $k_y=-\pi:\pi/30:\pi$, $L=10$, (b)  with $L_x=L_y=10$. \\
}
\label{fig:S42}
\end{figure}  
\end{comment}

\section{V. Generalizations to higher dimensions }\label{sec5}
Our approach for getting the correct 2D GBZs as well as their 1D constrained GBZs, if necessary, can be generalized to higher dimensions. In generic number of dimensions, the GBZ will first be symmetrically constructed in terms of relations between complex $z_x,z_y,z_z,...$ and real $k_1,k_2,k_3,...$. If there exist nonzero spectral windings, they will also have to be constrained to a lower dimension such that the all spectral windings in the constrained GBZ become zero. 

Specializing to 3 dimensions for definiteness (higher dimension cases can be analogously written down), suppose we have
  \begin{equation}\label{eqS51}
H_\text{3D}=\sum_{m,n,l} \sum_{a,b,c}t_{abc}|m,n,l\rangle\langle m+a,n+b,l+c|\ ,
\end{equation}
with hopping amplitudes $t_{abc}$ and $a,b,c$ generic integers which represent hopping directions and distances. In the same way as before,  we insert the ansatz $|\psi\rangle\propto\sum_{m,n.l}\psi_{mnl}|m,n,l\rangle$ with $\psi_{m,n,l}=z_x^mz_y^nz_z^l$ into the bulk equations to obtain
\begin{equation}\label{eqS52}
\begin{split}
E(z_x,z_y,z_z)=\sum_{a,b,c}t_{abc}z_x^az_y^bz_z^c\ .
\end{split}
\end{equation}
Like before, there are generally many sets of $(z_x,z_y,z_z)$ that corresponds to a particular energy $\bar E=E(z_x,z_y,z_z)$.\\

To incorporate the boundary conditions, we can express a wavefunction $|\Psi\rangle$ with eigenenergy $\bar E =E(z_x,z_y, z_z)$ as different 1D chains in the x,y or z-directions, with two other momenta as parameters:
    \begin{equation}\label{eqS53}
    \begin{split}
|\Psi\rangle=|\Psi^1\rangle&=|\Psi^2\rangle=|\Psi^3\rangle=\ \propto \sum_{m,n,l}\Psi_{mnl}|m,n,l\rangle\ ,\\
&\Psi_{mnl}=\Psi^1_{mnl}=\Psi^2_{mnl}=\Psi^3_{mnl}\ ,\\
\Psi^1_{mn}&=\sum_{z_y,z_z} \left(\sum_{j}f_{z_{x,j}}z_{x,j}^m\right)g_{z_y}  z_y^nh_{z_z}z_z^l\ ,\\
\Psi^2_{mnl}&=\sum_{z_x,z_z} f_{z_x} z_x^m \left(\sum_{j'}{g}_{z_{y,j'}}z_{y,j'}^n\right)h_{z_z}z_z^l\ ,\\
\Psi^3_{mnl}&=\sum_{z_x,z_y} f_{z_x} z_x^m g_{z_y} z_y^n \left(\sum_{j''}h_{z_{z,j''}}z_{z,j''}^l\right).\\
\end{split}
\end{equation}
Here $|\Psi^1\rangle,\ |\Psi^2\rangle,\ |\Psi^3\rangle$ are different direction decompositions of the same wave function $|\Psi\rangle$,  and the first summation in the equation $\Psi^{j}_{mnl}$, $j=1,2,3$ means we must consider all the sets $z_x,z_y,z_z$ which have same energy $\bar E$.\\

Taking the  different manifestations of the wave function $|\Psi\rangle$ into the boundary conditions along different directions (i.e.,${\bm x, \bm y,\bm z}$ direction), like boundary conditions along ${\bm x}$ direction, we have  $\Psi_{m'{n}}=\Psi^1_{m'{n}}=0 $ with special value $m'$ and any values $n$, like `1D' system with paramters $z_y,z_z$.
Thus, for a particular fixed energy, we can in general write down a relation between $z_x$ and $z_y,z_z$, i.e.  
\begin{equation}\label{eqS54}
\begin{split}
\mathscr{F}_1(z_x,z_y,z_z,k_1)=0\ ,
\end{split}
\end{equation}
with real $k_1$ ``momentum'' parametrizing one direction in the GBZ. 
In the same way, we can get other  relations along ${\bm y},\bm z$ directions
\begin{equation}\label{eqS55}
\begin{split}
\mathscr{F}_2(z_x,z_y,z_z,k_2)=0\ ,
\end{split}
\end{equation}
\begin{equation}\label{eqS56}
\begin{split}
\mathscr{F}_3(z_x,z_y,z_z,k_3)=0\ ,
\end{split}
\end{equation}
 with real $k_2,k_3$. By construction, the $z_x,z_y,z_y$ satisfy all the boundary conditions. %thus $z_x,z_y,z_z$ can be obtain by Eq.~\ref{eqS54},~\ref{eqS55},\ref{eqS56}, that is, 
Hence we have 3 unknowns $z_x,z_y,z_z$ with 3 independent relational equation (Eq.~\ref{eqS54},~\ref{eqS55},\ref{eqS56}), which can in principle be simultaneously solved to yield relations
 \begin{equation}\label{eqS57}
\begin{split}
z_i=\mathcal{F}_i(k_1,k_2,k_3)\ ,
\end{split}
\end{equation}
with $i=x,y,z$. Taking $z_x,z_y,z_z$ Eq.\eqref{eqS57} into the energy equation Eq.\eqref{eqS52}, we can equivalently get the GBZ-predicted OBC energy 
 \begin{equation}\label{eqS58}
\begin{split}
E(z_x,z_y,z_z)=\bar E(k_1,k_2,k_3)\ .
\end{split}
\end{equation}
If the winding number of $\bar E(k_1,k_2,k_3)$  Eq.\eqref{eqS52} is zero for all closed loops, the resultant GBZ of the 3D system will just be given by 
\begin{equation}\label{eqS59}
\begin{split}
\text{GBZ}_\text{3D}=\{z_x,z_y,z_z|z_i=\mathcal{F}_i(k_1,k_2,k_3),i=x,y,z\}\ .
\end{split}
\end{equation}

And if the spectral winding number of $E(k_1,k_2,k_3)$  Eq.\eqref{eqS58} is non-zero, we must determine what constraints we need on the parameters $k_1,k_2$ and $k_3$ such that the spectral winding of all possible paths is zero, and that the GBZs defined by $z_x,z_y,z_z$ are periodic as we vary any of the remaining $k_i$. The explicit construction of such constraints to obtain the correct GBZ depends on the model, and will be the subject of future work - in the next subsection, we give a minimal example in 3D. In general, the constraints can either reduce the correct GBZ to a 2D subspace, or even a 1D subspace.
 
\subsubsection{Illustrative simple 3D model} 
\begin{figure}[htb]
\centering
\includegraphics[width=5in]{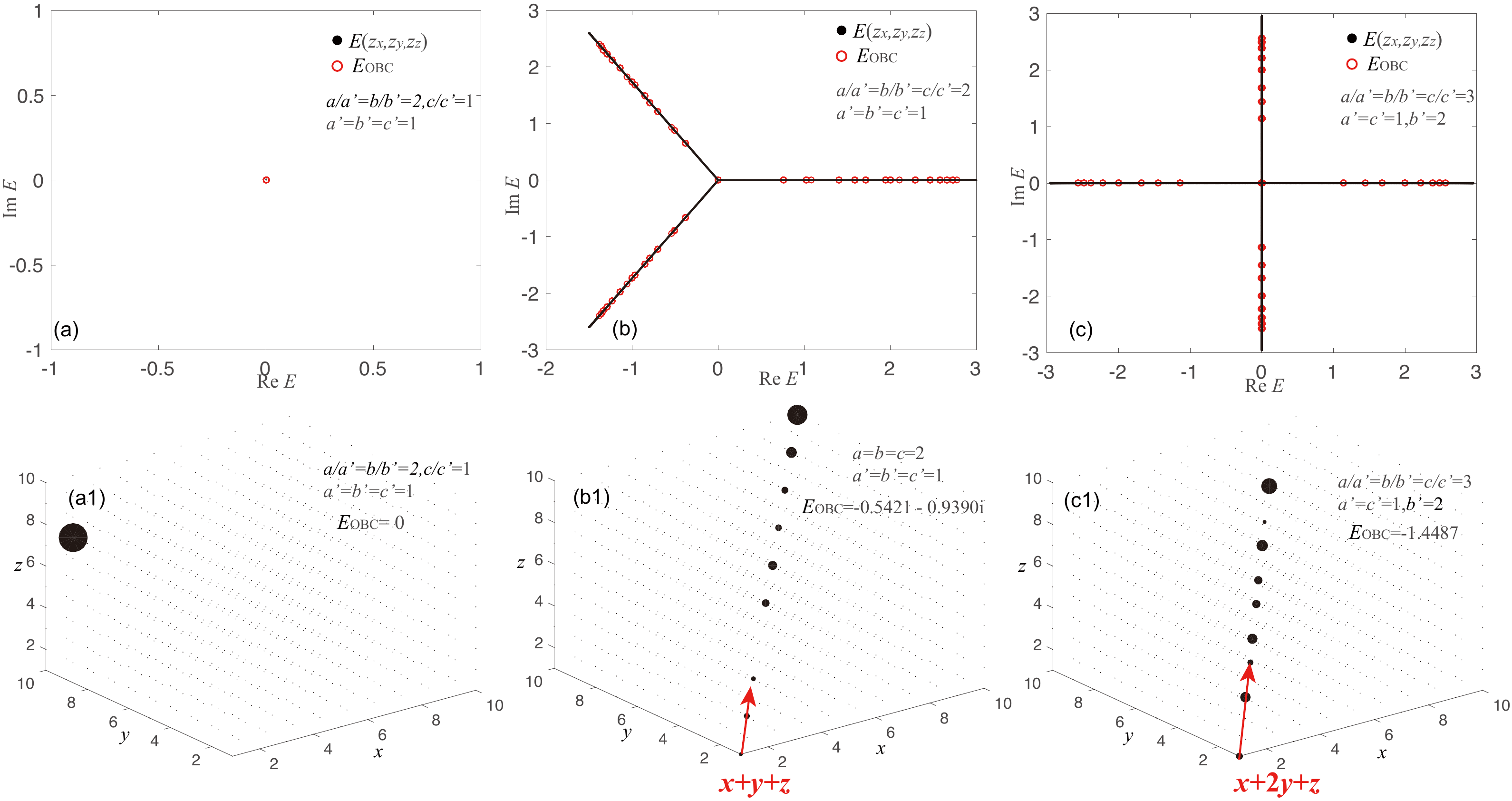}
\caption{Spectra and illustrative eigenstates for $H_\text{3D}$  with different $a,b,c,a',b',c'$ case. (a--c) Excellent agreement between the GBZ eigenenergies $E(z_x,z_y)$ of GBZ and OBC eigenenergies $E_{\text{OBC}}$ of Eq.\eqref{eqS510}.   (a1---c1) Spatial profiles for the $E_{\text{OBC}}$ eigenstate of a few other illustrative cases, clearly showing that the skin states are aligned along the ($a\bm x+b\bm y+c\bm z$) direction (b1,c1). The model parameters are $L_x=L_y=L_z=10$, $t_{abc}=1,t_{a'b'c'}=2$  and the specific values of $a,b,c,a',b',c'$ and  $E_{\text{OBC}}$ are indicated in the figure.}
\label{fig:S51}
\end{figure}

For a minimal example of the GBZ construction of a 3D model, we consider 2 hopping terms in 3D space i.e.\begin{equation}\label{eqS510}
H_\text{3D}=\sum_{m,n,l} t_{abc}|m,n,l\rangle\langle m+a,n+b,l+c|+t_{a'b'c'}|m,n,l\rangle\langle m-a',n-b',l-c'|\ ,
\end{equation}
with real $a,b,c,a',b',c'$. Assuming the wave function ansatz $|\psi\rangle\propto\sum_{m,n.l}\psi_{mnl}|m,n,l\rangle$ with $\psi_{m,n,l}=z_x^mz_y^nz_z^l$, we arrive at two possible cases. In the first case, $a/a'=b/b'=c/c'$ and previous derivations can be straightforwardly generalized to give 
 \begin{equation}\label{eqS511}
\begin{split}
&\text{GBZ}_\text{3D}=\left\{\left. z_x^{a+a'}z_y^{b+b'}z_z^{c+c'}=\frac{t_{a'b'c'}\sin(a'k)}{t_{abc}\sin (ak)}\text{e}^{i(a+a')k}\right|k\in \left(-\frac{\pi}{a+a'}\ ,\frac{\pi}{a+a'}\right] \right \}\ ,\\
&\qquad E(z_x,z_y,z_z)=E(k)=\left(\frac{t^{a'}_{abc}t_{a'b'c'}^{a}}{(\sin (a k))^{a}(\sin (a' k))^{a'}}\right)^{\frac{1}{a+a'}} \sin \left((a+a')k\right) \text{e}^{\frac{2i\pi  av}{a+a'}}\ ,
\end{split}
\end{equation}
with $E(z_x,z_y,z_z)=t_{abc}z_x^{a}z_y^bz_z^c+t_{a'b'c'}z_x^{-a'}z_y^{-b'}z_z^{-c'}$, as shown in FIG.~\ref{fig:S51} (b,b1,c,c1).  In this model Eq.\eqref{eqS511} with  $a/a'=b/b'=c/c'$, the GBZ is only determined by a condition on the combination $z_x^{a+a'}z_y^{b+b'}z_z^{c+c'}$  rather than $z_x$, $z_y$ and $z_z$ separately, and is akin to a 1D model  along the $a\bm x+b\bm y+c\bm z$ direction, which is  consistent with the results of numerical diagonalization, as in FIG.~\ref{fig:S21}(a1---a3). \\

In the other case where $a/a'=b/b'=c/c'$ does not hold, there is non-Bloch collapse due to uncompensated hoppings in certain directions, and the GBZ and energy are simply given by
 \begin{equation}\label{eqS512}
\begin{split}
\text{GBZ}&=\left\{z_x^{a+a'}z_y^{b+b'}z_z^{c+c'}=-1 \right\}\ ,\\
&\qquad E(z_x,z_y,z_z)=0\ ,
\end{split}
\end{equation}
as shown in FIG.~\ref{fig:S51} (a,a1). In both of these cases, the GBZ spectrum have zero winding number, and further dimensional reduction is not necessary.\\

In general, in arbitrarily high dimensional systems, the number of independent $z_i$ is always the same as the  number of independent relations $\mathcal{F}_i$, with zero spectral winding either automatically satisfied or fulfilled via dimensional reduction to a lower-dimensional GBZ. 

%\subsubsection{} 
%A unique feature of non-Hermitian systems is the skin effect, which is the extreme sensitivity to the boundary conditions. The spectrum between OBC and PBC can be totally distinct.It is well known that the spectrum of  periodic crystal can be characterized by the Bloch wave.The effective Brillouin zone (BZ), which is a generalization of Brillouin Zone (BZ) under the OBC in both Hermitian and non-Hermitian systems, can encode OBC Hamiltonian. The Hamiltonian and boundary conditions decide  effective Brillouin zone (BZ) that often has to be of a lower dimensionality than the physical lattice, in order to be consistent with vanishing winding numbers, as shown in Fig.\ref{fig:S61}.}

  \begin{figure}[htb]
\centering
\includegraphics[width=5in]{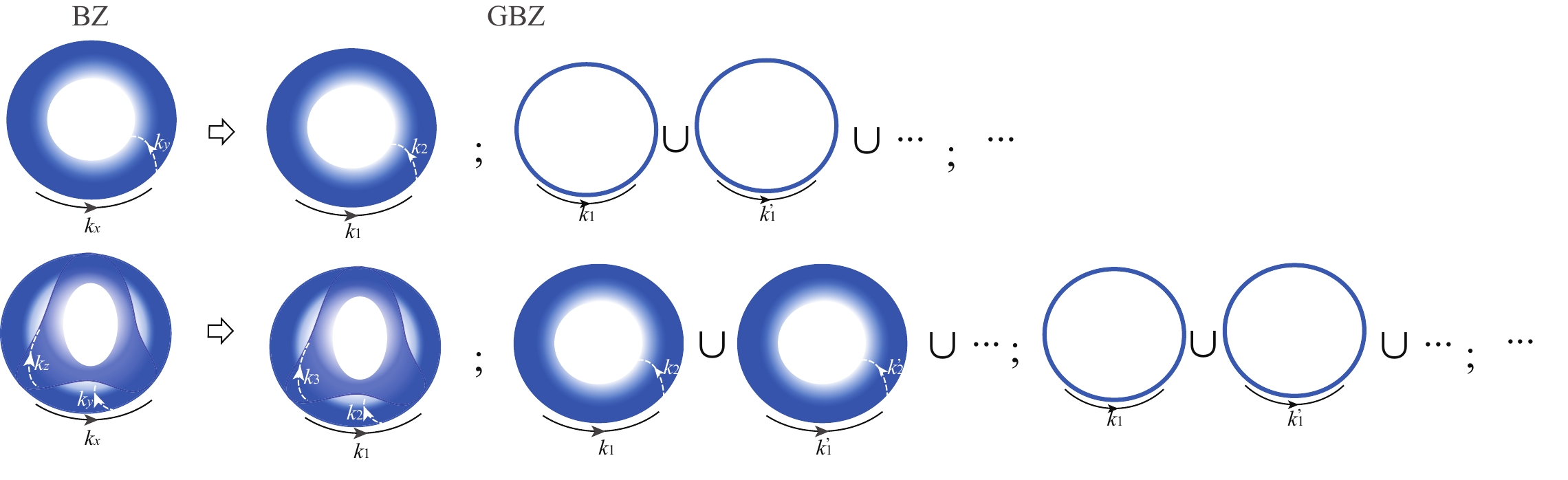}
\caption{Schematic illustration of effective BZ for higher dimensional systems.  (Top) In 2D lattices that are Hermitian or ``unentangled'', the BZ is simply a 2D torus; yet for ``entangled'' non-Hermitian cases, the effective BZ becomes the union of one or more 1D tori. (Bottom) Similarly, 3D lattices that are Hermitian or ``unentangled'' have 3D tori as their BZ, but for ``entangled'' non-Hermitian cases, the effective BZ is in general a union of various 1D and 2D tori. }
\label{fig:S61}
\end{figure} 

\end{document}